\documentclass[10pt]{article}

\usepackage{epsfig}
\usepackage{amsmath}
\usepackage{amssymb}
\usepackage{color}
\usepackage{url}

\usepackage{verbatim} 
\usepackage{epstopdf} 
\usepackage{graphicx} 
\usepackage[font=small]{caption} 

\usepackage{afterpage}

\usepackage{caption}
\usepackage{subcaption}
\usepackage{multicol}

\usepackage{multirow}

\usepackage{slashbox}

\renewcommand{\arraystretch}{1.2}

\begin{document}

\title{ Classification of Orbits in Poincar\'e Maps using Machine Learning}

\author{%
Chandrika Kamath\\[0.5em]
{\small\begin{minipage}{\linewidth}\begin{center}
\begin{tabular}{c}
Lawrence Livermore National Laboratory \\
7000 East Avenue, Livermore, CA 94551, USA\\
\url{kamath2@llnl.gov}\\
\hspace*{0.8in}
\end{tabular}
\end{center}\end{minipage}}
}

\date{April 17, 2022}
\maketitle

\begin{abstract}

  Poincar\'e plots, also called Poincar\'e maps, are used by plasma
  physicists to understand the behavior of magnetically confined
  plasma in numerical simulations of a tokamak. These plots are
  created by the intersection of field lines with a two-dimensional
  poloidal plane that is perpendicular to the axis of the torus
  representing the tokamak. A plot is composed of multiple orbits,
  each created by a different field line as it goes around the torus.
  Each orbit can have one of four distinct shapes, or classes, that indicate
  changes in the topology of the magnetic fields confining the plasma.
  Given the $(x,y)$ coordinates of the points that form an orbit, the
  analysis task is to assign a class to the orbit, a task that appears
  ideally suited for a machine learning approach.  In this paper, we
  describe how we overcame two major challenges in solving this
  problem - creating a high-quality training set, with few mislabeled
  orbits, and converting the coordinates of the points into features
  that are discriminating, despite the variation within the orbits of
  a class and the apparent similarities between orbits of different
  classes.  Our automated approach is not only more objective and
  accurate than visual classification, but is also less tedious,
  making it easier for plasma physicists to analyze the topology of
  magnetic fields from numerical simulations of the tokamak.

\end{abstract}

%
\section{Introduction}
%

The quest for low-cost fusion power has led to the construction of
experimental devices such as the DIII-D\cite{diiid09:web},  an
operational device for conducting magnetic fusion research, and
ITER~\cite{iter09:web}, an international project to help
make the transition from studies of plasma physics to
electricity-generating fusion power plants. These devices, called
tokamaks, use magnetic fields to confine the fusion fuel in the form
of a plasma, enabling physicists to perform experiments to determine
the best shape for the hot reacting plasma and the magnetic fields
necessary to hold it in place. To complement the experiments, computer
simulations are used to gain an understanding of the complex physics of the
plasmas, design new reactors, and select the parameters to be used in
experiments. Data from both the experiments and the simulations are
analyzed to provide the insights that will contribute to achieving the goal
of fusion power.
\begin{figure}[!htb]
\centering
\setlength\tabcolsep{1pt}
\vspace{-0.5cm}
\begin{tabular}{cc}
\raisebox{-0.5\height}{\includegraphics[trim = 0.0cm 0cm 3.0cm 17.0cm, clip = true, width=0.5\textwidth]{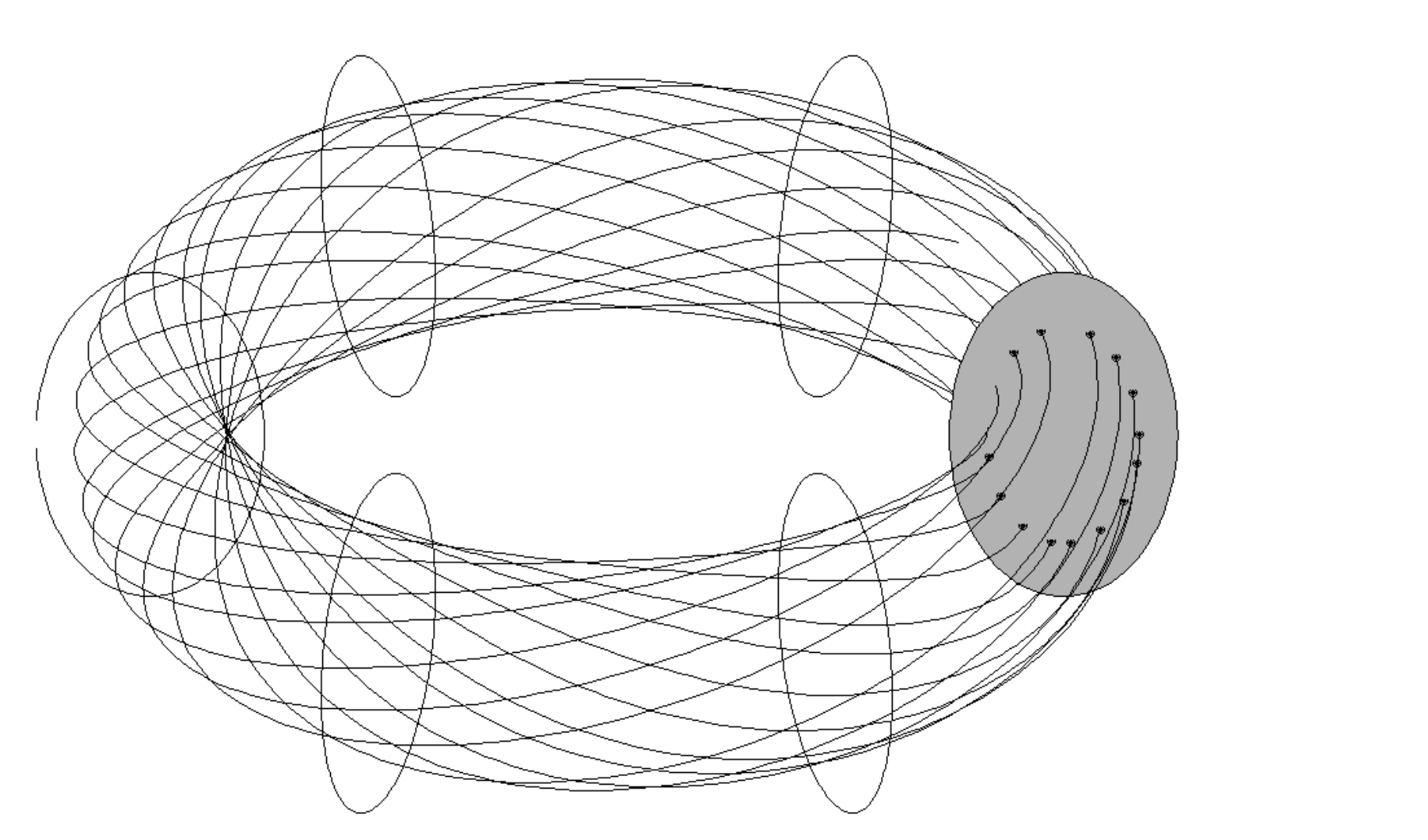}} &
\raisebox{-0.5\height}{\includegraphics[trim = 3.0cm 0cm 3.0cm 0.0cm, clip = true,width=0.4\textwidth]{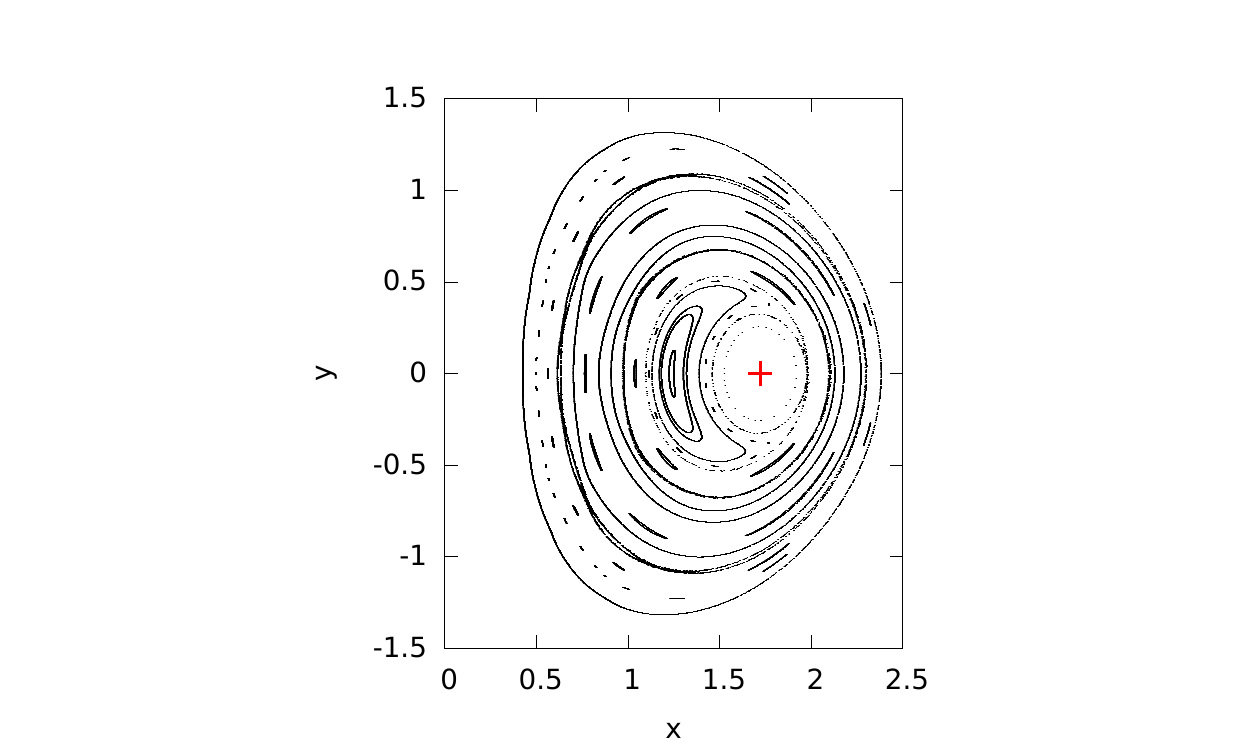}} \\
(a) & (b) \\
\end{tabular}
\vspace{-0.35cm}
\caption{ (a) Schematic view of a tokamak, in the form of a torus,
  showing the generation of the Poincare plot.  Six poloidal planes
  are shown as ellipses. An orbit, shown as points on the grey
  ellipse, is formed by the intersections of a field line with the
  ellipse as the line is traced around the torus. (b) A collection of
  orbits in a poloidal plane. The magnetic axis is indicated by a red
  cross.}
\label{fig:pplot_schematic}
\end{figure}

\begin{figure}[!t]
\centering
\begin{tabular}{cccc}
\includegraphics[trim = 3.5cm 0cm 3.9cm 0.0cm, clip = true,width=0.2\textwidth]{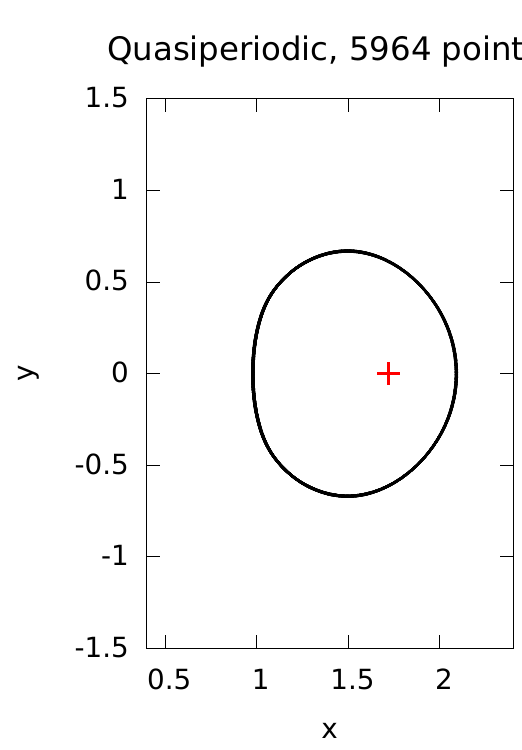} &
\includegraphics[trim = 3.5cm 0cm 3.9cm 0.0cm, clip = true,width=0.2\textwidth]{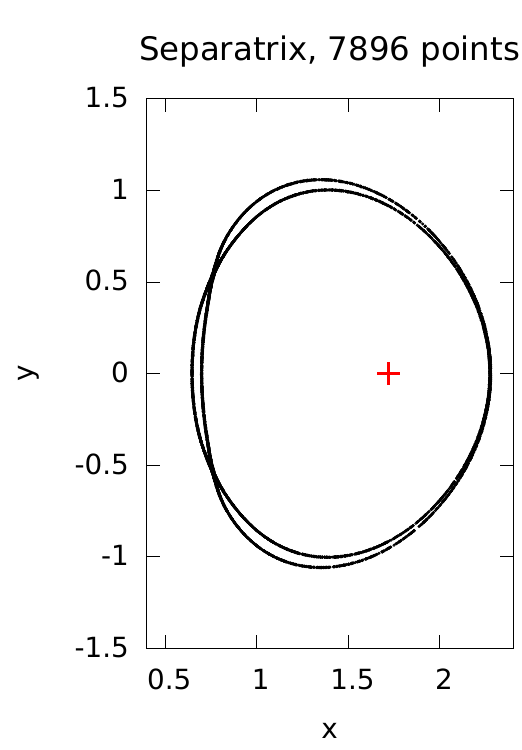} &
\includegraphics[trim = 3.5cm 0cm 3.8cm 0.0cm, clip = true,width=0.205\textwidth]{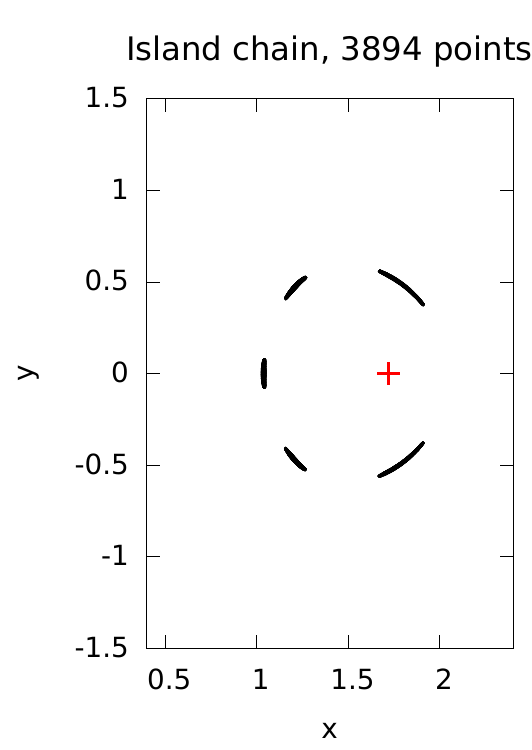} &
\includegraphics[trim = 3.5cm 0cm 3.9cm 0.0cm, clip = true,width=0.2\textwidth]{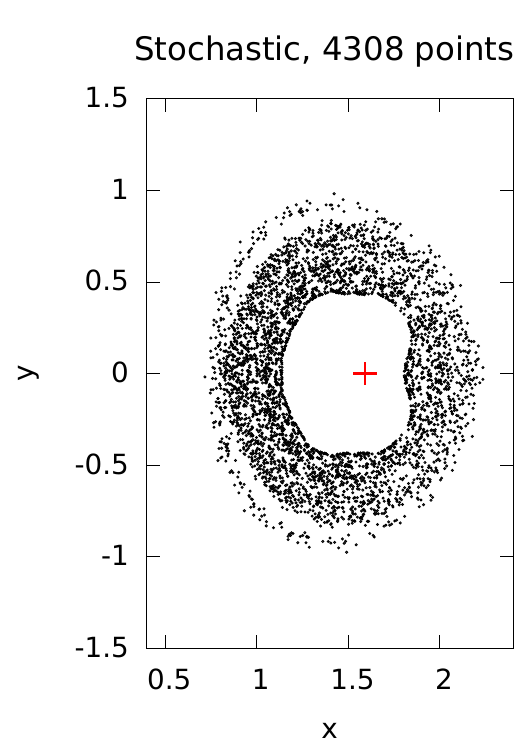} \\
(a) & (b) & (c) & (d) \\
\end{tabular}
\caption{Sample orbits illustrating the four classes - (a) a
  quasiperiodic orbit; (b) a separatrix orbit with three lobes, where
  a lobe is the region between the 'X'-points where the two curves of
  the separatrix cross; (c) an island orbit with five islands; and (d)
  a stochastic orbit. The number of points in the orbits range from
  4000 to 8000.} 
\label{fig:orbits}
\end{figure}

In this paper, we focus on a specific analysis problem that arises in
both simulation and experimental data, namely, the classification of
orbits in a Poincar\'e map, also called a Poincar\'e plot. These
two-dimensional plots are obtained for planes, called poloidal planes,
which intersect the torus-shaped tokamak perpendicular to the magnetic
axis, as shown in Figure~\ref{fig:pplot_schematic}(a). A plot consists
of several orbits, each composed of a number of points
(Figure~\ref{fig:pplot_schematic}(b)). For a given orbit, these points
are the intersections of a field line (the solid lines in
Figure~\ref{fig:pplot_schematic}(a)) with a poloidal plane, as the
field line is followed around the torus.  There are four distinct
shapes traced out by these points, leading to four classes of orbits:
quasi-periodic, separatrix, island chain, and stochastic, as shown in
Figure~\ref{fig:orbits}. In some cases, the orbit shows its
distinctive shape with just a few points, corresponding to the first
few intersections of the field line with the poloidal plane. In other
cases, an orbit may appear to be of one class, say island chain,
initially, but become a separatrix as additional points are added to
the orbit. When the data are generated by a computer simulation,
floating point errors can build up as the field lines are traced
around the torus, resulting in noise in the location of the later
points.

These Poincar\'e plots provide a convenient way to diagnose, at a
glance, changes in the topology of the magnetic fields confining the
plasma over time. The goal is to avoid undesirable changes that would result in
the hot core of the plasma escaping to mix with the cooler outer
regions.  Whether the hot plasma is confined or not is reflected in
the shape of the orbits: quasi-periodic orbits are indicative of
nested magnetic surfaces which provide good confinement; a separatrix
indicates reconnection, or rapid changes in the topology; while
islands and stochastic regions indicate progressively worse
confinement.

Classifying the orbits visually can be tedious, error-prone, and
subjective. When the number of orbits is large, as a result of many
simulations run over many time steps, it becomes clear that there is a
need for an automated approach. An obvious solution is to create a
training set by extracting features representing each orbit, followed
by a machine learning classifier for prediction. However, the
implementation of this solution is not straight-forward due to
challenges in labeling the orbits and in identifying relevant
features for orbits of different classes
(Section~\ref{sec:challenges}). In this paper, we describe how we
address these challenges (Section~\ref{sec:approach}), and discuss the
results of our experiments with i) decision trees for predicting the class of
an orbit, and ii) feature selection for understanding which features are
more relevant in discriminating among the classes
(Section~\ref{sec:results}). We conclude the paper with a summary of
our work in Section~\ref{sec:conc}.

Our contributions in this paper are as follows: given this rather
unusual data set, where an instance, or orbit, is a collection of
points in two-dimensions, we show how we can transform the $(x,y)$
coordinates of the points into a representation of the visual
structure of the orbit. Using simple, but interpretable,
classifiers, we show how we can generate a high-quality training data
set by refining both the assignment of correct class labels to the orbits and
the representative features extracted for each orbit. This approach
allows us to capture the subtle differences between orbits of
different classes that appear similar when viewed as a whole, as well
as the similarities within orbits of a class despite the variation in
their shape. As a result, we achieve higher accuracy than that
obtained using visual classification.  By automating the
step of classification of the orbits in a Poincar\'e plot, we provide
plasma physicists improved capability to understand their simulations.

%
\section{Challenges to the analysis}
\label{sec:challenges}
%

At first glance, this problem of classification of the orbits appears
relatively straightforward, and one that can be solved easily as the
four orbits shown in Figure~\ref{fig:orbits} have distinctive
characteristics. The quasiperiodic orbit appears as points on a single
closed curve; the separatrix appears as two intertwined curves; the
island chain has angular gaps; and there is no structure to the points
in a stochastic orbit. Therefore, by assigning a class label to each
orbit and representing it with a suitable set of features, we can
create a training data set that we could use to build a machine
learning model for classification of orbits.  However, a closer
inspection of the data indicates two main challenges, which we
discuss in detail next.

%
\subsection{Generating a correctly-labeled training set}
\label{sec:generate_labels}
%

To generate the training set, we started by visually assigning one of
the four class labels to each orbit.  Further, since we had a limited
set of orbits (four sets of 66 orbits each), many with several
thousands of points, we could create a series of orbits from each
orbit by using only the initial intersections of the field line with
the poloidal plane.  Thus, if an orbit of a specific class was defined
by $m$ points, $m > 1000$, we created separate {\it derived} orbits by
using the first $1000, 1500, 2000, ...$ points and assigned them
the same class.

\begin{figure}[!htb]
\centering
\setlength\tabcolsep{1pt}
\begin{tabular}{cccc}
\includegraphics[trim = 3.6cm 0cm 3.6cm 0.0cm, clip = true,width=0.20\textwidth]{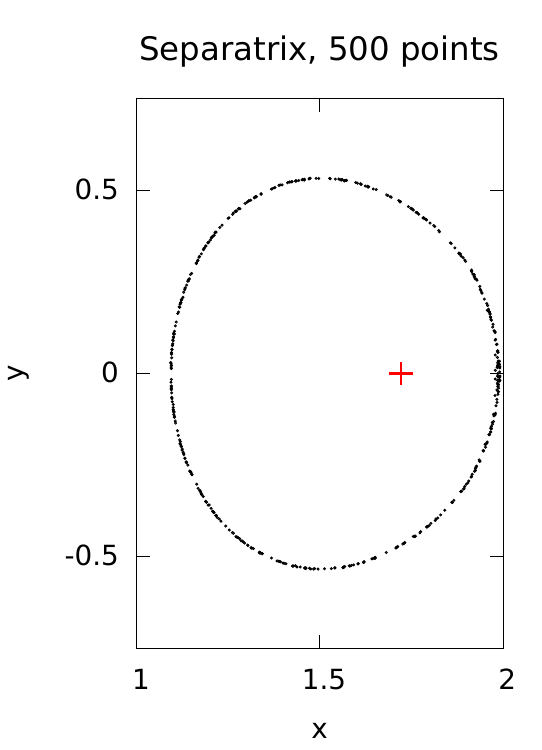} &
\includegraphics[trim = 3.6cm 0cm 3.6cm 0.0cm, clip = true,width=0.20\textwidth]{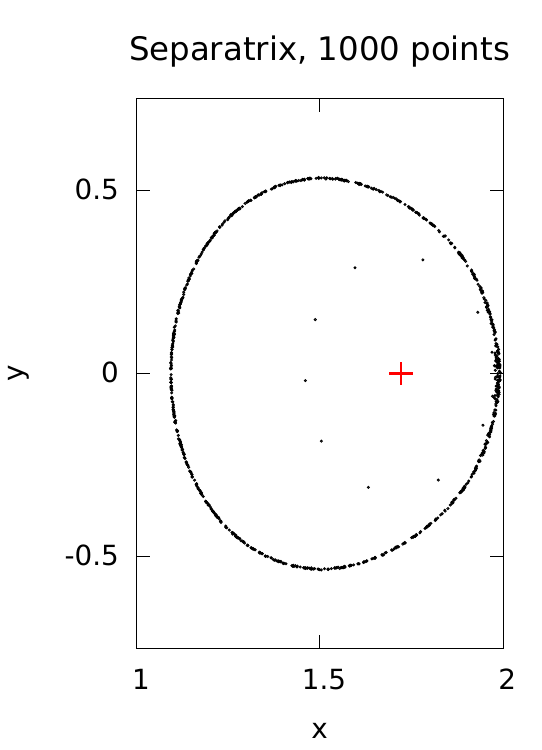} &
\includegraphics[trim = 3.6cm 0cm 3.6cm 0.0cm, clip = true,width=0.20\textwidth]{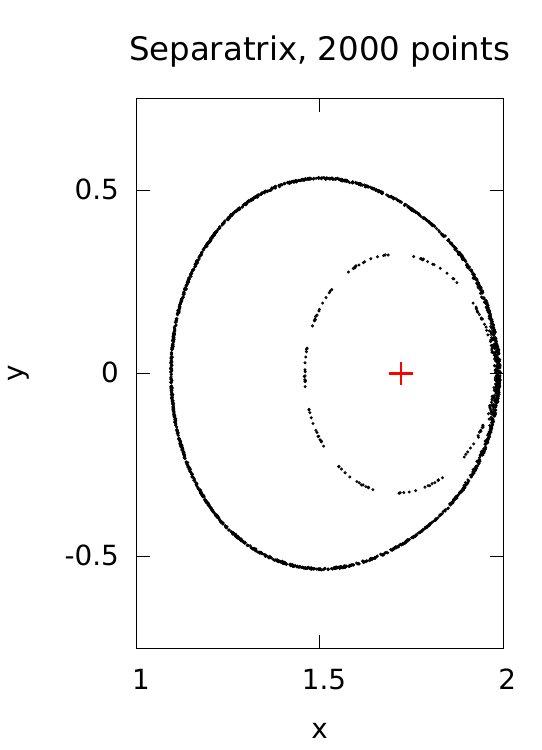} &
\includegraphics[trim = 3.6cm 0cm 3.6cm 0.0cm, clip = true,width=0.20\textwidth]{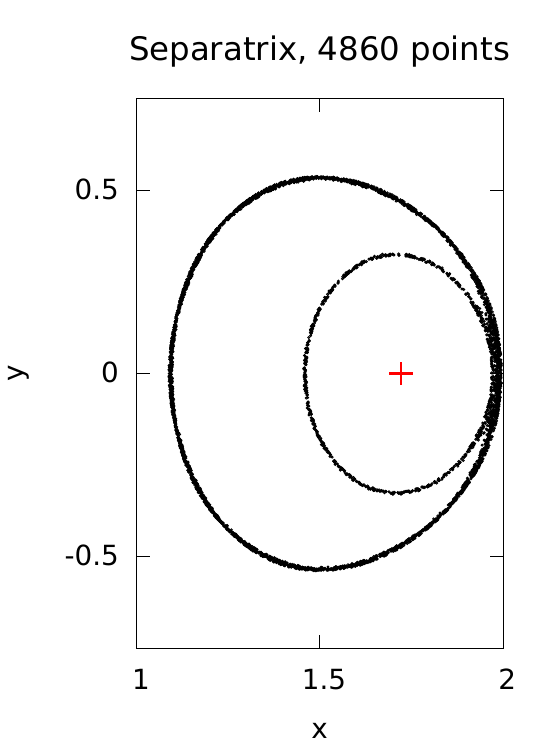} \\
(a) & (b) & (c) & (d) \\
\end{tabular}
\vspace{-0.2cm}
\caption{Views of a separatrix orbit as the number of points is
  increased from (a) 500, showing a quasiperiodic to (b) 1000 points,
  which is a separatrix with the inner curve made up of very few
  points, to a clear separatrix at (c) 2000 and finally, (d) 4860
  points.  }
\label{fig:variation_1}
\end{figure}

\begin{figure}[!htb]
\centering
\setlength\tabcolsep{1pt}
\begin{tabular}{cccc}
\includegraphics[trim = 3.1cm 0cm 3.0cm 0.0cm, clip = true,width=0.20\textwidth]{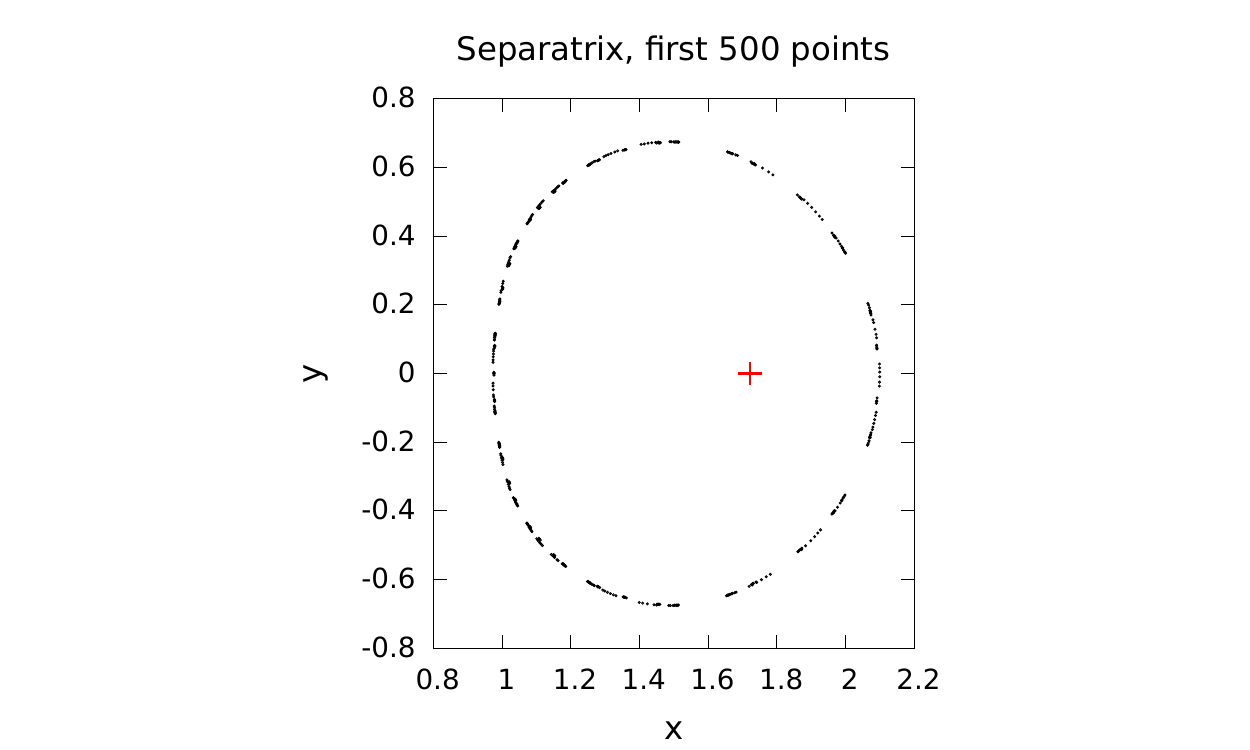} &
\includegraphics[trim = 3.1cm 0cm 3.0cm 0.0cm, clip = true,width=0.20\textwidth]{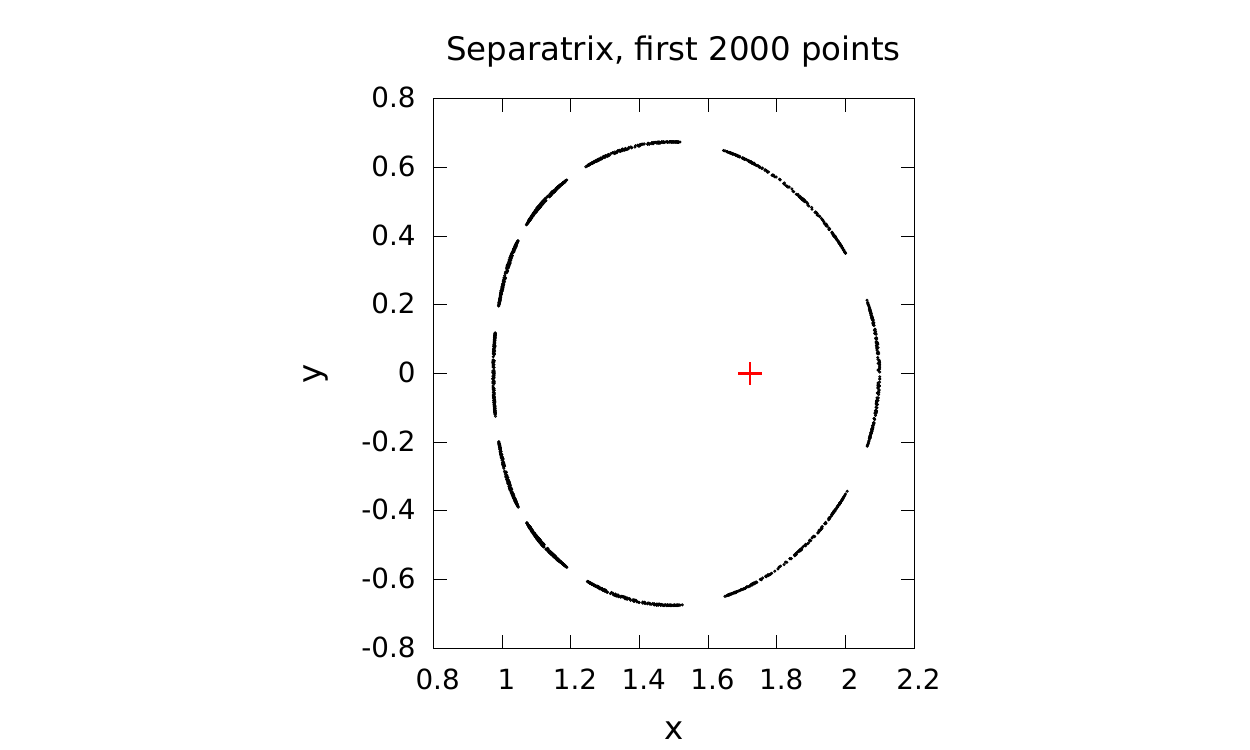} &
\includegraphics[trim = 3.1cm 0cm 3.0cm 0.0cm, clip = true,width=0.20\textwidth]{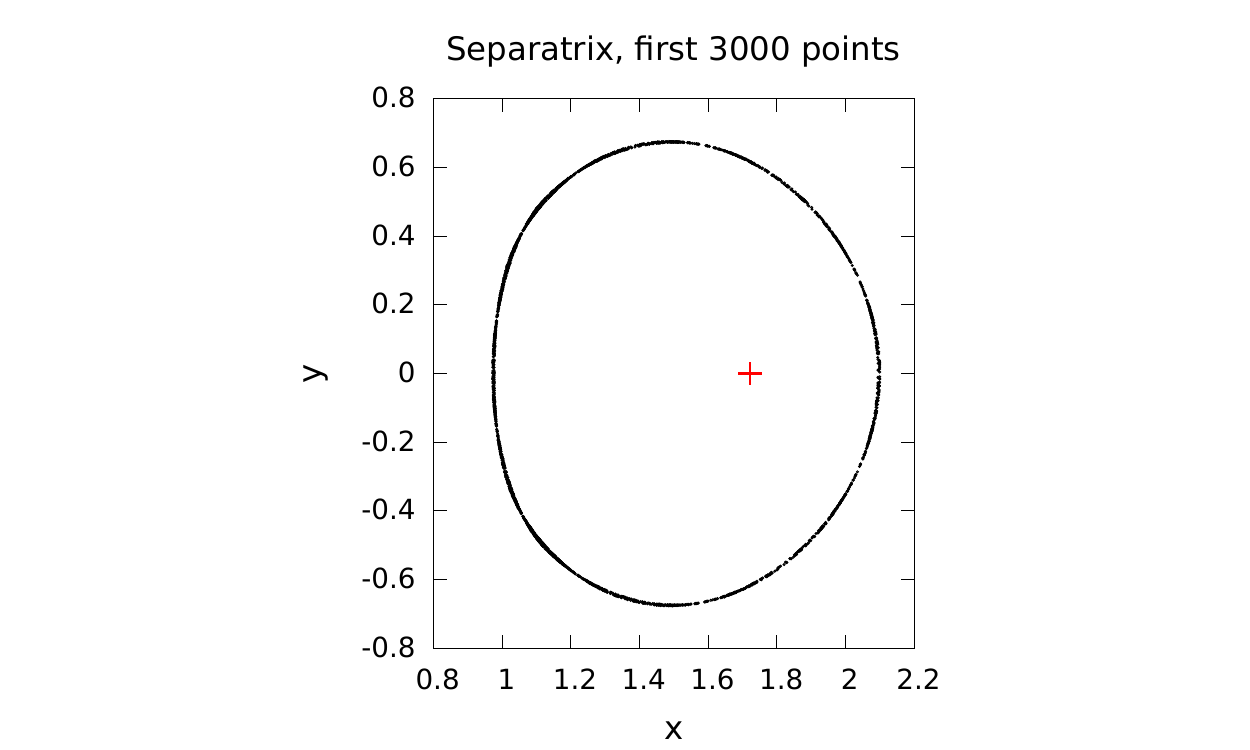} &
\includegraphics[trim = 3.1cm 0cm 3.0cm 0.0cm, clip = true,width=0.20\textwidth]{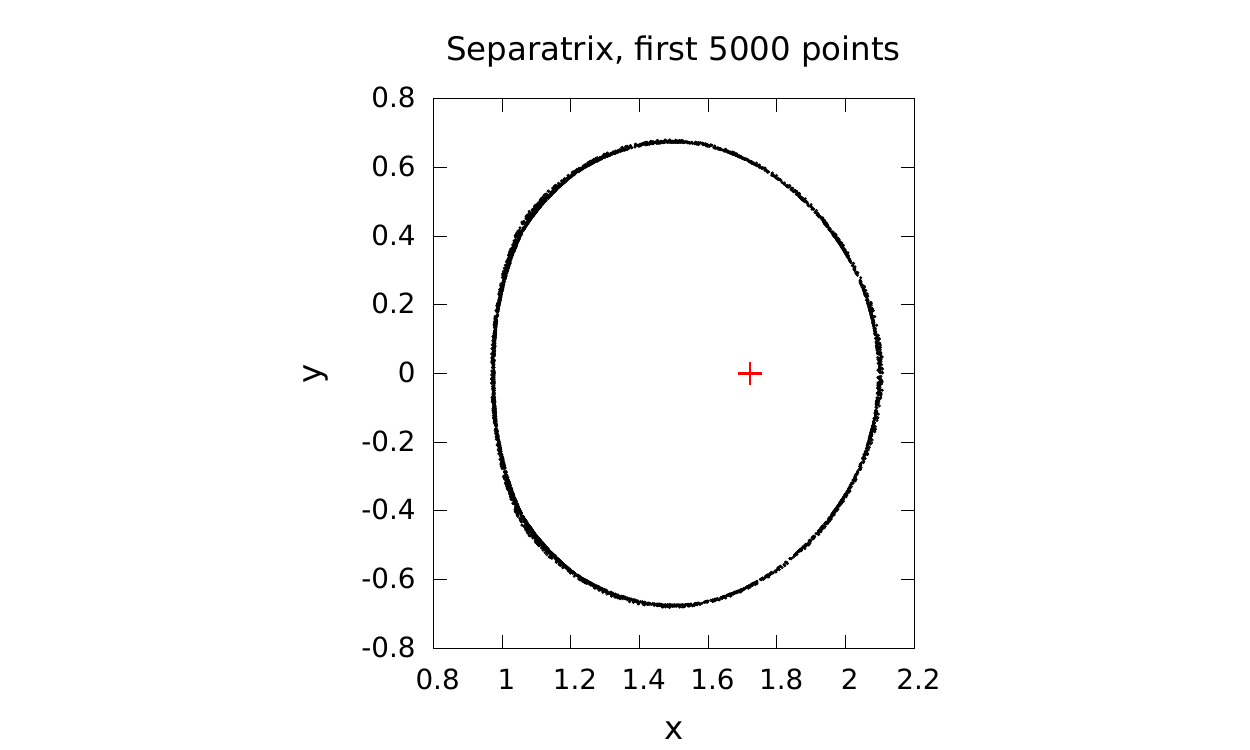} \\
\includegraphics[trim = 3.0cm 0cm 3.0cm 0.0cm, clip = true,width=0.20\textwidth]{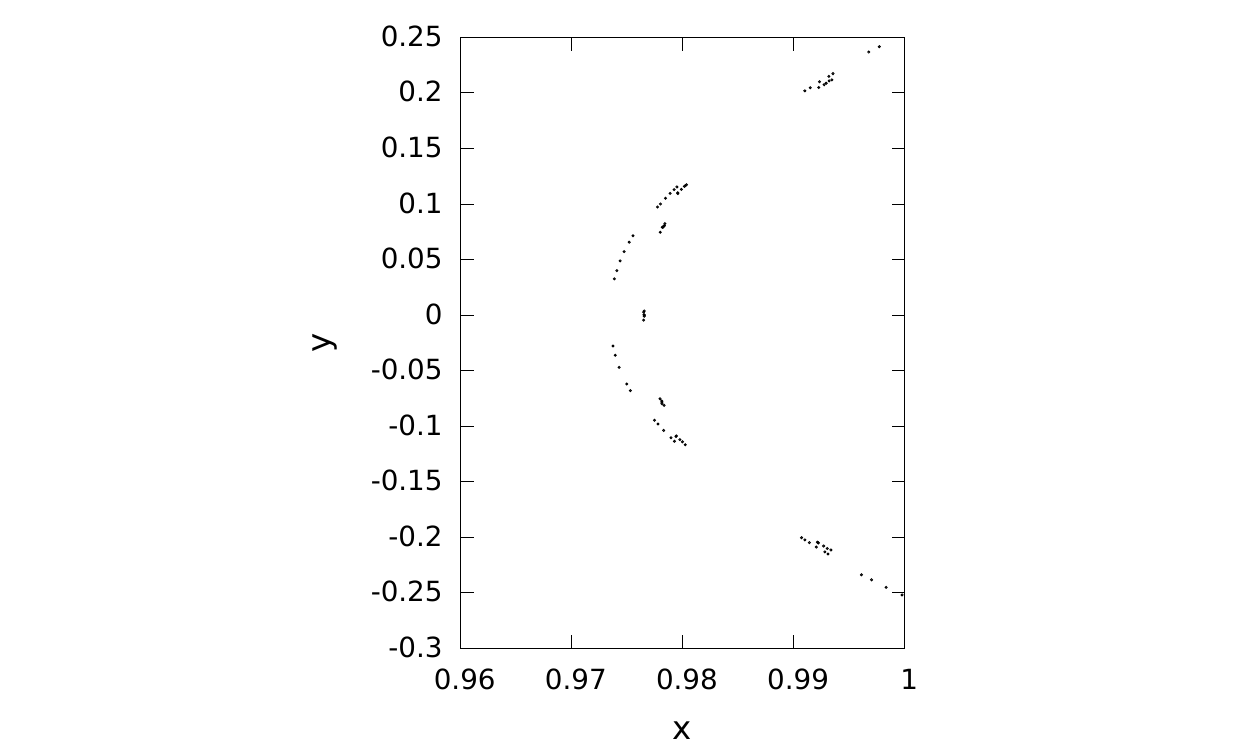} &
\includegraphics[trim = 3.0cm 0cm 3.0cm 0.0cm, clip = true,width=0.20\textwidth]{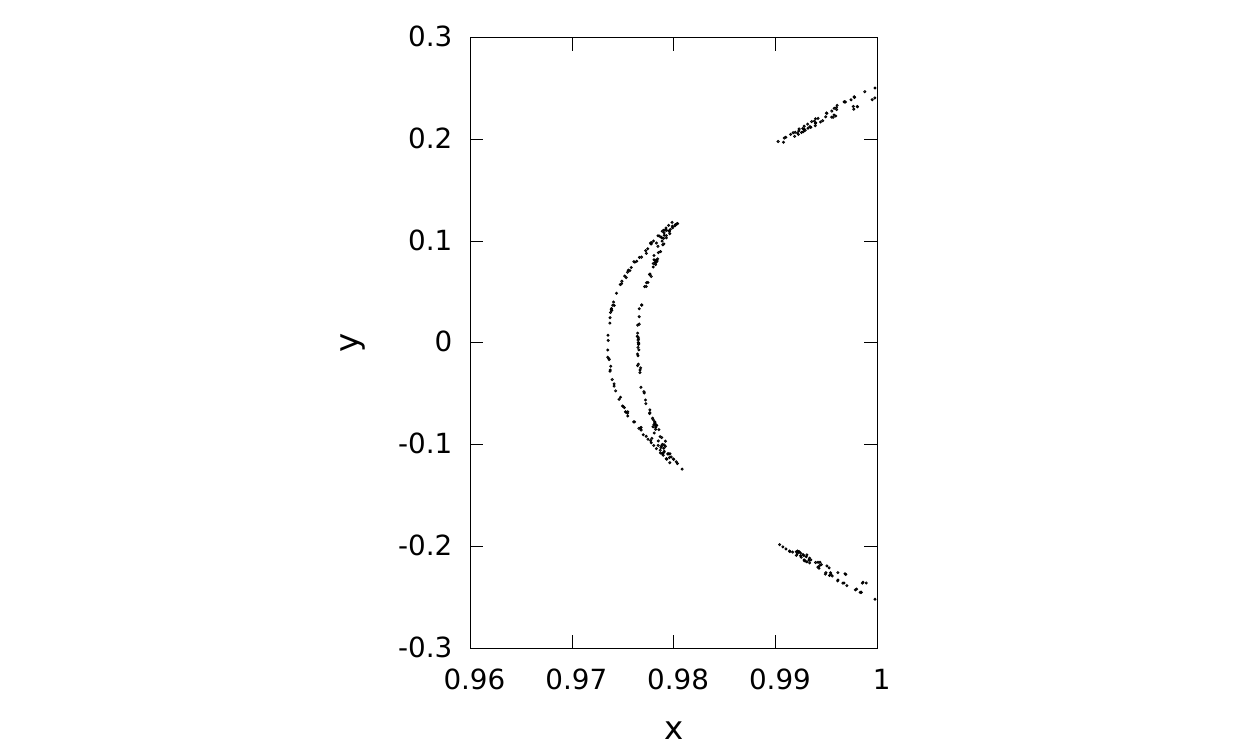} &
\includegraphics[trim = 3.0cm 0cm 3.0cm 0.0cm, clip = true,width=0.20\textwidth]{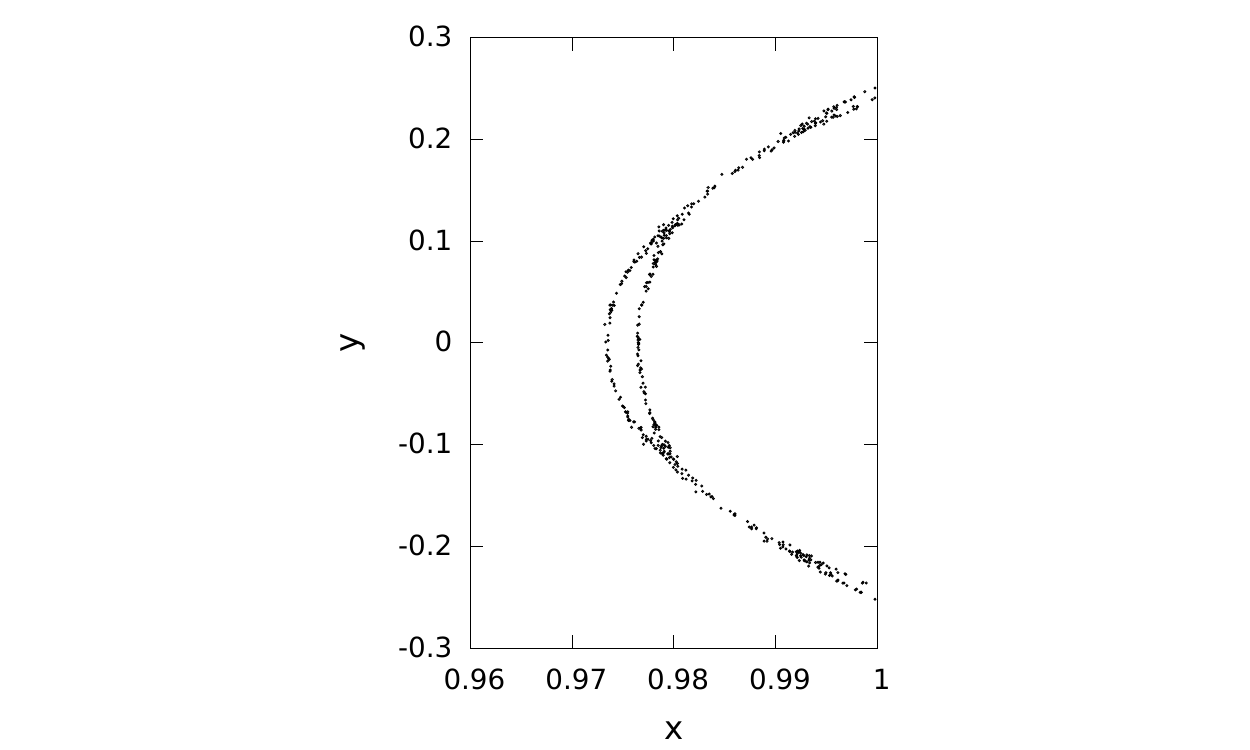} &
\includegraphics[trim = 3.0cm 0cm 3.0cm 0.0cm, clip = true,width=0.20\textwidth]{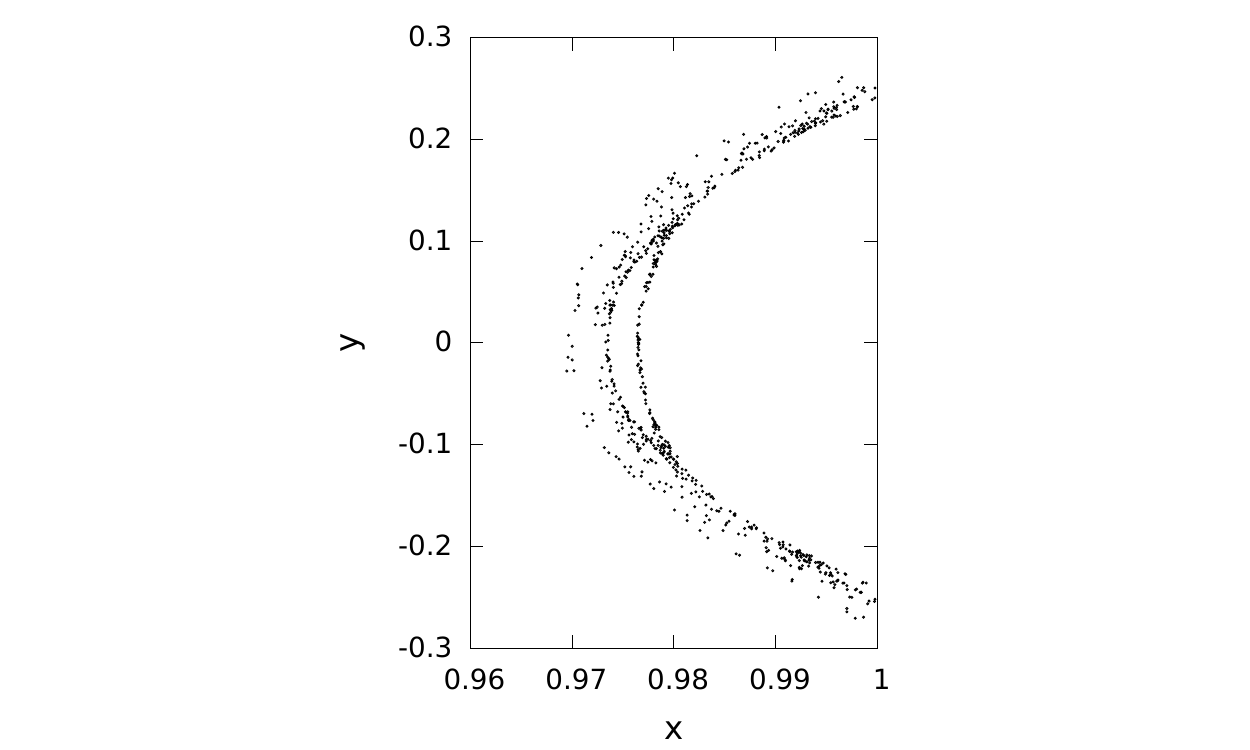} \\
(a) & (b) & (c) & (d) \\
\end{tabular}
\vspace{-0.2cm}
\caption{ A separatrix orbit is shown using the first (a) 500; (b)
  2000; (c) 3000; and (d) 5000 points, with the bottom row showing the
  details of the left part of the orbit.  The orbits have a
  multi-scale structure, as they appear to be quasiperiodic at a
  coarse scale (top row), but the fine scale detail indicates that
  as the number of points is increased, the class changes from
  island chain, to separatrix, to mildly stochastic.}
\label{fig:variation_2}
\end{figure}

However, we then realized that the class of the derived orbits could
vary as the number of points was increased.  For example, in
Figure~\ref{fig:variation_1} panel (a), an orbit appears as a
quasiperiodic when we plot the first 500 points. At 1000 points, we
see a handful of points forming an inner curve, points that could be
easily overlooked in a visual labeling of the orbit. This inner curve
takes shape only when many more points are added, clearly indicating a
separatrix orbit in panels (c) and (d). A similar example is shown in
Figure~\ref{fig:variation_2}, where the class of the orbit changes
from island chain to separatrix to mildly stochastic as the number of
points in the orbit is increased. This meant that we needed to
visually inspect each of the derived orbits to assign a class label;
we could not simply assign them to the same class.

We observed that some orbits, especially the separatrix, have some
stochasticity, especially for a large number of points. This 
results from the way in which the field lines are traced along the torus.
A true stochastic orbit has the points spread out over a large radial
distance as in Figure~\ref{fig:orbits}(d).

\begin{figure}[!htb]
\centering
\setlength\tabcolsep{1pt}
\begin{tabular}{ccc}
\includegraphics[trim = 3.0cm 0cm 3.5cm 0.0cm, clip = true,width=0.2\textwidth]{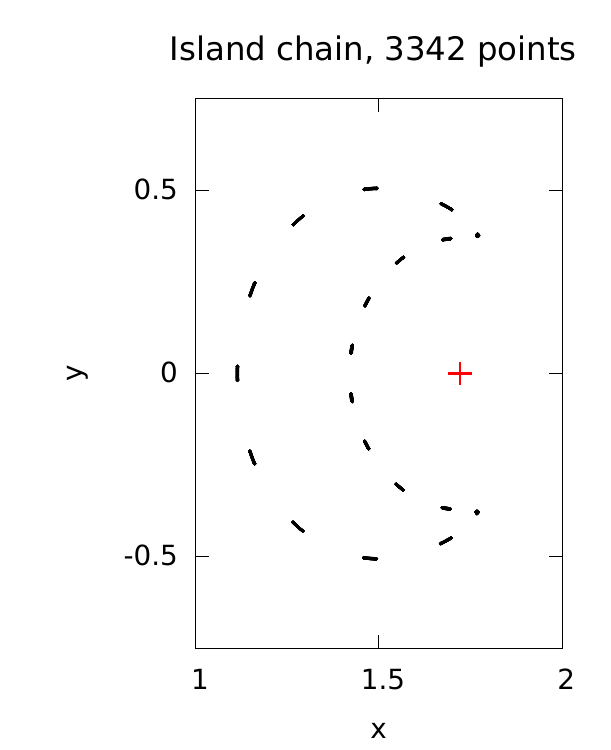} &
\includegraphics[trim = 3.0cm 0cm 3.5cm 0.0cm, clip = true,width=0.2\textwidth]{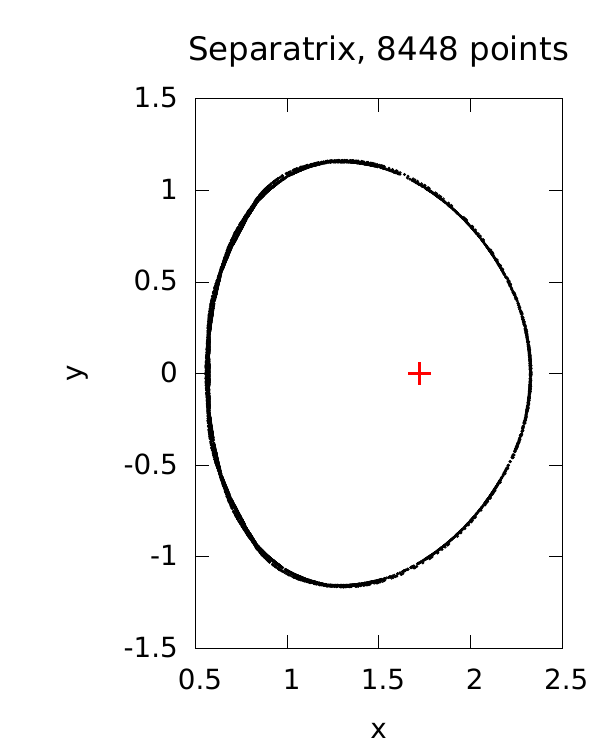} &
\includegraphics[trim = 3.0cm 0cm 3.5cm 0.0cm, clip = true,width=0.2\textwidth]{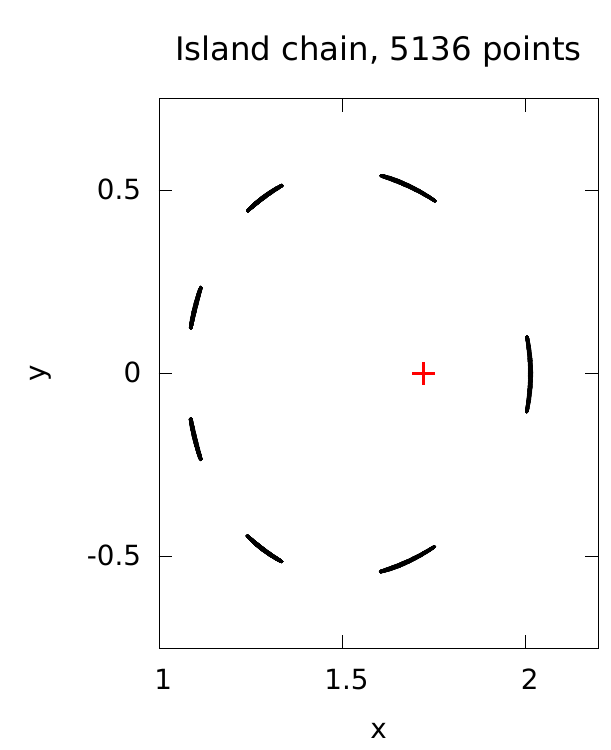} \\
\includegraphics[trim = 3.5cm 0cm 3.5cm 0.0cm, clip = true,width=0.2\textwidth]{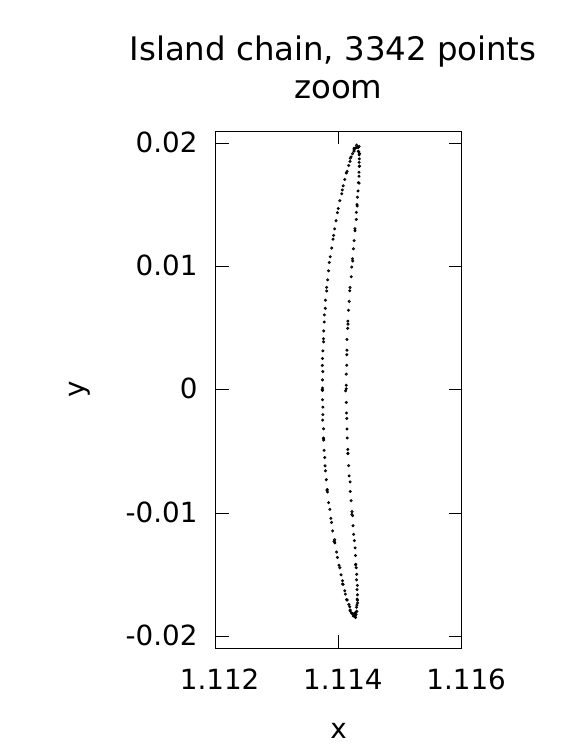} &
\includegraphics[trim = 3.5cm 0cm 3.5cm 0.0cm, clip = true,width=0.2\textwidth]{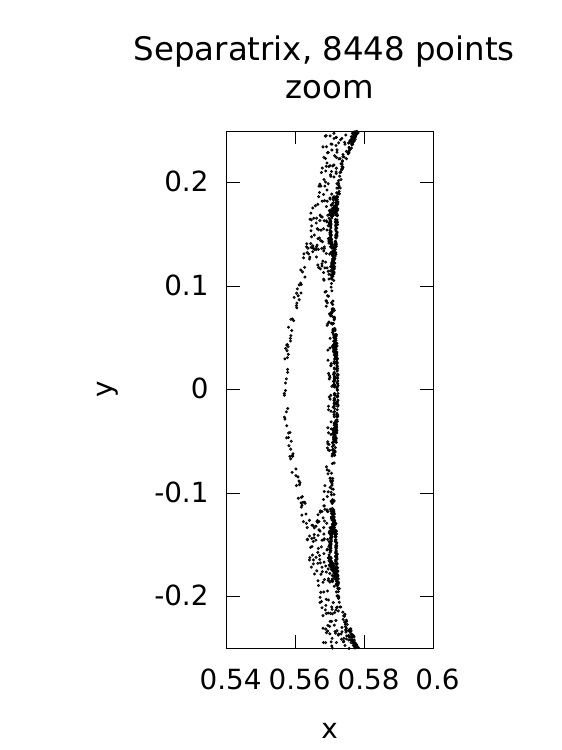} &
\includegraphics[trim = 3.5cm 0cm 3.5cm 0.0cm, clip = true,width=0.2\textwidth]{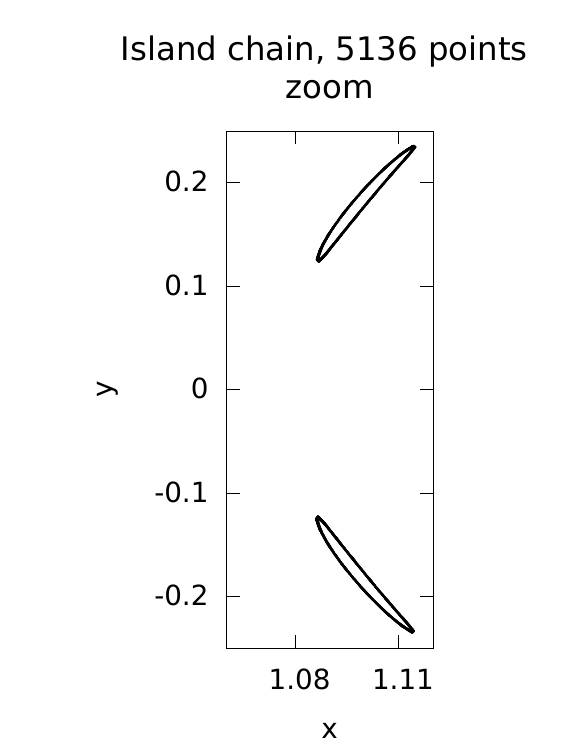} \\
(a) & (b) & (c)\\
\vspace{-0.5cm}
\end{tabular}
\caption{Thin island chains and narrow separatrix orbits can be
  difficult to classify visually. The details of
  the region on the left of each orbit (shown in the bottom row) help in 
  classification.  (a) A single island in the form of a crescent
  is found to be composed of several segments that are islands
  themselves. (b) A very thin separatrix that appears to be a
  quasiperiodic orbit. (c) An island chain with seven islands that
  appears as an incomplete quasiperiodic orbit.  }
\label{fig:variation_3}
\end{figure}

\begin{figure}[!htb]
\centering
\setlength\tabcolsep{1pt}
\begin{tabular}{ccc}
\includegraphics[trim = 2.5cm 0cm 3.0cm 0.0cm, clip = true,width=0.2\textwidth]{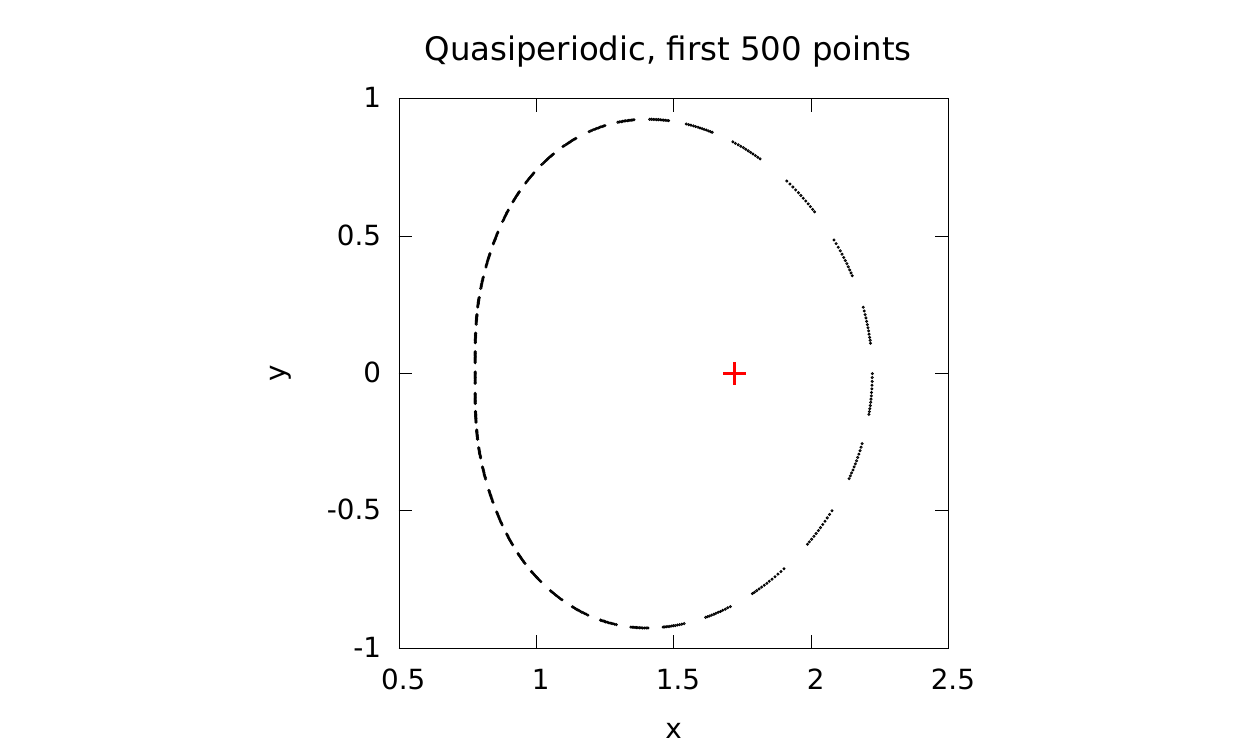} &
\includegraphics[trim = 2.5cm 0cm 3.0cm 0.0cm, clip = true,width=0.2\textwidth]{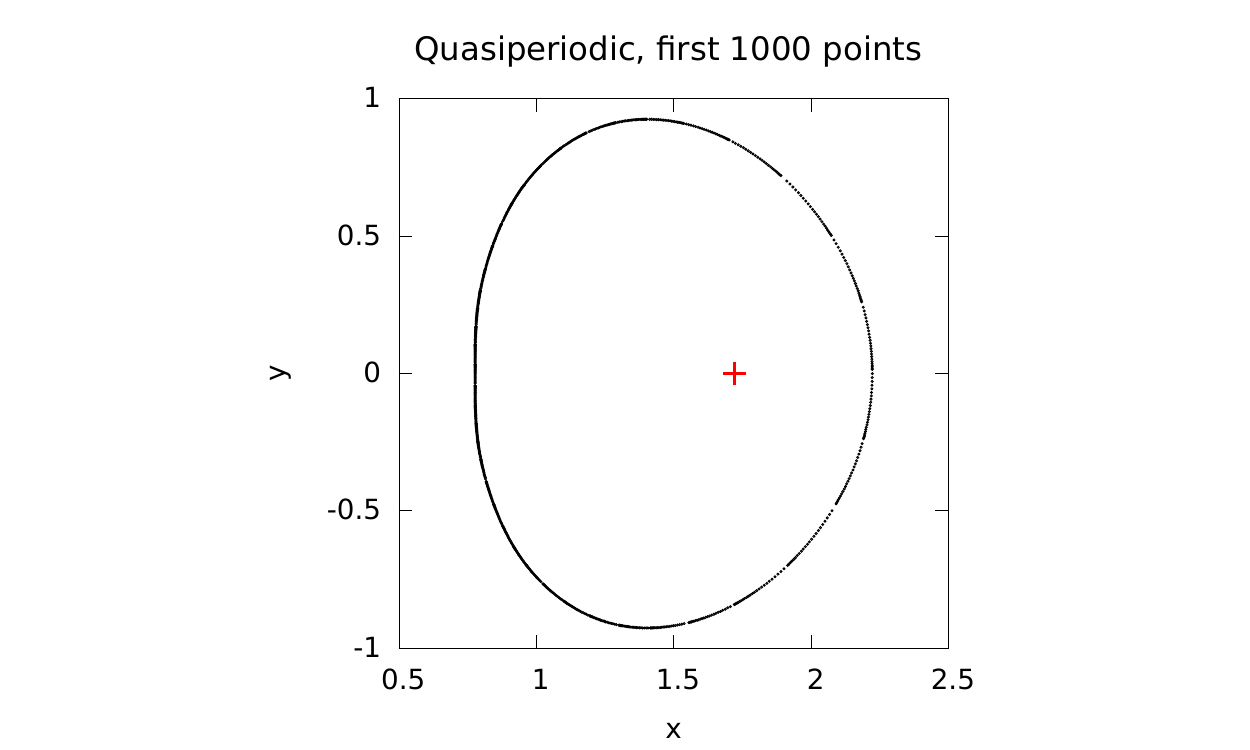} &
\includegraphics[trim = 2.5cm 0cm 3.0cm 0.0cm, clip = true,width=0.2\textwidth]{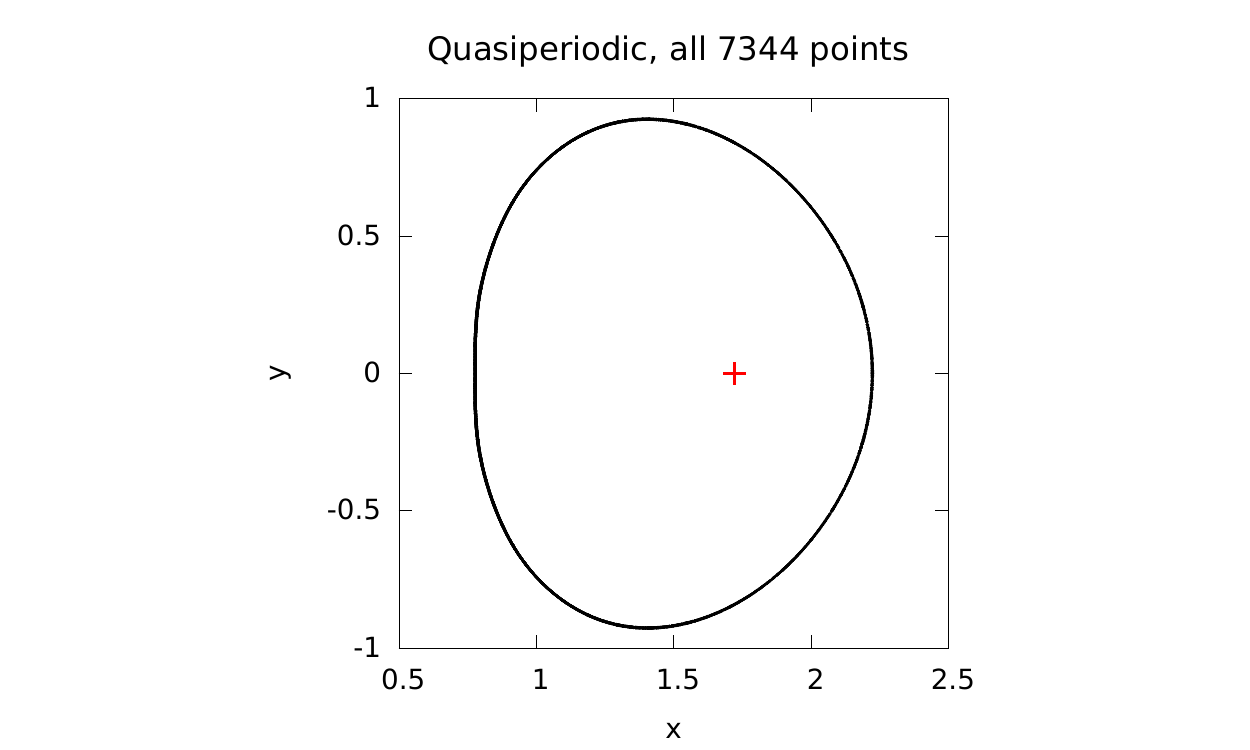} \\
\includegraphics[trim = 3.0cm 0cm 3.0cm 0.0cm, clip = true,width=0.2\textwidth]{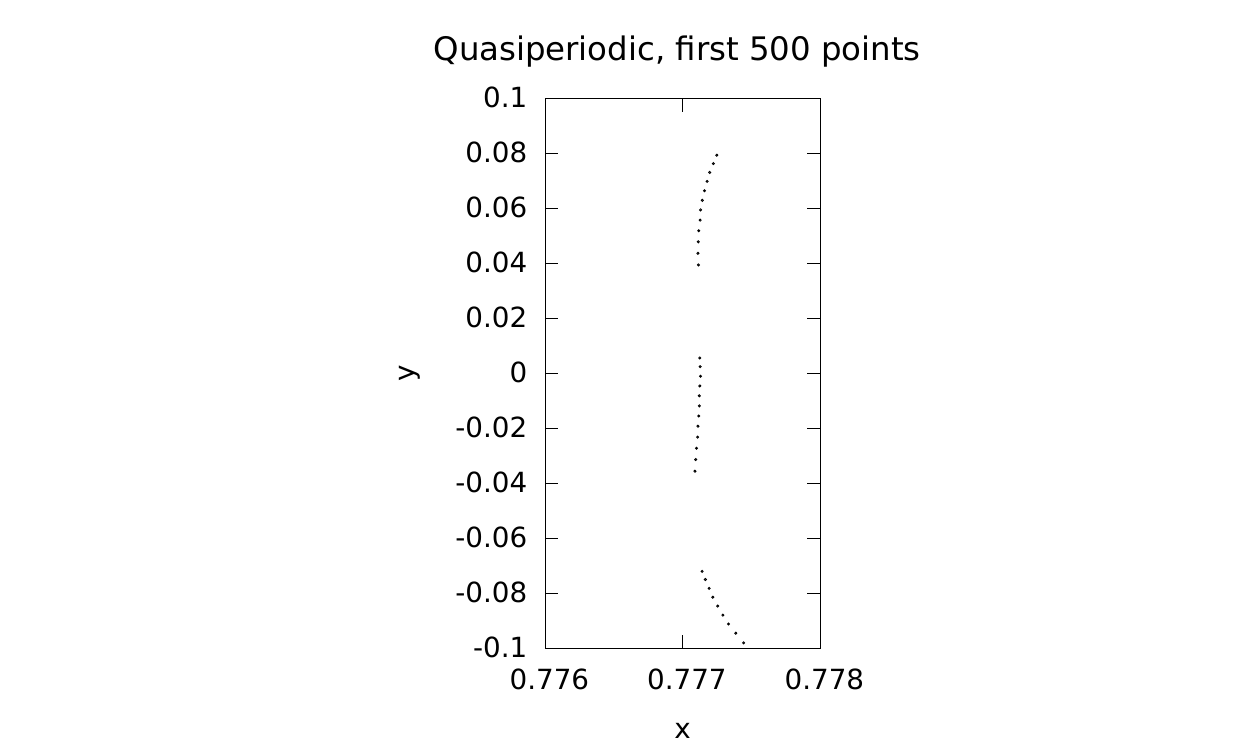} &
\includegraphics[trim = 3.0cm 0cm 3.0cm 0.0cm, clip = true,width=0.2\textwidth]{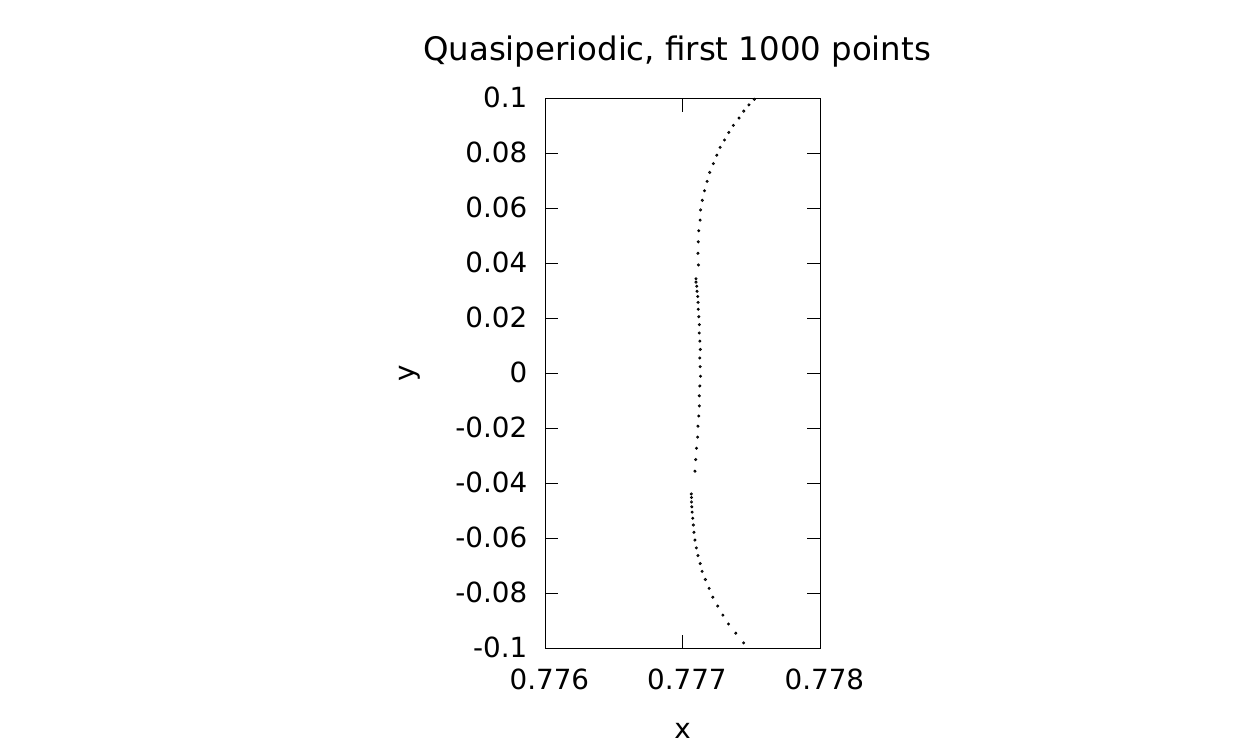} &
\includegraphics[trim = 3.0cm 0cm 3.0cm 0.0cm, clip = true,width=0.2\textwidth]{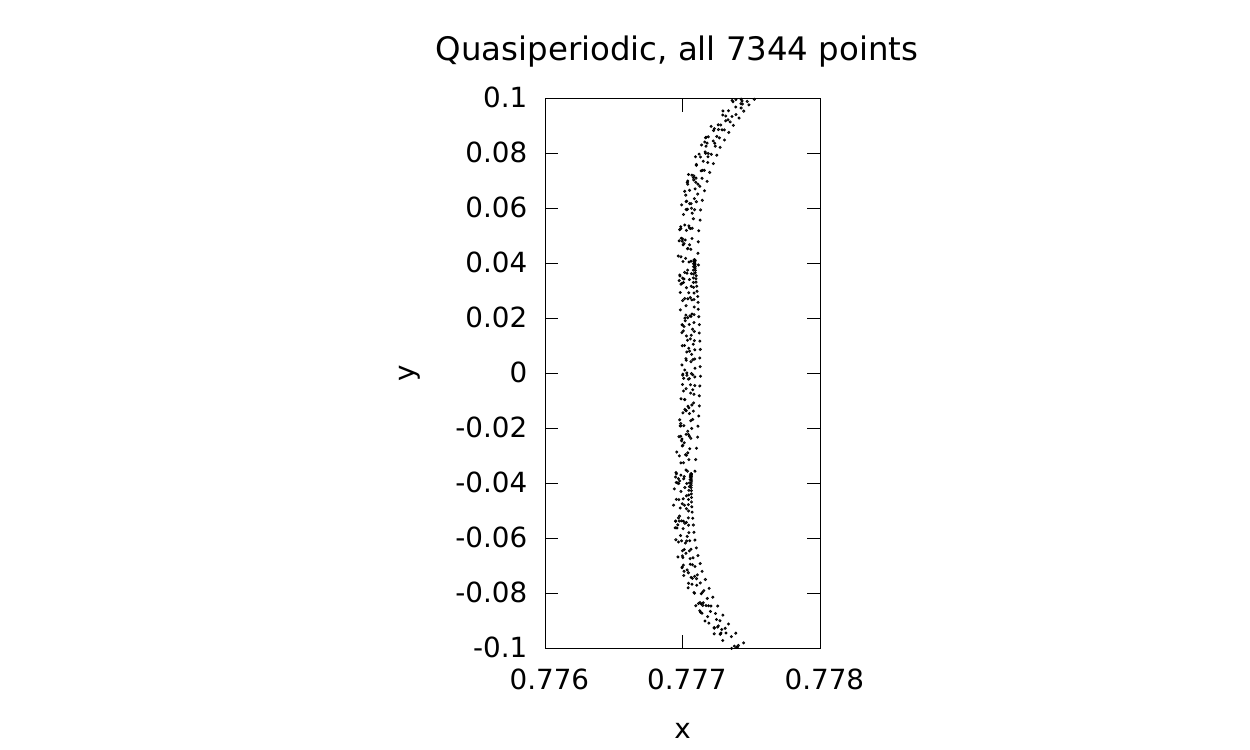} \\
(a) & (b) & (c)\\
\end{tabular}
\caption{A quasiperiodic that could be a thin island chain at 500
  points (a) is confirmed to be a quasiperiodic when more points are
  added in (b) and (c) and a detailed view is observed (bottom row). The
  detailed view in (c) appears to indicate a stochastic orbit, but the
  radial range of values is very small, unlike a stochastic orbit.}
\label{fig:variation_4}
\end{figure}

Figure~\ref{fig:variation_2} also illustrates the second challenge in
labeling orbits --- this orbit, regardless of the number of points,
appears to be quasiperiodic when viewed as a whole; we need to look at
the detail to determine the correct class.
Figure~\ref{fig:variation_3} shows other examples where a visual
inspection of the full orbit is insufficient to assign the correct
class label. Panel (a) appears to be a single island in
the form of a crescent, but is actually an island composed of thin
islands, that is, the crescent shape will always be composed of
incomplete segments. In panels (b) and (c), we see that, what appear
to be a quasiperiodic orbit and an incomplete quasiperiodic orbit, are
actually a separatrix and an island chain, respectively, where the
lobes of the separatrix and the islands are very thin, with a very
small radial variation.  Based on this example, the orbit in
Figure~\ref{fig:variation_4}(a) could be either an incomplete
quasiperiodic or an island chain; as more points are added, it could
be a quasiperiodic or a very thin separatrix. The details in the
bottom row correctly classify this orbit as quasiperiodic, with a very
small amount of stochasticity when the number of points is large.

This multi-scale nature of some of the orbits, where an orbit appears
to be of one class when viewed at a coarse scale, but is of a
different class when viewed at a finer scale (that is, a zoomed-in
view), further makes it challenging to label the orbits correctly.  We
can no longer rely solely on a visual inspection of the entire orbit to
assign a class; requiring a closer inspection of the points makes the
class assignment even more tedious. This provides a strong motivation
for our work and, as we show later in Section~\ref{sec:label}, we can
use the process of automating the assignment of labels to bootstrap the
correct assignment of class labels.

%
\subsection{Generating the features for the training set}
\label{sec:generate_features}
%

The second challenging aspect of the problem of orbit classification
is the identification and extraction of representative features for
each orbit. Obviously, the features selected must be discriminating so
they can be used to differentiate among the orbits of different
classes.  However, as we have seen, the orbits within a class can
appear quite different, while orbits from different classes may appear
very similar.  
As a result, obvious features, such as the angular gaps that
characterize the island orbits, are not sufficient to differentiate
them from incomplete quasiperiodic orbits, while a small radial
variation of the points within a small angular window could indicate a
quasiperiodic, a thin island chain, or a thin separatrix.
There are also other considerations that influence the definition and
extraction of representative features:

\begin{itemize}

\item The data are available as $(x,y)$ coordinates of
  the points in an orbit and must be converted to represent the
  structure we see when we visualize the orbit, either as a whole, or
  a zoomed-in view.

\item These features must be invariant to scale, rotation, and
  translation, as the class of the orbit does not change if it is
  scaled, rotated, or translated.

\item The features must be robust to noise as small changes in the
  coordinates of the points in an orbit do not change
  the orbit class.

\item The features must not depend on the number of points in the
  orbit, or how far out the orbit is relative to the magnetic axis.

\end{itemize}

Transforming the $(x,y)$ coordinates of an orbit into a representation
of the visual structure we see is what makes this problem unusual and
challenging.

%
\section{Related work}
\label{sec:related}
%

We next briefly review existing work related to the classification of
orbits in Poincar\'e maps to place our work in context.  The topic of
Poincar\'e plots, or Poincar\'e maps, is of interest to several
communities, including plasma physics~\cite{sanderson2006:poincare}
and spacecraft trajectory design~\cite{tricoche2021:poincare}. A
common theme in these classical physics systems, where the dynamics is
described in terms of Hamilton's equations, is identifying regular and
chaotic regions in phase space.  This specific problem has been
addressed using box counting to determine the fractal dimension of a
set of points~\cite{albert2020:stellarator} and by calculating the
spectrum of Lyapunov exponents using fast algorithms, such as the
structure-preserving Gaussian process surrogate~\cite{rath2021:orbit}.

The analysis of Poincar\'e maps in magnetic confinement fusion, where
they are referred to as puncture plots, has been driven by two class
of methods. The visualization community has focused on topology-based
methods. For example, Sanderson et al.~\cite{sanderson2010:recurring}
have used field lines with near minimal lengths to determine the
topology of a magnetic field, while Tricoche, Garth, and
Sanderson~\cite{tricoche2011:pmapvis} have combined a six-stage
topological approach, specifically tailored to the data, with
visualization of scalar maps to provide context to the topological
visualization. These analyses have also included island chain and
separatrix orbits, in addition to quasiperiodic and chaotic orbits.

In contrast, the data mining community has taken a feature-based
approach.  Yip~\cite{yip91:kam} represented an orbit as a minimal
spanning tree, a tree that contains all nodes of the graph where the
sum of the edge lengths is minimal. Next, edge lengths greater than a
specified threshold were removed, splitting the tree into distinct
subgraphs, and features were obtained for the full tree and the
subgraphs. For example, the diameter of a graph, which is the longest
shortest path between any two nodes in the graph was identified,
followed by the ``branches'' of the graph that have only one node in
the diameter.  Then, each branch was assigned properties, such as a
shallow or a deep branch based on its length.  The entire orbit was
represented by features such as the number of subgraphs and the
fraction of nodes that are on a shallow branch.  These features were
then used in rules to identify the class of an orbit.

In our previous
work~\cite{bagherjeiran2005:graph,bagherjeiran2006:pplot}, we started
by applying the rule-based approach of Yip to our data sets. We found
that we could improve the results by including additional features and
by using standard machine learning classifiers instead of a rule based
system. While we obtained nearly 90\% accuracy rate for
classification, our approach was not robust enough to be applied in
practice. We found that an orbit had to be represented by a large
number of points (2000-2500) for accurate classification and it was
challenging to set the values of the many thresholds used in defining
the features. In addition, the graph structure was difficult to
implement in software, depended on the number of points in an orbit,
and was prone to giving incorrect results if any of the key points
determining the structure were perturbed slightly.  This present work
is an attempt to address these deficiencies.

%
\section{Solution Approach}
\label{sec:approach}
%

We used a three-step, machine-learning approach to classifying the
orbits, starting with the labeling of the orbits and the generation of
features for each orbit to create a training data set, followed by
building a classification model with the training set to discriminate
among the different classes. We used decision trees as our model of
choice as it allows us to understand how the decision to assign a
specific class is made, allowing us to iteratively refine both the
labels assigned to an orbit, as well as the features used to describe
it, thus enabling the creation of a high-quality training data set.

\subsection{Labeling the orbits}
\label{sec:label}

We started by labeling the orbits using a visual inspection to assign
each orbit to one of four classes --- quasi-periodic, separatrix,
island chain, and stochastic.  As discussed earlier, given the limited
number of orbits, we increased the number of instances in our data set
by creating derived orbits composed of the first $k$ points in an
orbit, with $k = 1000, 1500, \ldots$.  The lower bound on $k$ was
chosen as we required a certain minimum number of points in an orbit
to extract meaningful features. As the class of an orbit can change
with increasing number of points, the use of these derived orbits also
allows us to increase the diversity of the instances in the training
data set.

Initially, we assigned to each derived orbit, the same class as the
original orbit, which was determined using a visual inspection of all
points at coarse-scale. Next, we created an initial training data set
by extracting obvious features for the orbits, such as angular gaps or
the radial spread of points, and built a decision tree model. We then
found that we needed to improve the robustness of the features
extracted as detailed in Section~\ref{sec:localfeat2}. As we
iteratively improved the quality of the features, the prediction error
reduced.

We then reached a point where we were unable to reduce the prediction
error any further. A closer look at the decision tree model indicated
that it would often incorrectly predict quasiperiodic orbits with gaps
as island chains, or ones without gaps as separatrix orbits.  When we
followed the path the features for these orbits took through the
decision tree, we found that at the leaf node, all training instances
were of the same class as the one predicted for the orbit.  This
indicated that in the feature space, the orbit being labeled was in a
region where the class was different from the class we had assigned.
We next checked the features for the orbit to determine whether we had
extracted them correctly. In the process, we inspected each orbit
closely, and realized that the features extracted were correct, but,
as described in Section~\ref{sec:generate_labels},
either the orbits had a multiscale nature and we had assigned them to
the wrong class, or that the class of the derived orbit, with fewer
points, was different from that of the orbit with all the points.

This accidental discovery indicated that we needed to look closely at
the details of each orbit to assign the correct class, making the
labeling even more tedious. Instead, we used the decision tree model
to our advantage and checked the details only for the orbits with
incorrect predictions to determine whether we should change the label
assigned. This approach reduced some of the tediousness of labeling
the orbits and allowed us to bootstrap our way to a higher quality
training data set.

\subsection{Feature Extraction}
\label{sec:features}

The identification and extraction of discriminating features was
performed iteratively - we selected a few sample orbits from each
class, extracted a set of features and analyzed them to determine if
they were representative enough to account for the variation we saw
among the orbits. As the quality of the features improved, we used
them to build a decision tree classifier and evaluated its accuracy.
If certain types of orbits were consistently mis-classified by the
model, we followed the features for such orbits through the decision
tree, identified the aspect of the orbit that was missing in the
existing features, and refined the features, continuing the process
until sufficient accuracy was obtained.

The feature extraction consists of three phases - the initial
preprocessing of the data, the extraction of local features
representing the local region around the points, and the calculation
of global features representing the orbit as a whole. We next discuss
these phases in detail.

\subsubsection{Preprocessing the data}
\label{sec:preprocess}

\begin{figure}[!htb]
\centering
\setlength\tabcolsep{1pt}
\begin{tabular}{ccc}
\includegraphics[trim = 2.5cm 0cm 3.0cm 0.0cm, clip = true,width=0.21\textwidth]{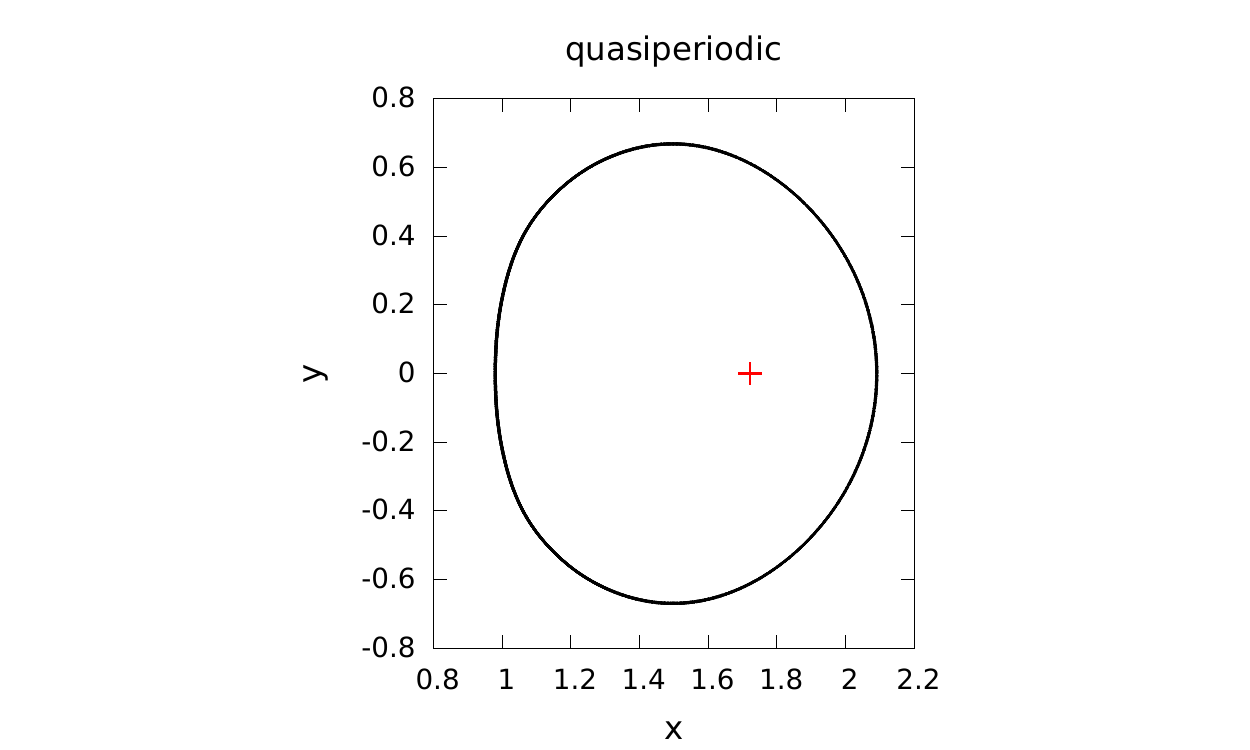} &
\includegraphics[trim = 2.5cm 0cm 3.0cm 0.0cm, clip = true,width=0.21\textwidth]{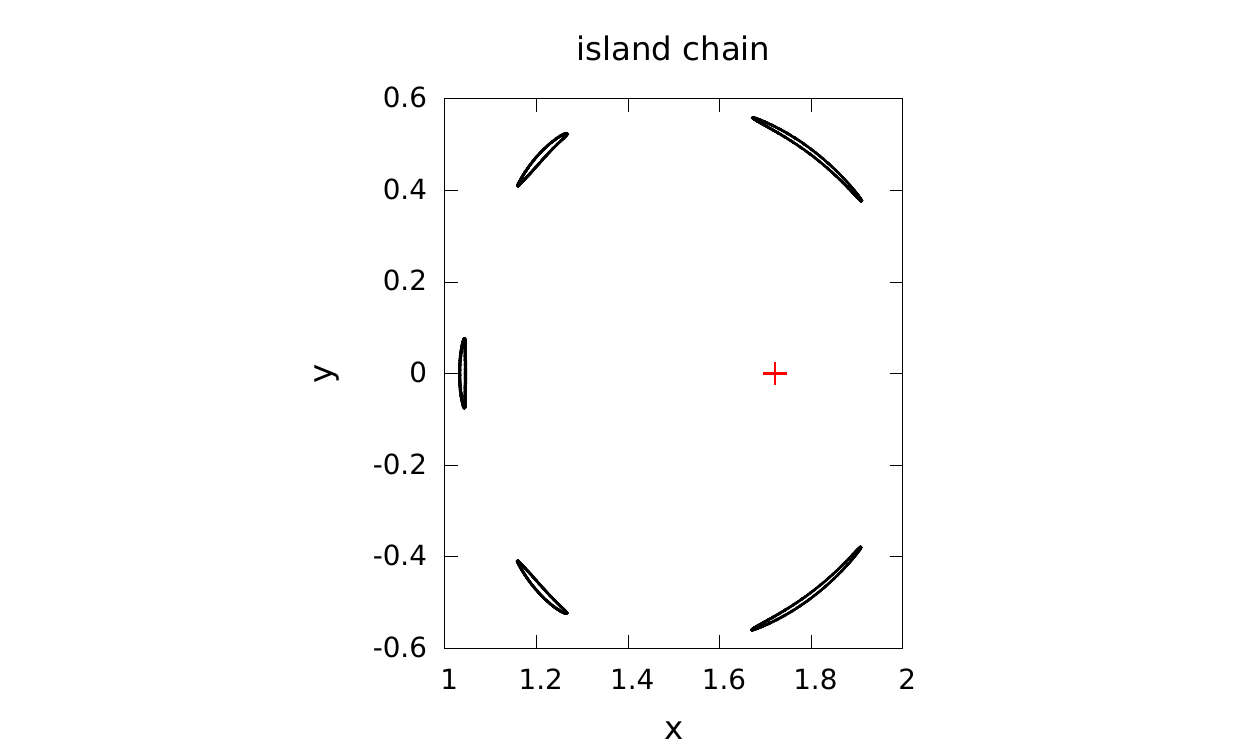} &
\includegraphics[trim = 2.5cm 0cm 3.0cm 0.0cm, clip = true,width=0.21\textwidth]{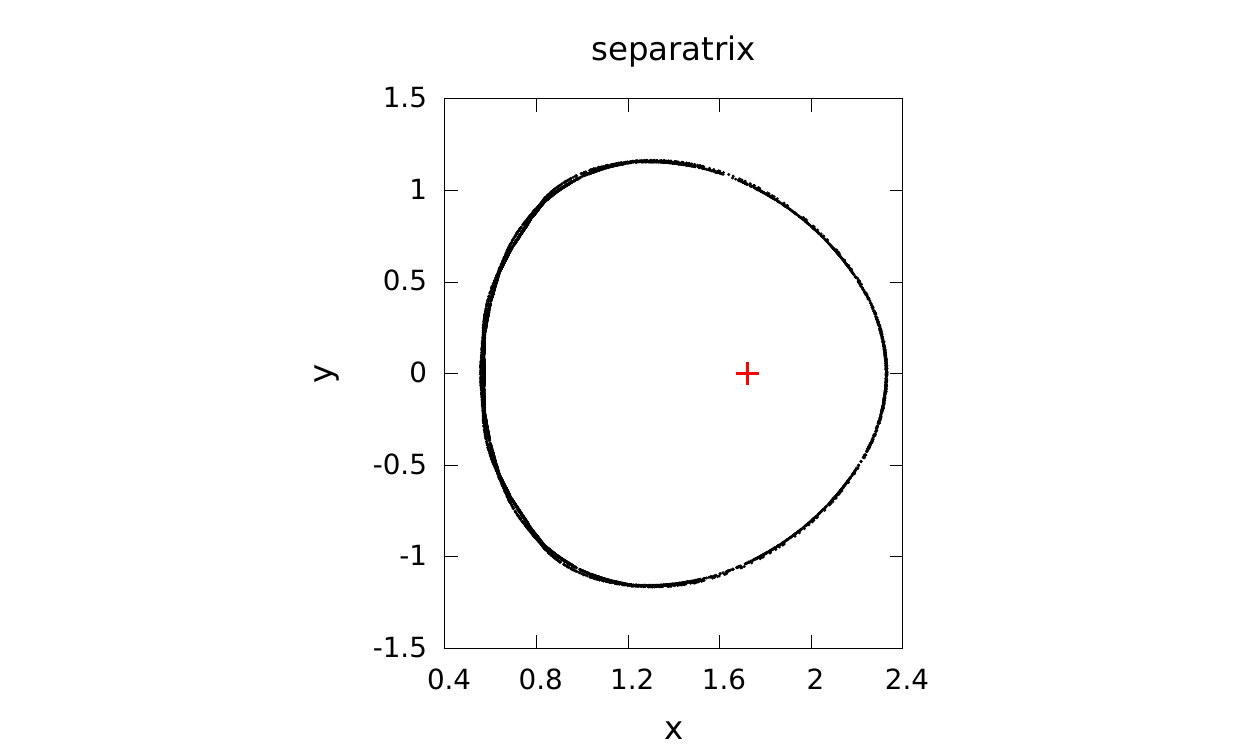} \\
\includegraphics[trim = 2.5cm 0cm 2.5cm 0.0cm, clip = true,width=0.21\textwidth]{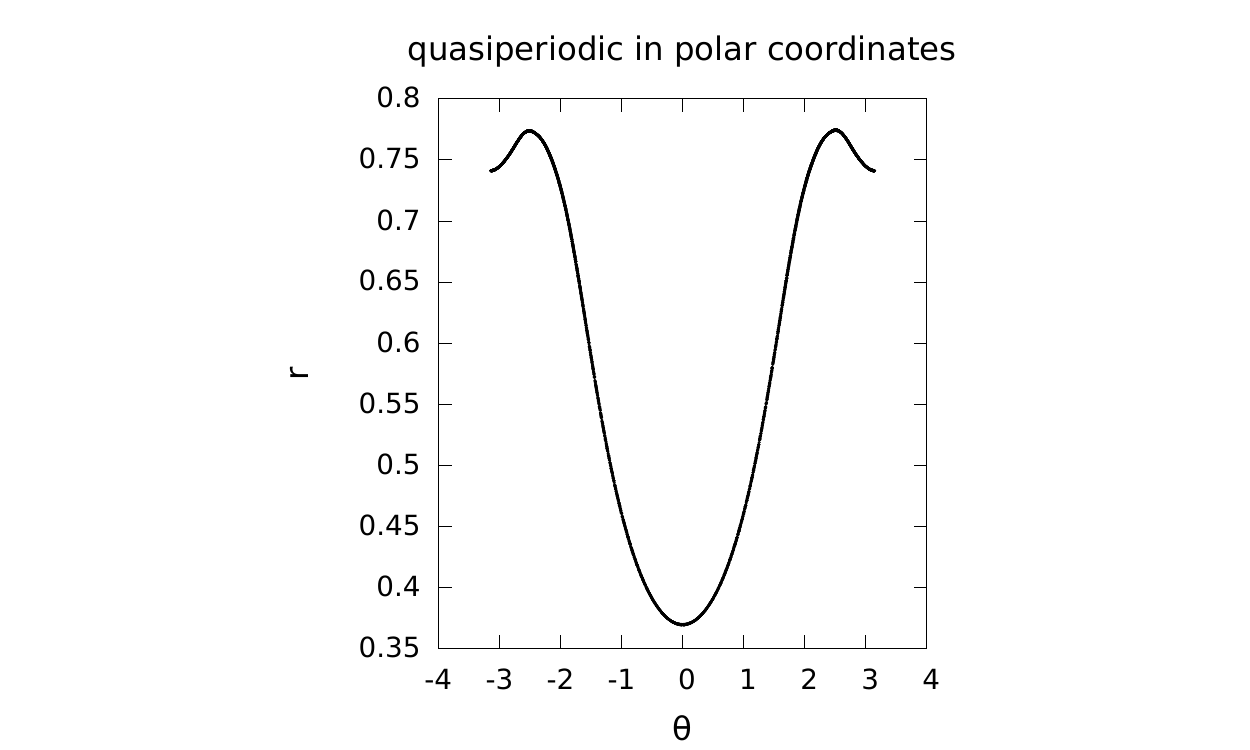} &
\includegraphics[trim = 2.0cm 0cm 2.0cm 0.0cm, clip = true,width=0.24\textwidth]{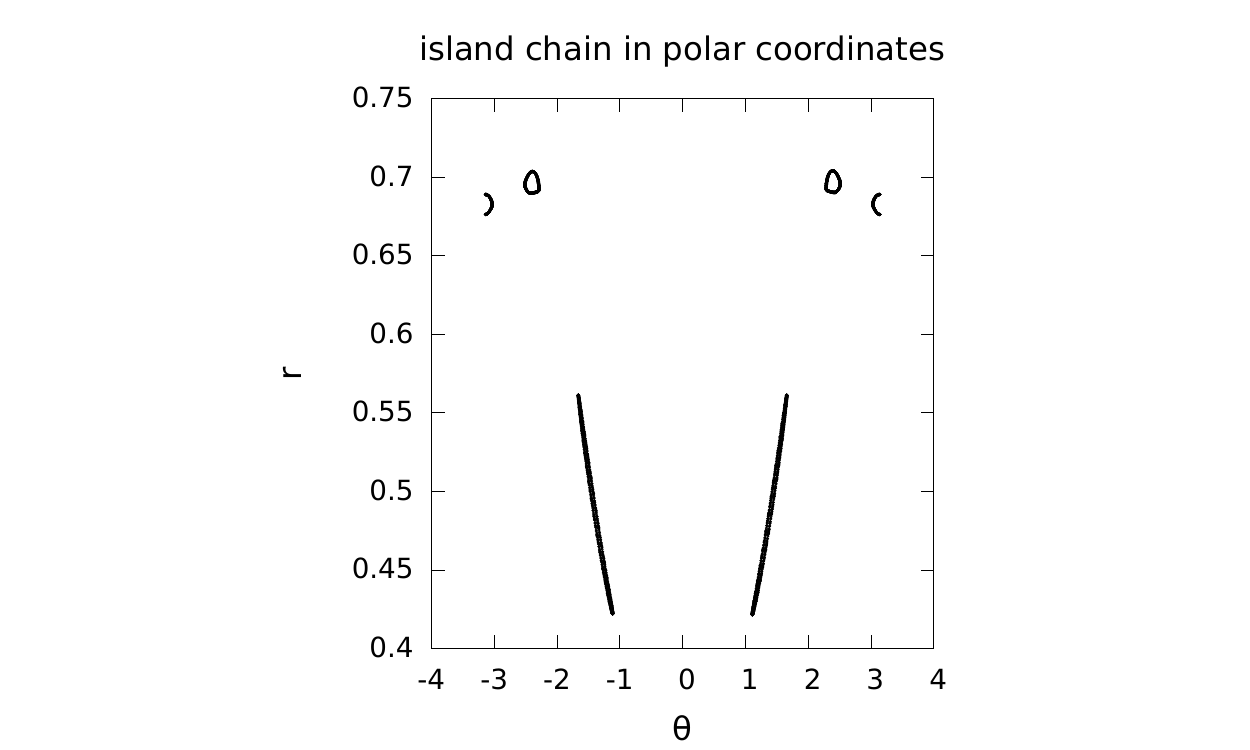} &
\includegraphics[trim = 2.5cm 0cm 2.5cm 0.0cm, clip = true,width=0.21\textwidth]{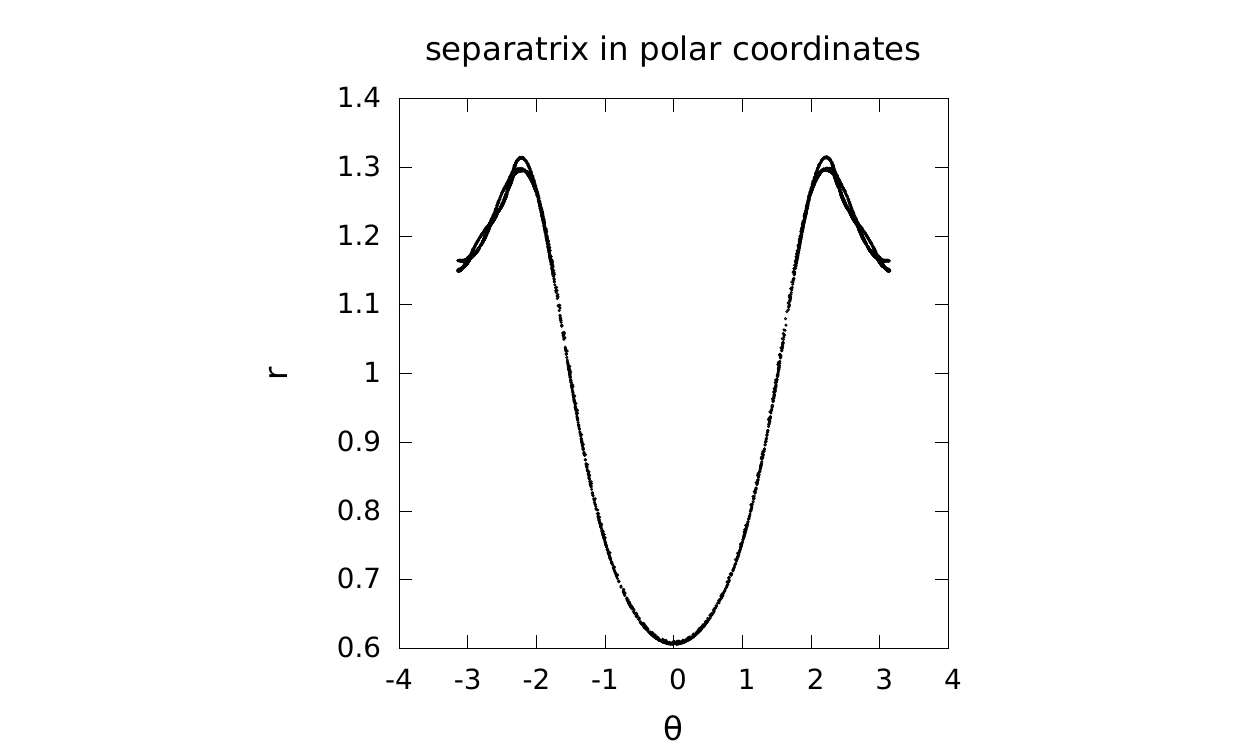} \\
\includegraphics[trim = 2.5cm 0cm 2.5cm 0.0cm, clip = true,width=0.21\textwidth]{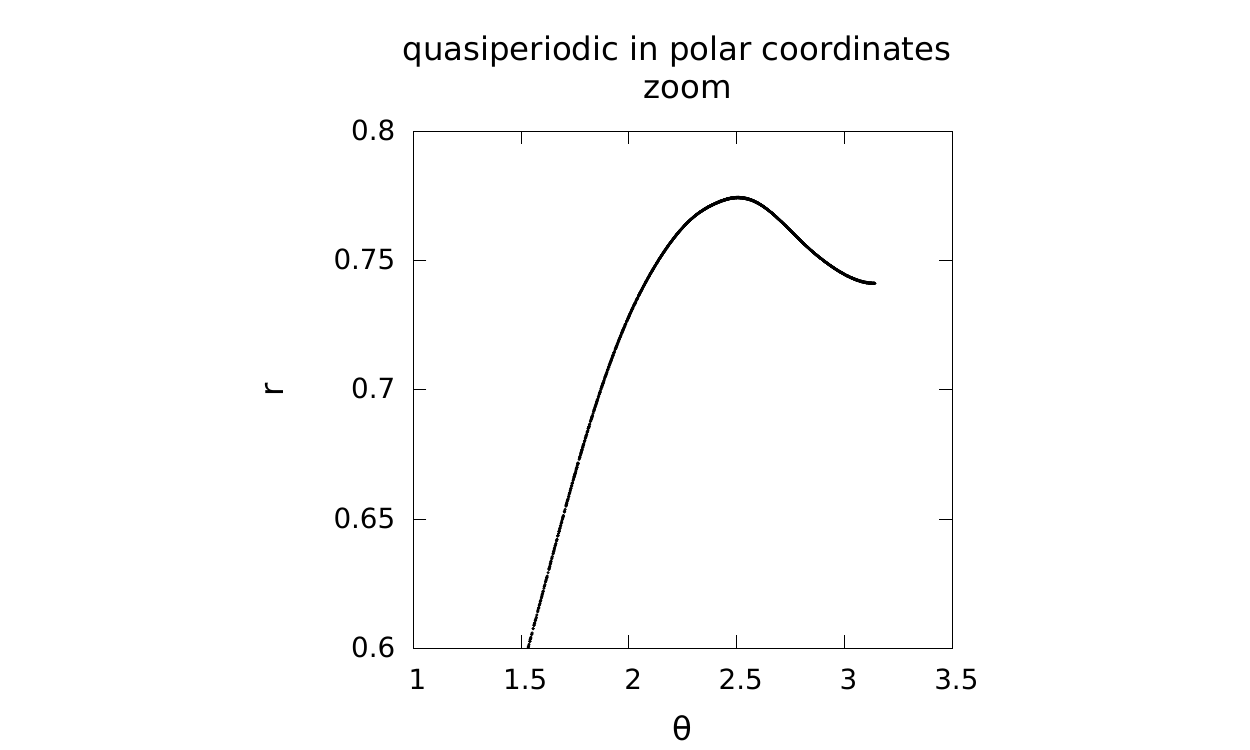} &
\includegraphics[trim = 2.0cm 0cm 2.0cm 0.0cm, clip = true,width=0.24\textwidth]{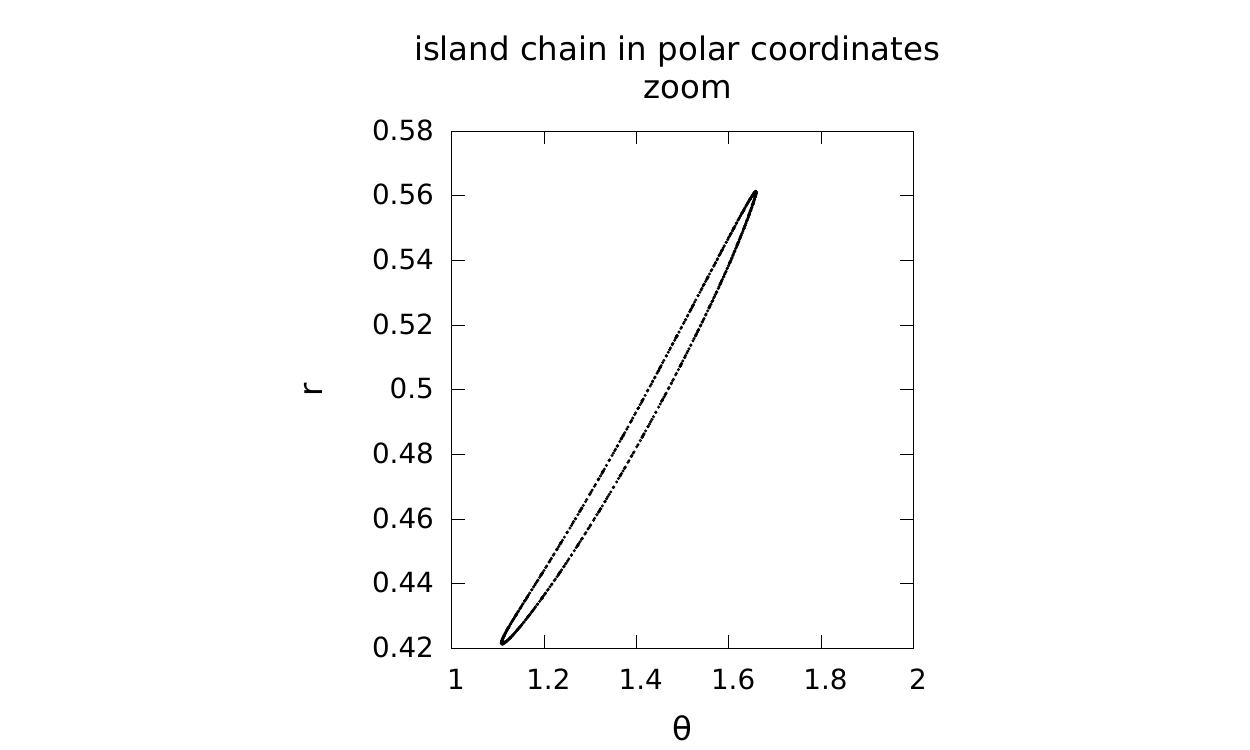} &
\includegraphics[trim = 2.5cm 0cm 2.5cm 0.0cm, clip = true,width=0.21\textwidth]{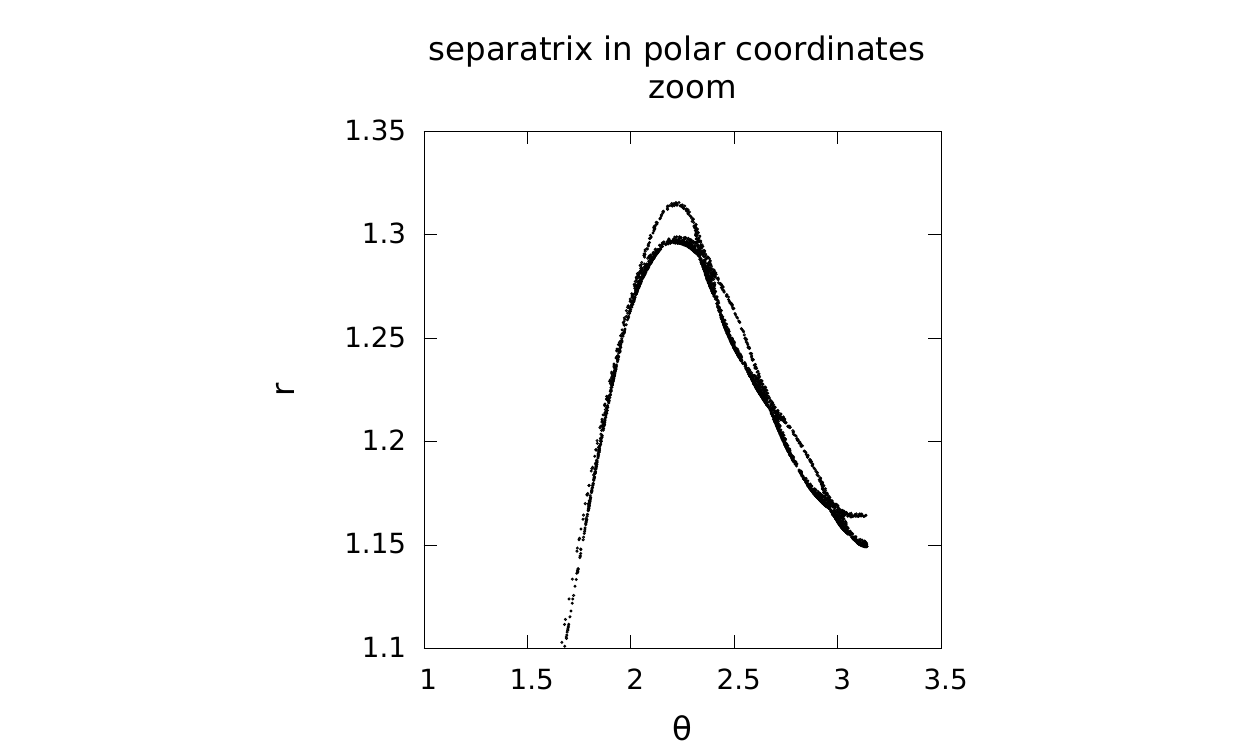} \\
(a) & (b) & (c) \\
\end{tabular}
\caption{The orbits in polar coordinates, with the angle, $\theta$, on
  the $x$-axis and the radius, $r$, on the $y$-axis. (a) The
  quasiperiodic orbit from Figure~\ref{fig:orbits}(a); (b) the island
  chain orbit from Figure~\ref{fig:orbits}(c); and (c) the separatrix
  orbit from Figure~\ref{fig:variation_3}(b). Top row: $(x,y)$
  coordinates, middle row: $(r,\theta)$ coordinates, and bottom row: a
  zoomed-in view of the polar coordinates that highlights the
  difference between the quasiperiodic and thin separatrix orbits.}
\label{fig:polar}
\end{figure}

It is possible to use the original $(x,y)$ coordinates of the points
that form an orbit to extract the features. However, we found that
converting the data to polar coordinates, using the magnetic axis as
the origin, exaggerated the radial variation in an orbit, making it
easier to differentiate among the orbits. For example,
Figure~\ref{fig:polar}, using $(r,\theta)$ coordinates, clearly shows
the difference between the quasiperiodic orbit from
Figure~\ref{fig:orbits}(a) and the very thin separatrix orbit from
Figure~\ref{fig:variation_3}(b).

\subsubsection{Initial extraction of local features}
\label{sec:localfeat1}

Once we converted the data into polar coordinates, we wanted to
extract {\it local} features for the orbit, that is, features that
represent the structure of the points in a small $\theta$ window.
We started by defining these windows based on the $\theta $ value of
the points in an orbit. If we created $n$ windows, each subtending an
angle $\delta_n = 360^\circ /n$ or $\delta_n = 2 \pi /n$ radians at
the magnetic axis, the points in the $i$-th window would satisfy 
\begin{equation}
-\pi+(i-1)\delta_n \le \theta <  -\pi + i\delta_n.
\end{equation}
Each point was assigned to one and only one window and all orbits
have the same number of windows, $n$. 

Next, for the points in each window, we obtained $r_{max}$ and
$r_{min}$, which are the largest and smallest values of $r$,
respectively, and calculated ${\Delta r}_{max} = \max (r_{max} -
r_{min})$, which is the largest radial width across all windows.  The
points in each window were then shifted radially by $r_{min}$ so that
the minimum $r$ in each window became zero, and scaled by ${\Delta
  r}_{max}$, so that the new largest radial width across all the
windows in an orbit became 1.0. This scaling normalizes the radial
variation in the points within the windows across all orbits, so that
separatrix and island orbits with wide lobes are treated similar to
those with thin lobes.

After the normalization, we extracted local features for each window,
such as the variation in the $r$ values of the points in a window so
we could distinguish quasiperiodic orbits from separatrix orbits, as
well as island orbits from quasiperiodic orbits with gaps.  We also
included a simple count of the points in a window, to reflect the
observation that the X points of the separatrix and the end regions of
islands have a higher concentration of points.  These local features
for each of the $n$ windows were converted to features for the entire
orbit by taking the mean, minimum, and maximum values of the features
across the windows. We also included global features, such as the
number of windows with no points.

However, our early experiments indicated two problems with this
approach to extracting representative features. The first was the
definition of a window. The features extracted were very dependent on
the angular width of the window, $\delta_n$, and the location of the
points in an orbit relative to the boundaries between the windows.
This meant that the features were not rotation invariant, as a
feature, such as the maximum density of points across all the windows,
could vary substantially if we just rotated the orbit.

\begin{figure}[!htb]
\centering
\begin{tabular}{c}
\includegraphics[width=3.25in]{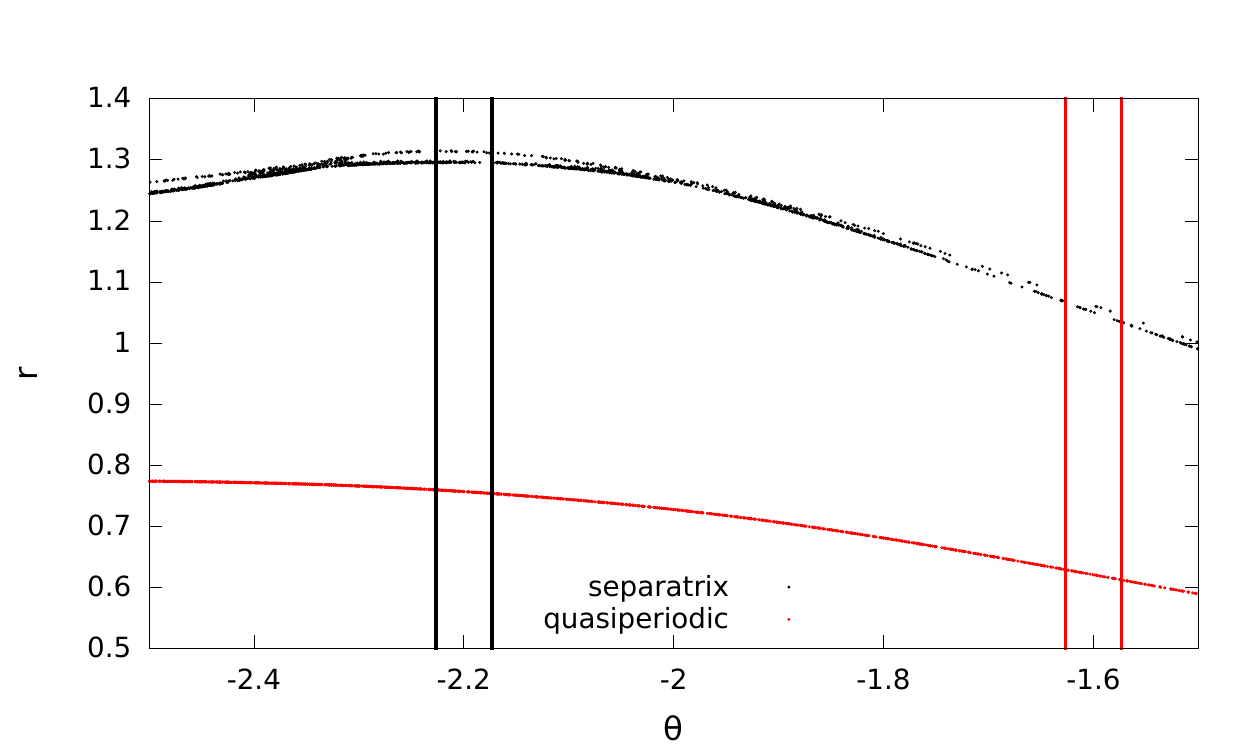} \\
\end{tabular}
\caption{A subset of the orbits in polar coordinates, with $\theta$ on
  the x-axis and $r$ on the y-axis. Top: a separatrix orbit; bottom: a
  quasiperiodic orbit. The change in $r$ over a $3^\circ$ window
  centered at -2.2 for the separatrix (in black) is similar to the
  change in $r$ over a $3^\circ$ window centered at -1.6 for the
  quasiperiodic orbit (in red). Thus, using the range of $r$ values in
  a window does not indicate whether the points in the window can
  be fit by one curve or two.}
\label{fig:delta_theta}
\end{figure}

The second issue was that using just the range of $r$ values within a
window was not sufficient to determine if the points in a window lay
on one curve, as in a quasiperiodic orbit, or two, as in a separatrix
or island chain orbits. As shown in Figure~\ref{fig:delta_theta}, when
we consider the variation in $r$ in a $3^\circ$ window, we can get
similar results for a separatrix (top curve, window centered at
$\theta=-2.2$ radians) as we would for a quasiperiodic orbit (bottom
curve, window centered at $\theta=-1.6$ radians). This holds whether
we considered the difference between the largest and smallest $r$ for
the points in the window, that is, $| r_{max} - r_{min} |$, or the
maximum difference between $r$ values of consecutive points, that is,
$\max_j | r_j - r_{j+1} |$.

\subsubsection{Robust extraction of local features}
\label{sec:localfeat2}

\begin{figure}[!t]
\centering
\begin{tabular}{c}
\includegraphics[width=3.25in]{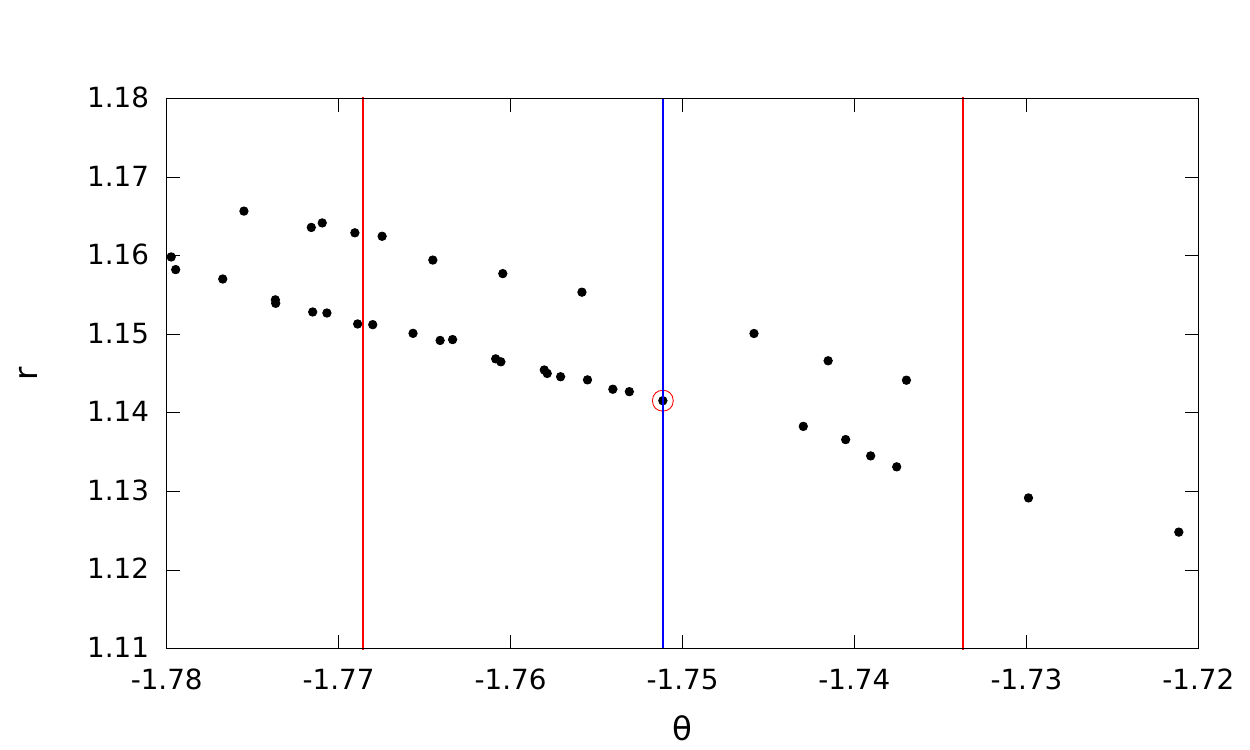} \\
\end{tabular}
\caption{A zoomed in view of a separatrix orbit showing a window
  centered at the point with $\theta = 1.75115$, indicated by the
  middle vertical line.  The two vertical lines on either side
  indicate the boundaries of a window which is $2^\circ =$ 0.0349
  radians wide. All points in this window are used to generate the
  features associated with the central point. }
\label{fig:delta_theta1}
\end{figure}

These observations prompted us to consider more robust ways to extract
local features.  To make the features rotation invariant, we
re-defined a window to be centered around {\it each} of the $m$ points
in an orbit, as shown in Figure~\ref{fig:delta_theta1}. Thus, instead
of a fixed number, $n$, of non-overlapping windows for all orbits, we
now had a variable number, $m$, of overlapping windows, each centered
at a point and subtending an angle of $\delta$, which we set to
$2^\circ $. We continued to use the normalization of the $r$ values as
explained earlier so the features are not influenced by the distance
of the orbit from the magnetic axis or the width of island and
separatrix lobes.

Next, to determine if the points in a window lay on one curve (as in
the case of quasiperiodic orbits) or two curves (as would be the case
for island and separatrix orbits), we obtained a least squares fit of
a second order polynomial to the points in a window and calculated the
maximum error across all points in this window; we refer to this
maximum error as {\it lserr}, or the least square error for the
window. This error is small for quasiperiodic orbits, but large for
the other classes, especially for windows where is a radial spread in the
spatial distribution of the points.

As the least squares fit, and other features explained next,
require a certain minimum number of points within a window, we only
consider windows with greater than four points in the rest of the
discussion. The points associated with such windows are referred to as
{\it valid} points.

Next, we required a feature that would represent the
spatial distribution of points within each window to help
differentiate the random scattering of the points in a stochastic
orbit from the more organized distribution of points on two curves
seen in the island chain and separatrix orbits.  We considered a
feature derived from quadrat counts in spatial
statistics\cite{cressie93:book}. A quadrat method essentially divides
a region into non-overlapping subsets, often rectangular in shape and
then counts the number of points in each subset. These counts can then
be used to determine if there is any spatial pattern in the data. A
completely random spatial pattern will have a Poisson distribution and
can be identified using the property that its mean is equal to its
variance. In our problem, we used a $4\times4$ grid in $(r, \theta)$
over each window, and since we have relatively few points in each
window, we used the number of non-zero cells in the grid to represent
the ``spread'' of the points. We expected that if a window has a
larger spread, it will have a more stochastic distribution of points.

We calculated two additional features to represent the distribution of
points across the orbit. The first was, {\it conc}, the concentration
of points for each valid window, defined as the number of points in
the window, scaled by the number of points in the orbit, so orbits
with larger number of points do not result in biased features. The
second feature was the difference in the $\theta$ values between a
point and the point to its left, which we refer to as {\it ldtheta}.
Note that we need to consider all points for {\it ldtheta}, not just
the valid points (that is, those with more than four points in the
window) because points with large {\it ldtheta} are often in windows
with few points.

\subsubsection{Calculation of derived features}
\label{sec:derivedfeat}

Thus far, we have considered an angular window in $\theta$ around each
point in the orbit and calculated quantities that characterize the
distribution of the points in the window. These local features are
calculated for windows with more than 4 points and, for each such
window, include {\it lserr}, which is the maximum error in fitting a
second order polynomial to the points; the spread of the points based
on quadrat counts; and {\it conc}, the concentration of points in the window,
which is the scaled number of points.  In addition, the
feature {\it ldtheta} at each point represented angular gaps in the orbit.

These vector-valued, local features describe  the
region around each valid point. To derive global
features, in the form of scalars, that can represent the orbit as a
whole, we calculated the mean, maximum,
minimum, and standard deviation of the local features across all the
valid windows in an orbit. In addition, for {\it ldtheta}, we defined
a new binary feature, {\it ldtheta\_large} that was set to 1 if an
orbit had any value of {\it ldtheta} greater than $10^\circ$, and 0
otherwise.

However, we found that these global quantities did not capture the
subtle variations among the orbit classes, such as changes in the
local features as we traversed along the orbit or any
patterns when two of the local features were considered together. For
example, a closer look at the separatrix orbits indicated that
the {\it conc} value in windows near an X point is higher than
in windows near the middle of a lobe, while the {\it lserr} value
is higher near the middle of the lobe and tapers off
near the X points. A similar behavior is seen between the center and
the two corners of each island in an island chain.

\begin{figure}[!htb]
\centering
\setlength\tabcolsep{1pt}
\begin{tabular}{cccc}
\includegraphics[trim = 3.5cm 0cm 3.9cm 0.0cm, clip = true,width=0.2\textwidth]{quasiperiodic.pdf} &
\includegraphics[trim = 3.5cm 0cm 3.9cm 0.0cm, clip = true,width=0.2\textwidth]{separatrix.pdf} &
\includegraphics[trim = 3.5cm 0cm 3.9cm 0.0cm, clip = true,width=0.2\textwidth]{islandchain.pdf} &
\includegraphics[trim = 3.5cm 0cm 3.9cm 0.0cm, clip = true,width=0.2\textwidth]{stochastic.pdf} \\
\vspace{0.2cm}
\includegraphics[width=0.24\textwidth]{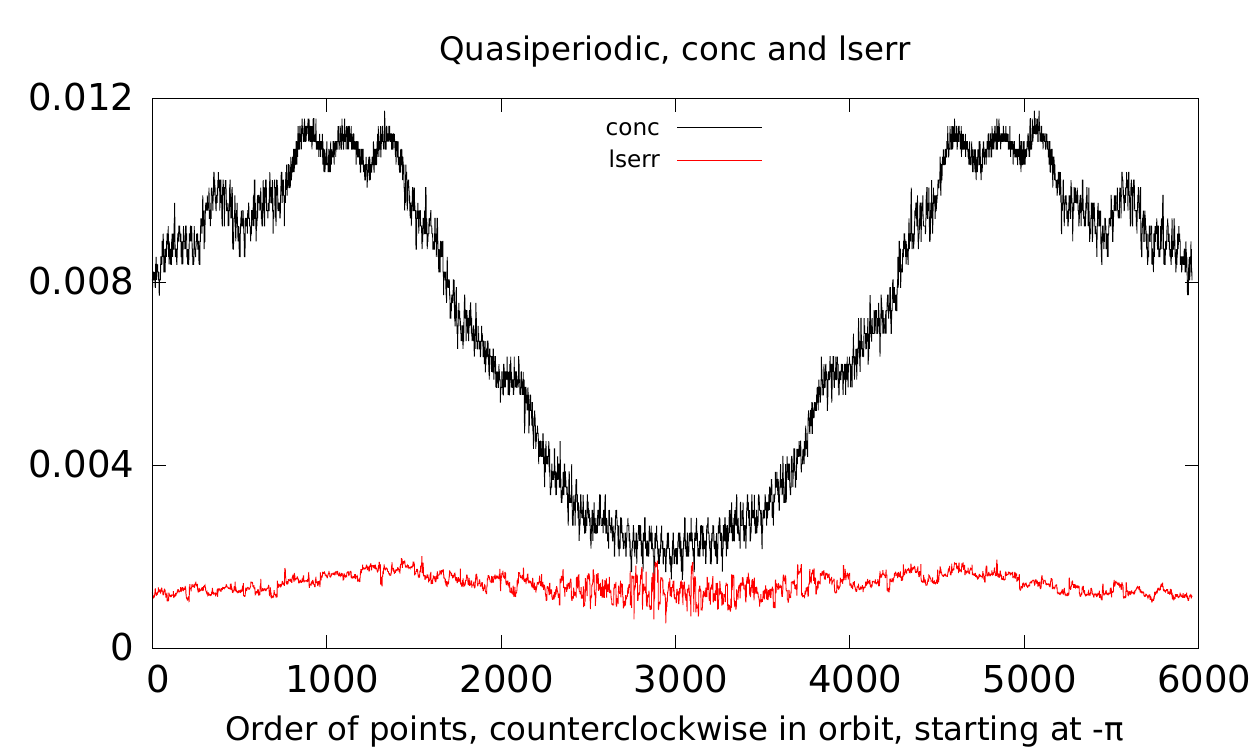} &
\includegraphics[width=0.24\textwidth]{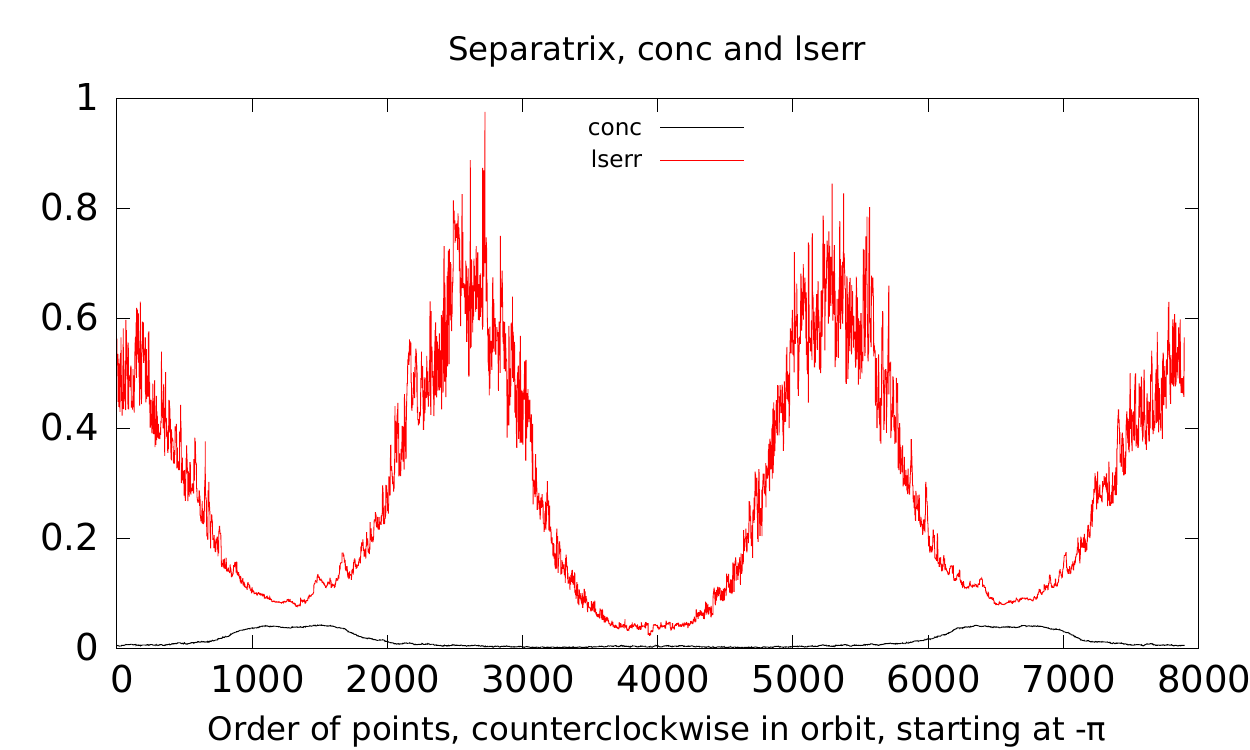} &
\includegraphics[width=0.24\textwidth]{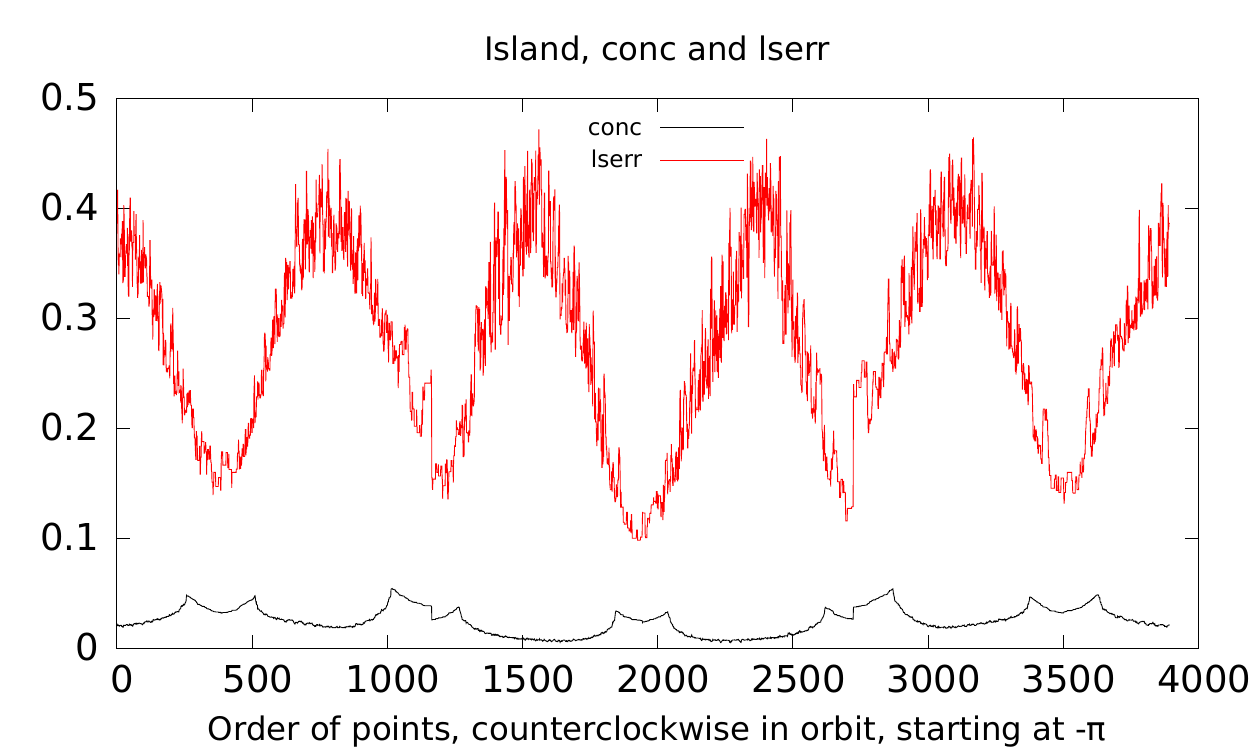} &
\includegraphics[width=0.24\textwidth]{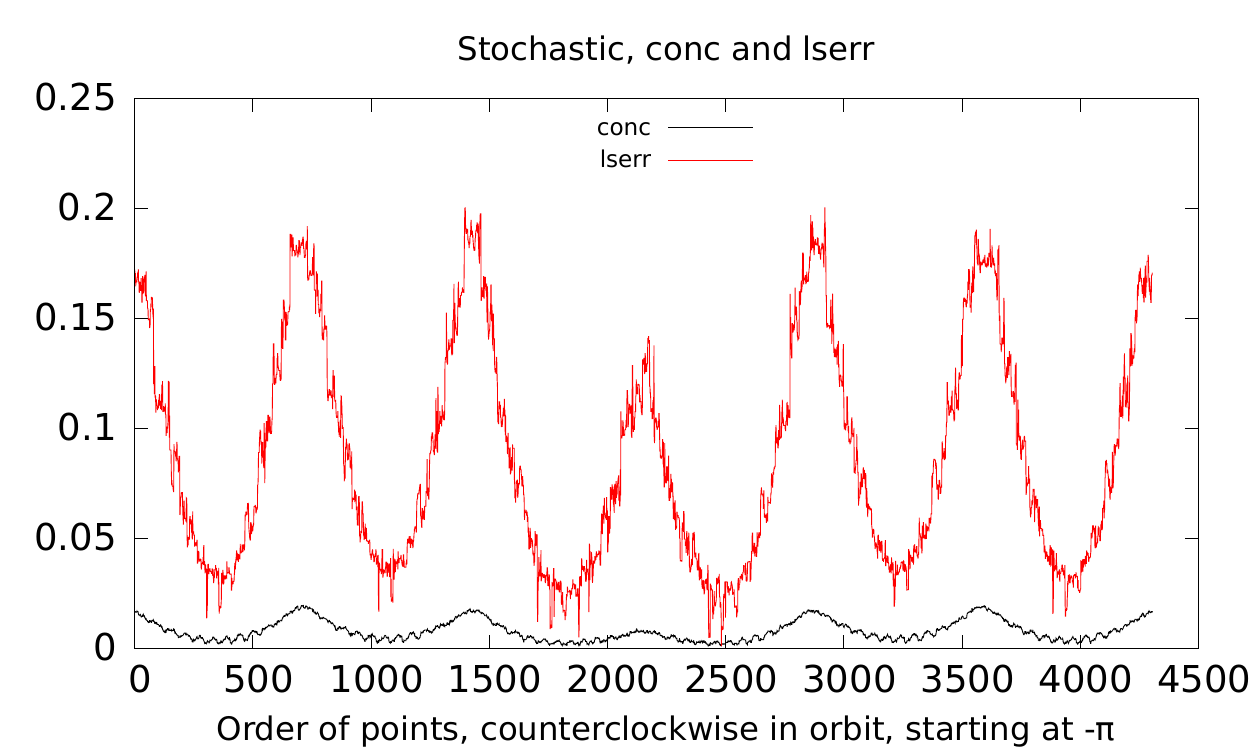} \\
\includegraphics[width=0.24\textwidth]{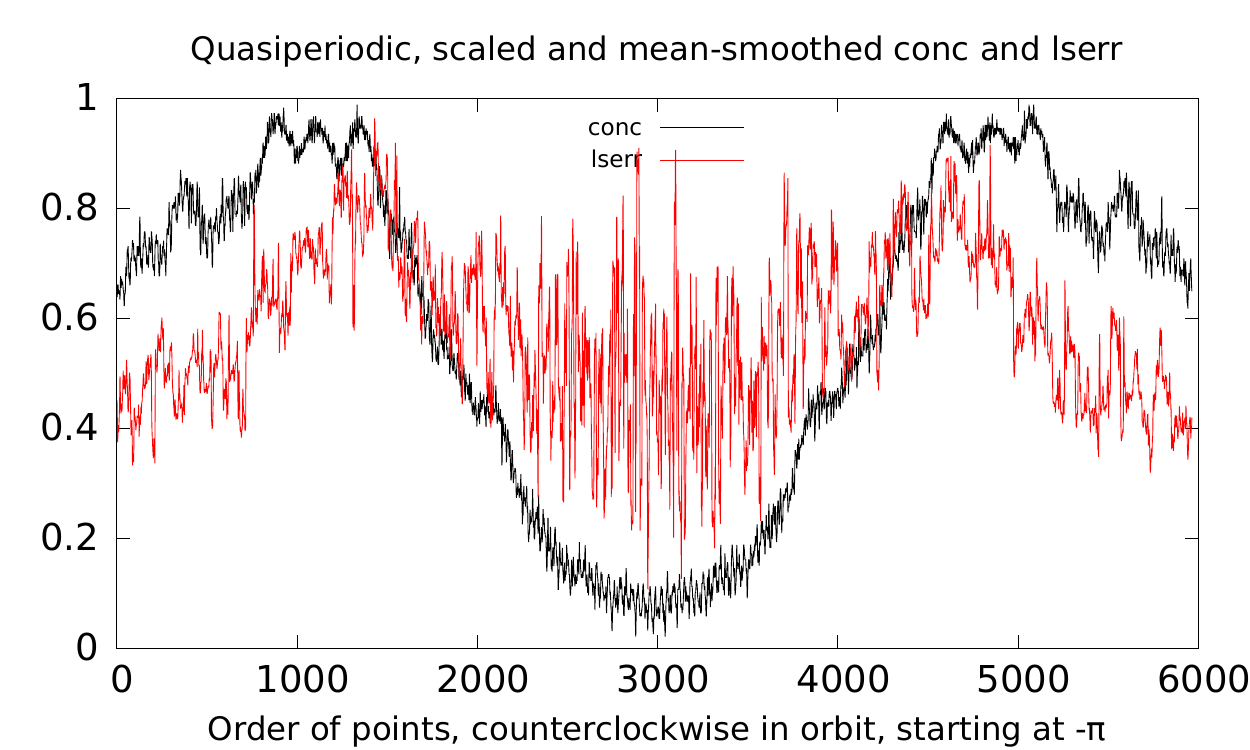} &
\includegraphics[width=0.24\textwidth]{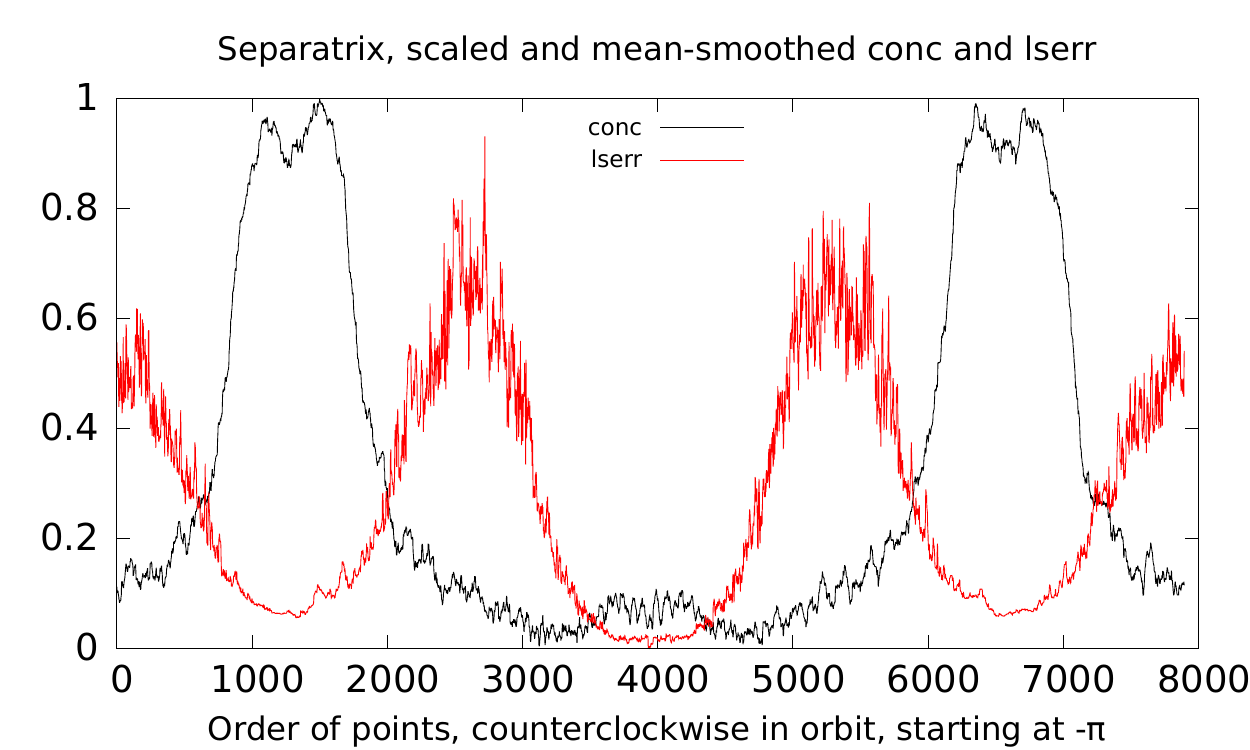} &
\includegraphics[width=0.24\textwidth]{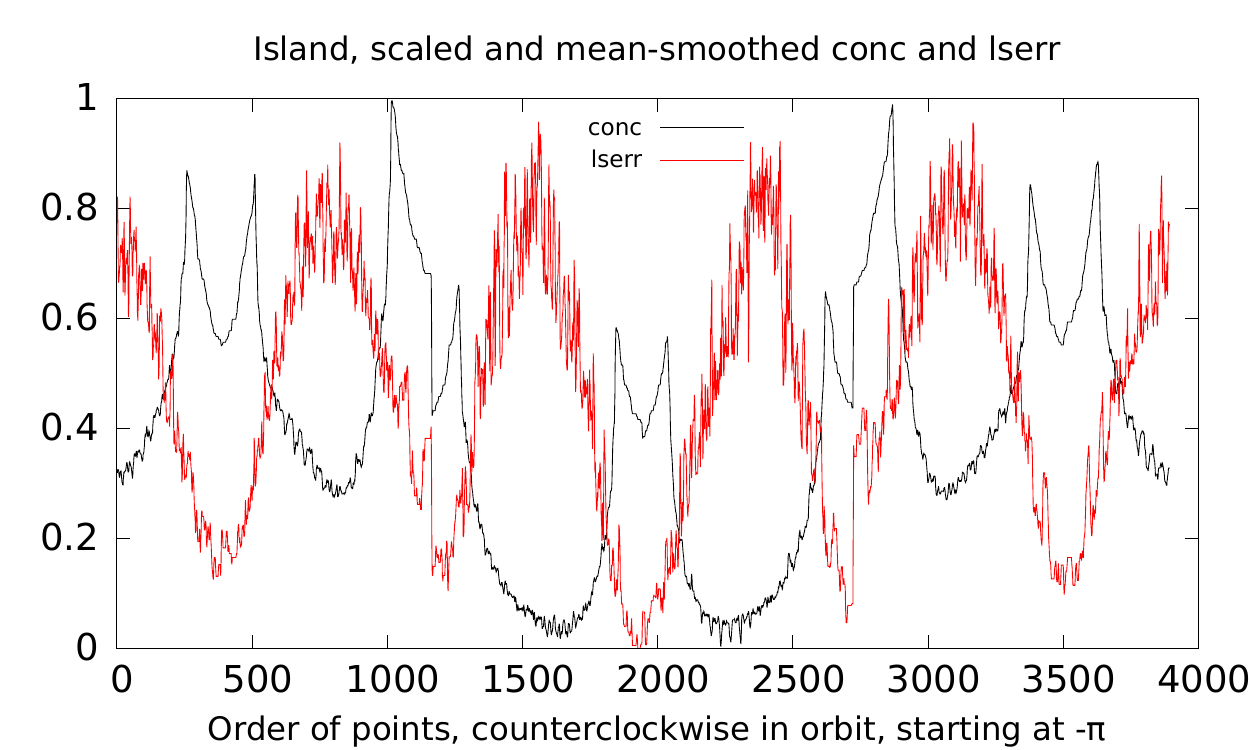} &
\includegraphics[width=0.24\textwidth]{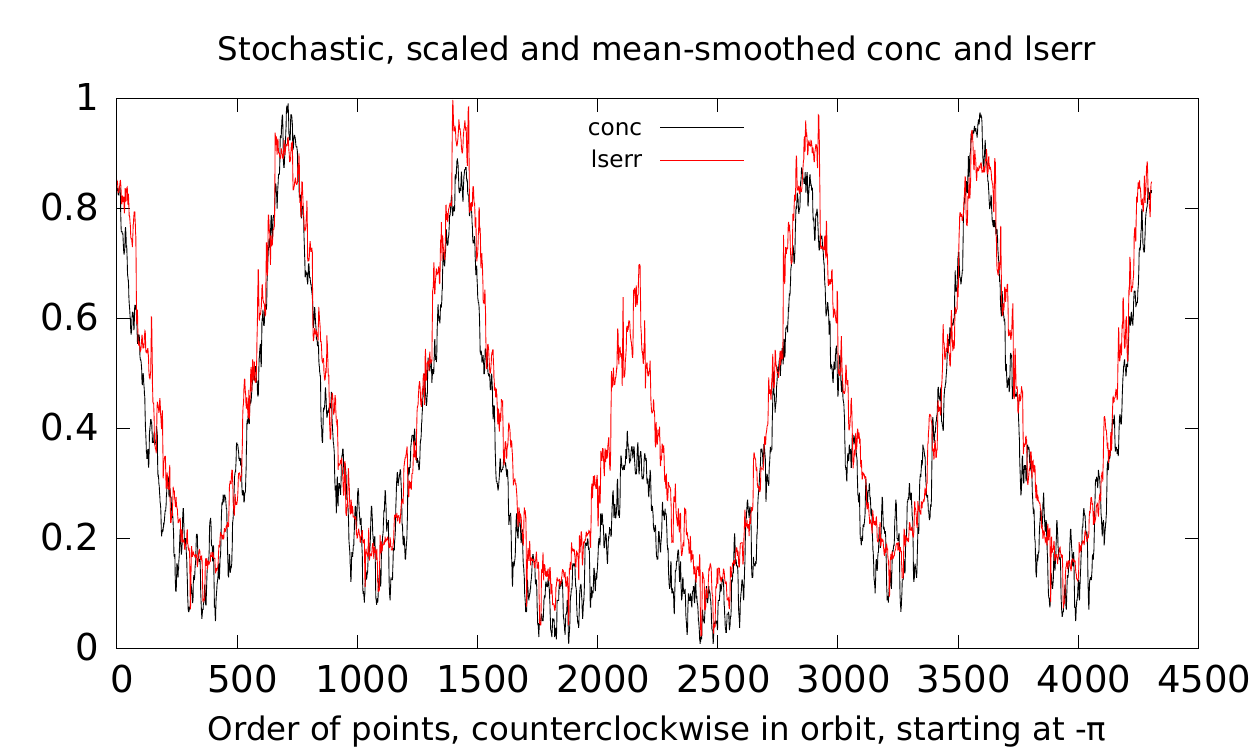} \\
\vspace{0.2cm}
\includegraphics[width=0.24\textwidth]{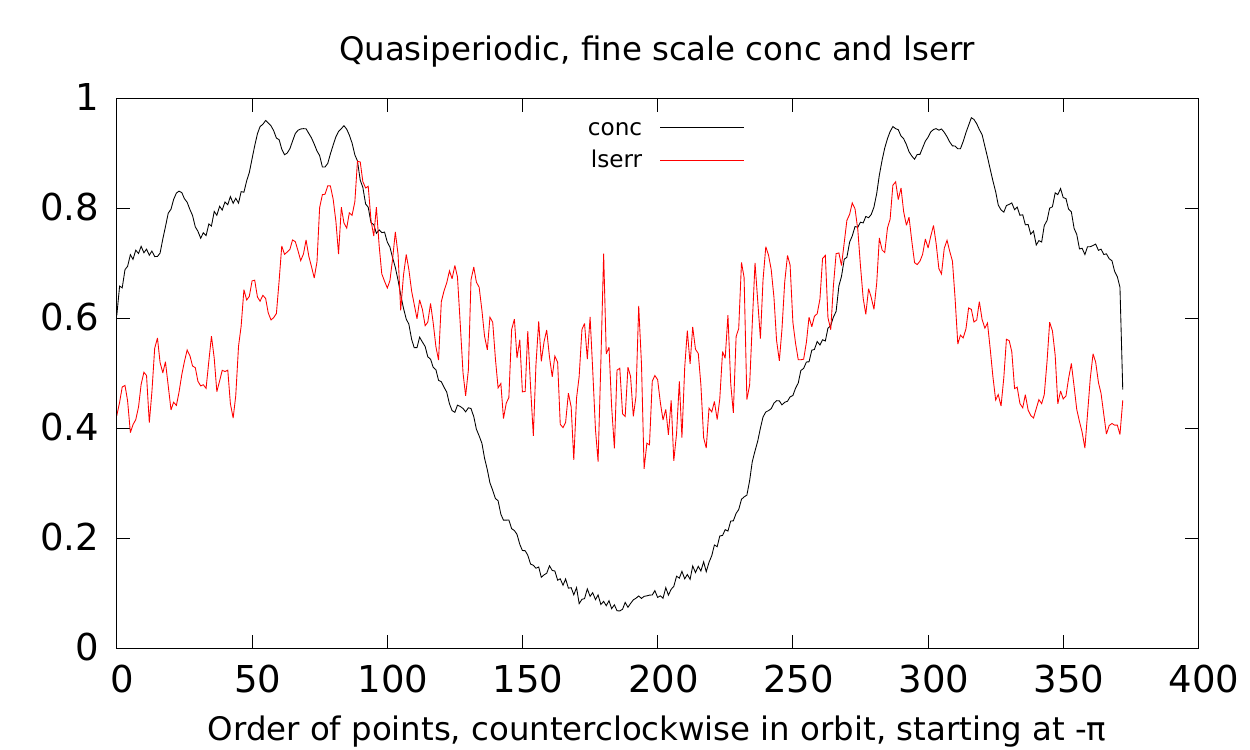} &
\includegraphics[width=0.24\textwidth]{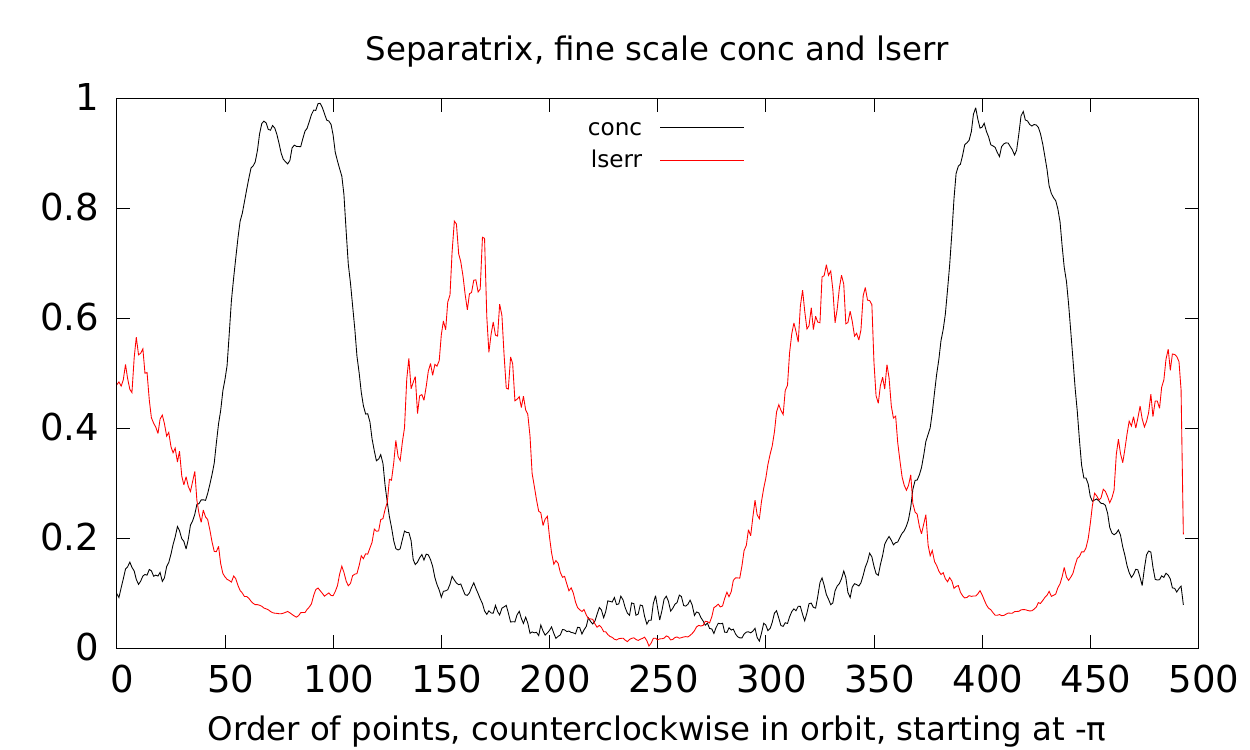} &
\includegraphics[width=0.24\textwidth]{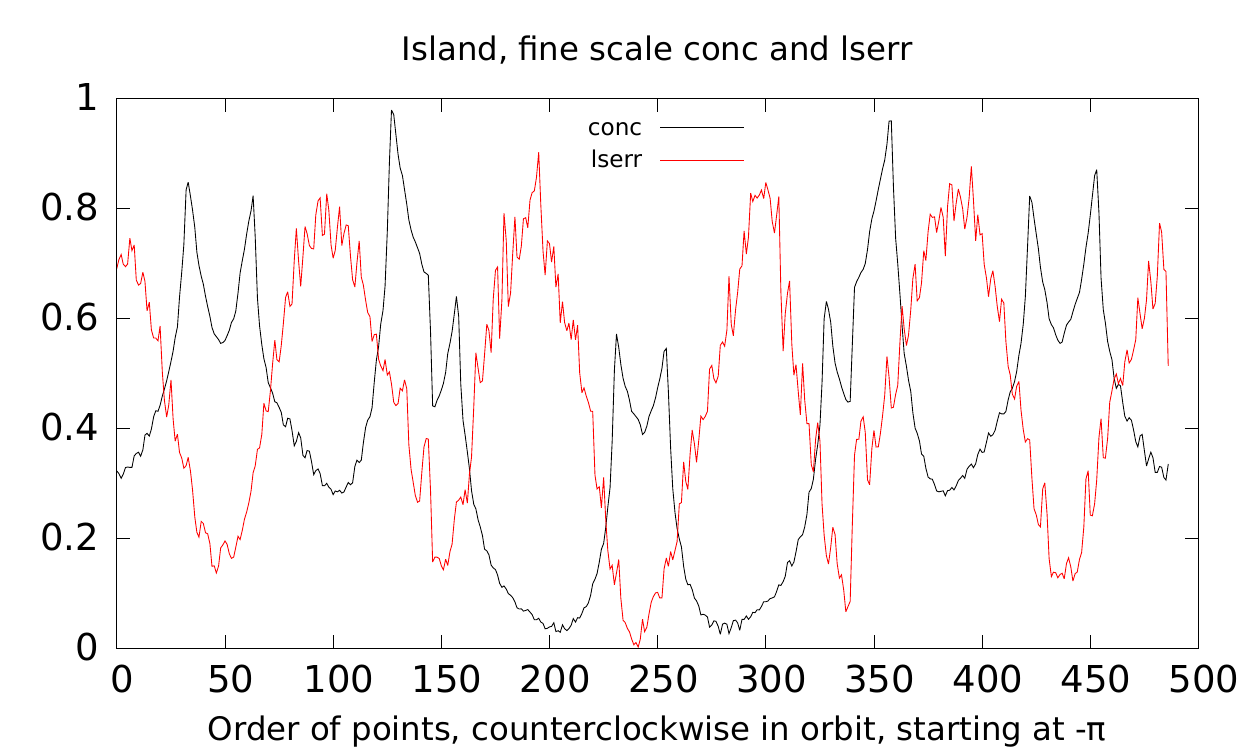} &
\includegraphics[width=0.24\textwidth]{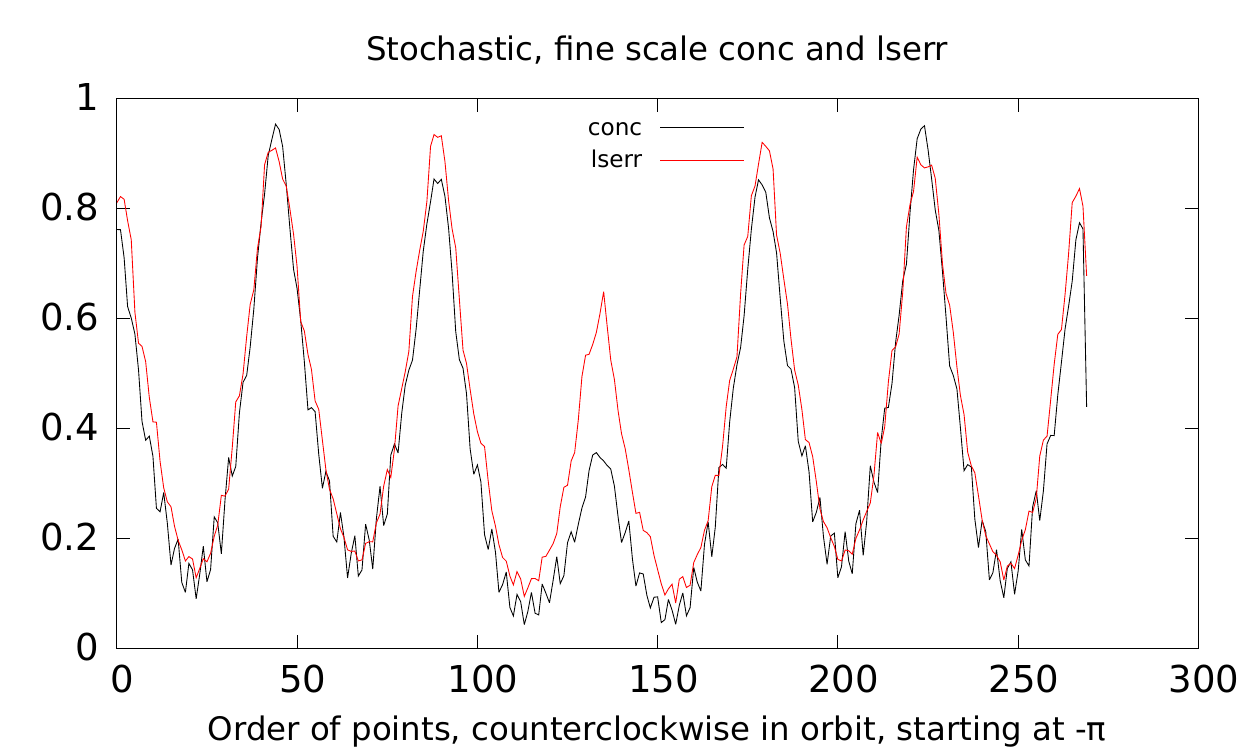} \\
\vspace{0.2cm}
\includegraphics[width=0.24\textwidth]{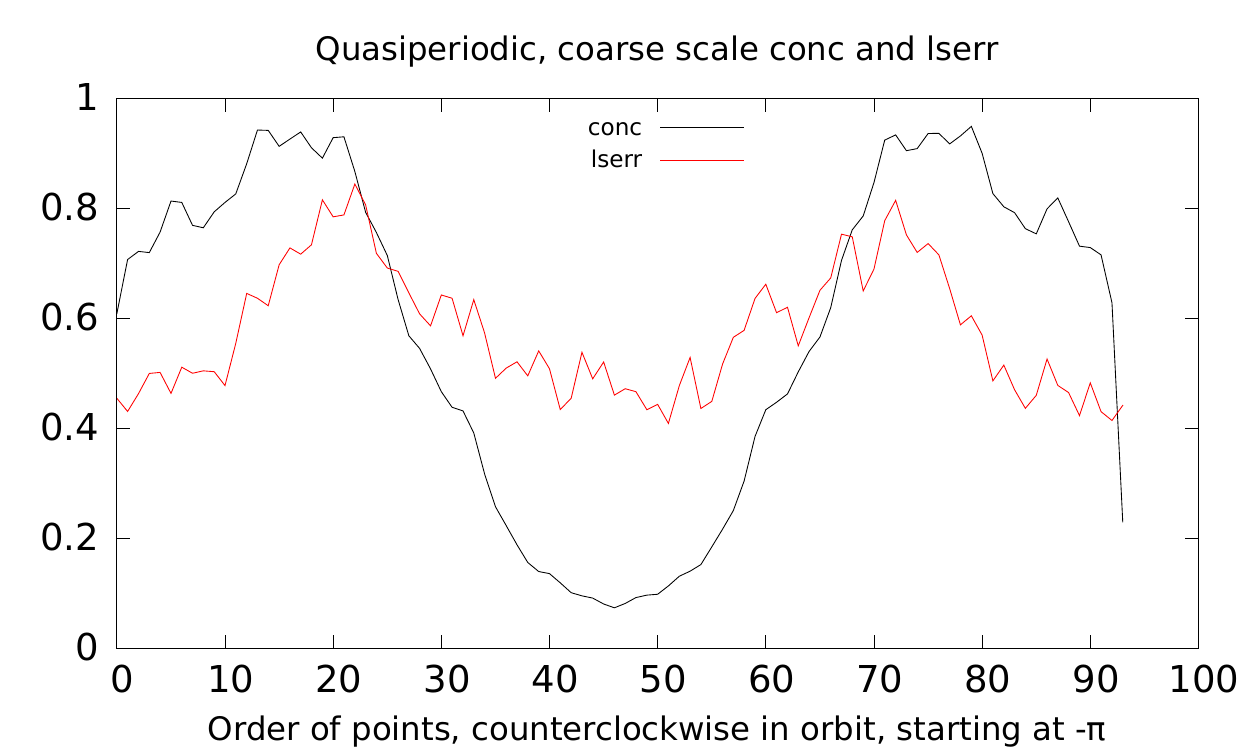} &
\includegraphics[width=0.24\textwidth]{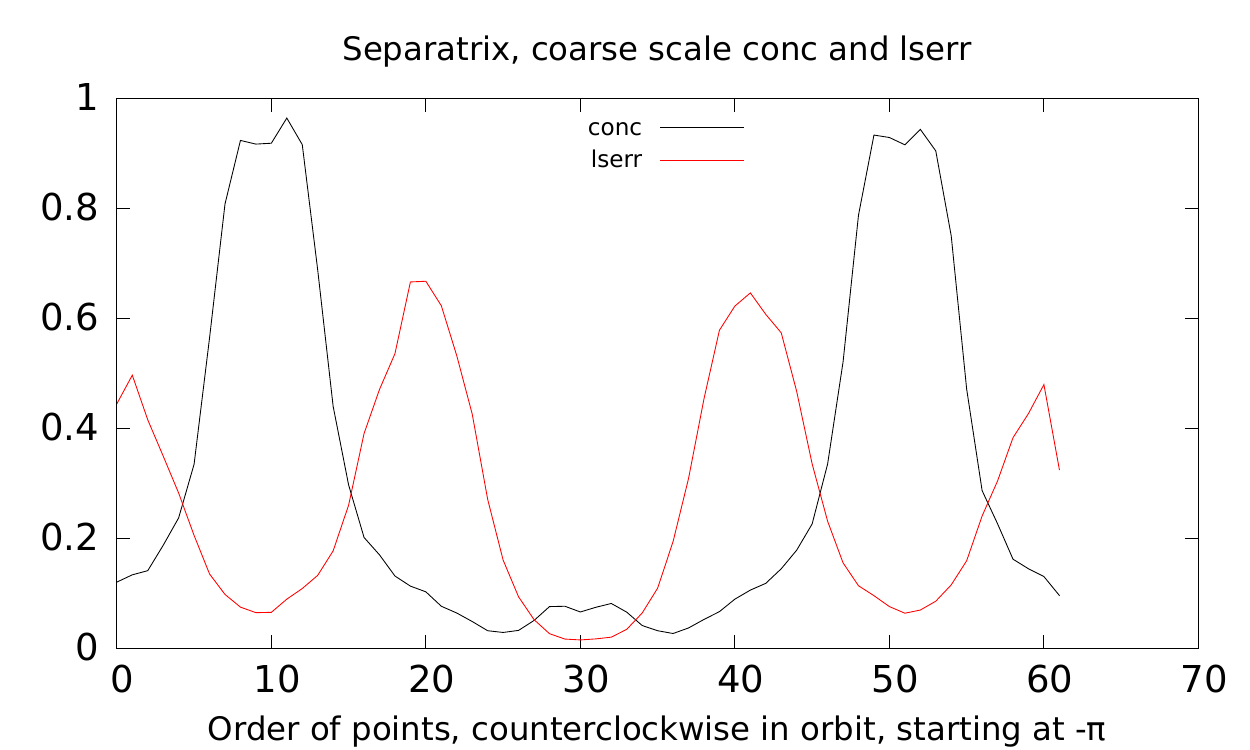} &
\includegraphics[width=0.24\textwidth]{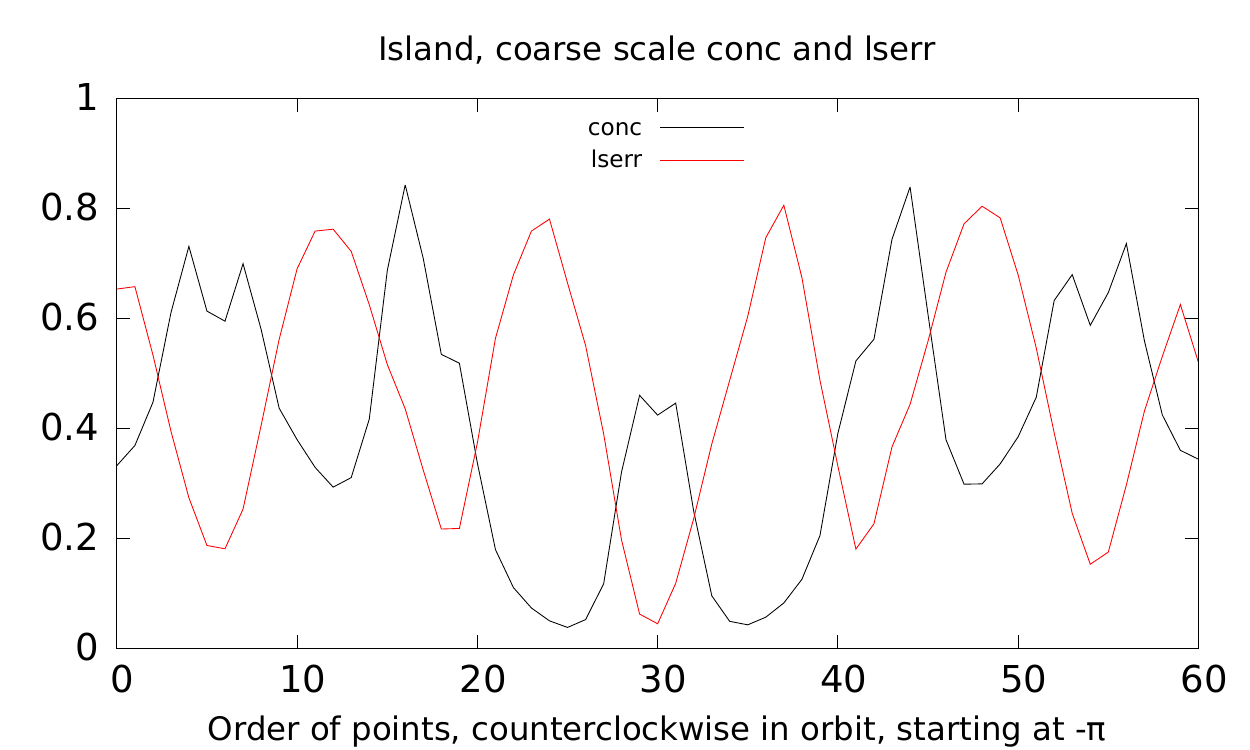} &
\includegraphics[width=0.24\textwidth]{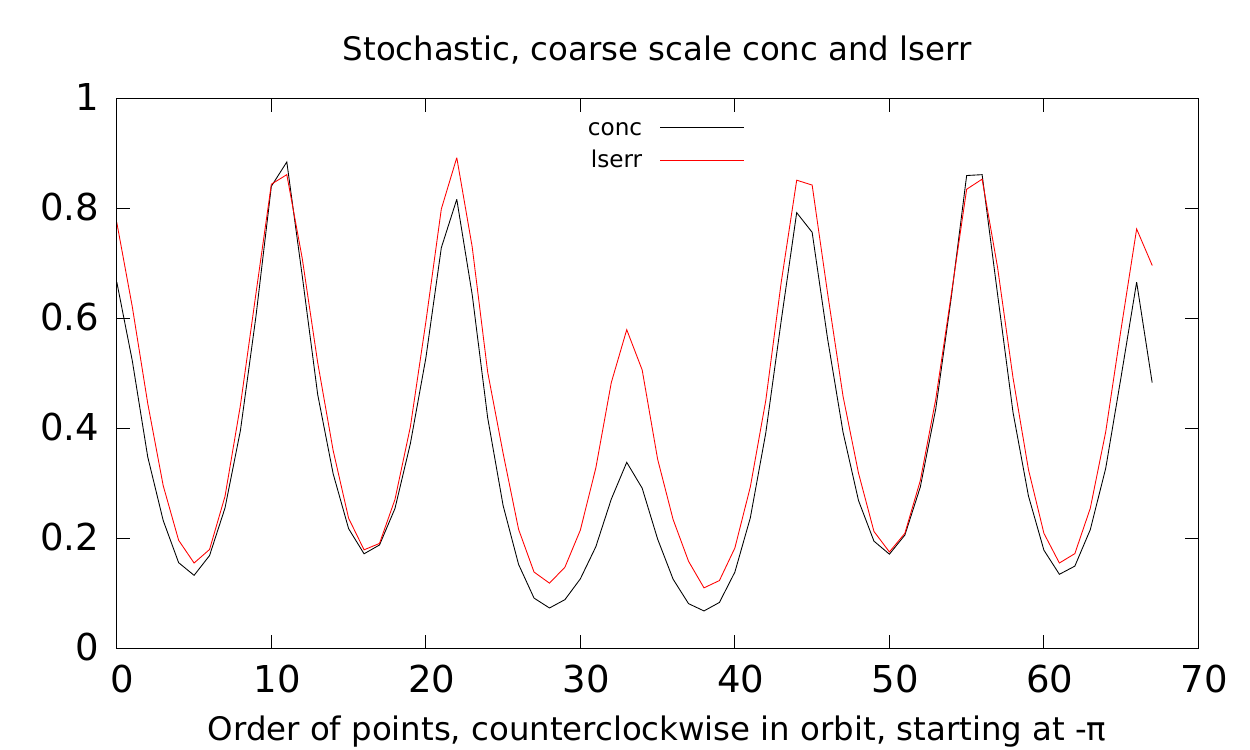} \\
\vspace{0.2cm}
\includegraphics[width=0.24\textwidth]{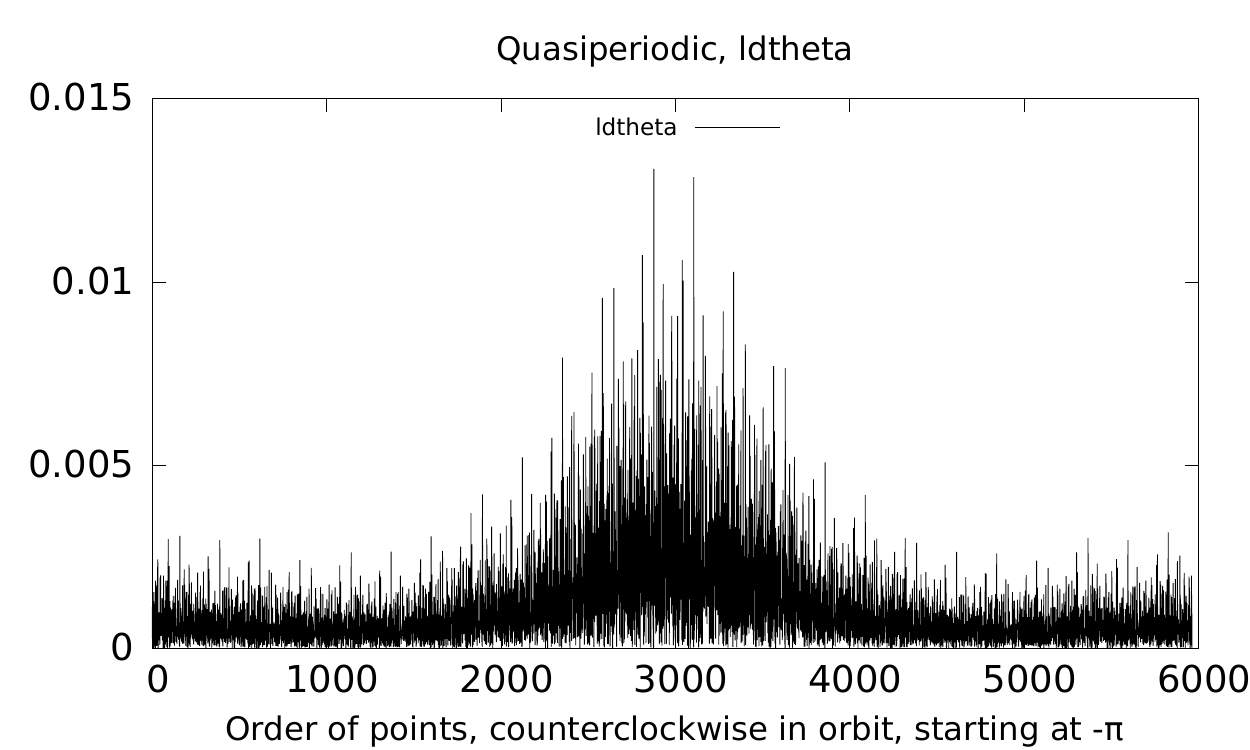} &
\includegraphics[width=0.24\textwidth]{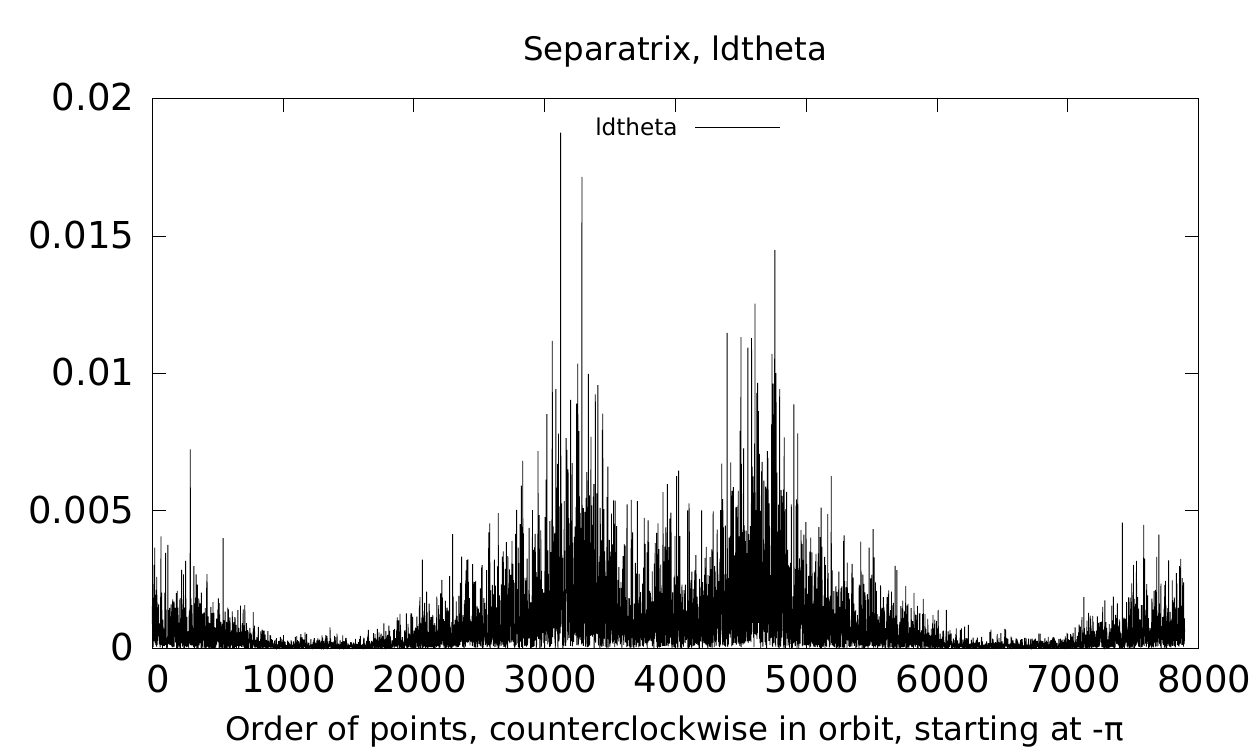} &
\includegraphics[width=0.24\textwidth]{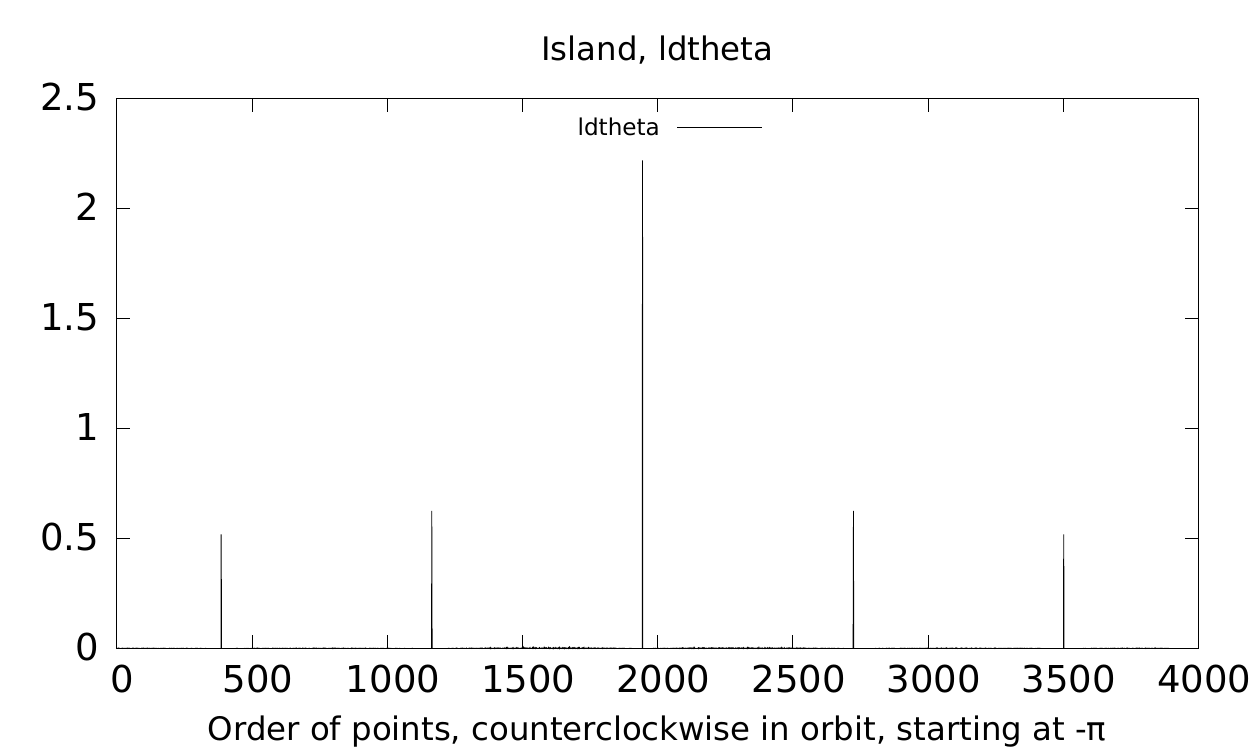} &
\includegraphics[width=0.24\textwidth]{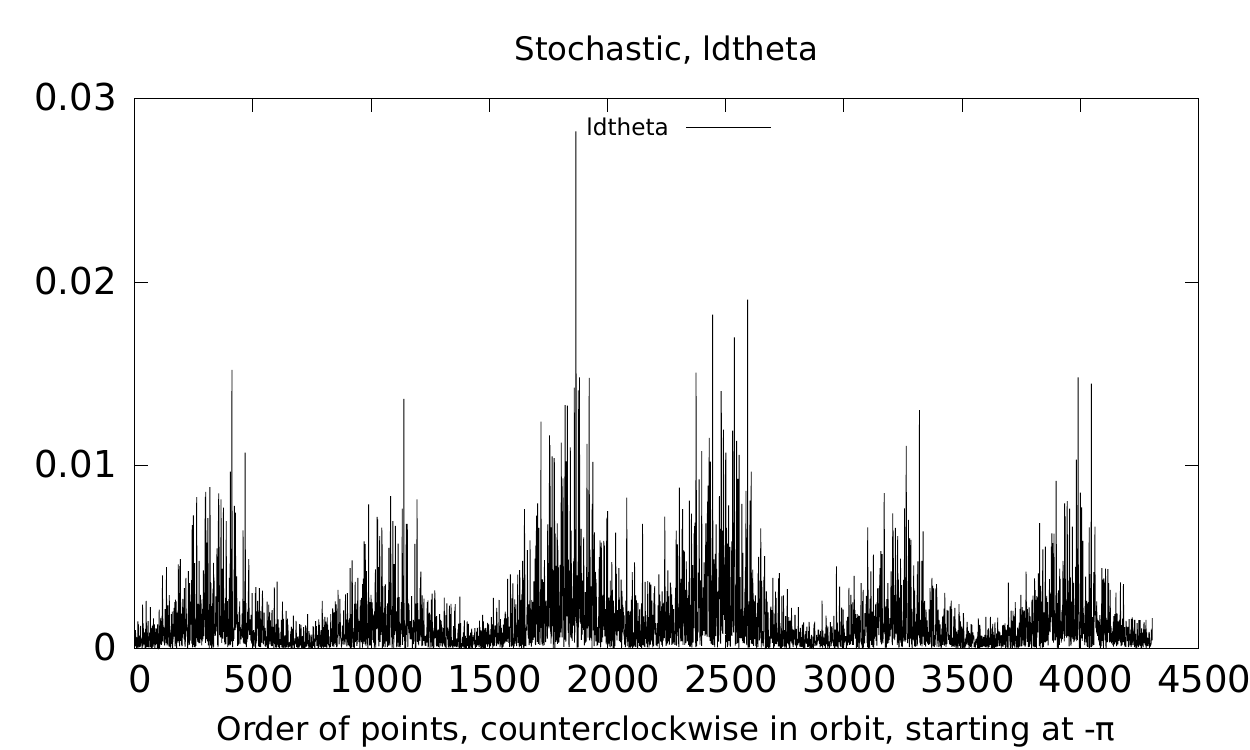} \\
\end{tabular}
\caption{The {\it lserr} and {\it conc} features, in red and black,
  respectively, for the four sample orbits from
  Figure~\ref{fig:orbits}.  Top row: original orbits, from left to
  right: quasiperiodic, separatrix, island chain, and stochastic.
  Second row: the {\it lserr} and {\it conc} features in the window
  around each valid point. Third row: the features scaled to [0,1]
  and smoothed using a simple mean filter of width 3.  Fourth row:
  Smoothed features obtained after decimated
  wavelet transform until the number of points in an orbit is between
  250 and 500. Fifth row: the features after additional smoothing with
  the decimated wavelet transform until fewer than 100 points are left
  in the orbit. Rows four and five are referred to as the fine and
  coarse versions of the two features after wavelet smoothing. Last
  row: value of the {\it ldtheta} variable. The points in rows two to
  six are ordered by value of $\theta$, going counterclockwise from
  $-\pi$ to $\pi$ radians.}
\label{fig:lserr_conc1}
\end{figure}

This observation prompted us to consider the {\it lserr} and {\it
  conc} features at each window together.
Figure~\ref{fig:lserr_conc1}, top row, shows the sample orbit of each
class from Figure~\ref{fig:orbits}. The second row shows the values of
{\it lserr} and {\it conc} features, in red and black, respectively,
where the x axis indicates all the valid points in the orbit, going counter
clockwise from $ - \pi$ to $\pi$ radians.
The plots in the second row indicate that the values of {\it lserr}
for the quasiperiodic orbit are much smaller than for the other three
orbits, which is expected.  Also, as expected, we see clear peaks and
valleys in {\it lserr} and {\it conc} for the island chain and
separatrix orbits, though surprisingly, similar peaks and valleys are
also present for the quasiperiodic and stochastic orbits.

Figure~\ref{fig:lserr_conc1}, third row shows the {\it lserr} and {\it
  conc} features after both have been scaled to lie between [0,1] and
after minimal smoothing with a mean filter of width 3.  We make two
main observations on these plots. First, the number of peaks is equal
to the number of lobes in the separatrix or the number of islands in
the island chain, but, for the quasiperiodic and stochastic orbits, it
is not possible to relate the number of peaks to any structures in the
orbits.  Second, there is a clear pattern in the locations
of the peaks and valleys for {\it lserr} and {\it conc}.  Stochastic
orbits have the peaks and valleys of the two curves aligned, both in
magnitude and location. In island chain, the peaks and valleys are of
different magnitude, with the peak of one aligned with the valley of
the other, and vice versa. For the separatrix orbit, the
magnitudes of the two features are different, with the valleys in {\it
  lserr} aligned with the peaks in {\it conc}. In the quasiperiodic
orbit, the two features have similar magnitudes and their peaks and
valleys are aligned, with the latter more so than the former.

To translate these observations into scalar features we wanted to
identify the peaks and valleys in {\it lserr} and {\it conc} and
evaluate their alignment.  However, both curves are quite noisy and the
minimal smoothing used was insufficient. To further reduce the high
frequency components, we applied wavelet smoothing
with the biorth 3,1 wavelet~\cite{burrus98:book,hubbard98:book}, as it
is a symmetric wavelet with a small support.  However, we could not
use a fixed number of wavelet levels to apply to each orbit as the
length of the two curves varies across orbits.  Instead, we applied
the decimated wavelet transform until the number of points in the
smooth, low-frequency part of the data was in the range [250,500) (for
the fine level), and less than 100 (for the coarse level), as shown in
Figure~\ref{fig:lserr_conc1}, third and fourth rows, respectively. We
chose two levels to identify the peak-valley alignment in {\it lserr}
and {\it conc} in case the fine level still had some remaining noise,
as seen in some of the orbits.

Finally, the last row in Figure~\ref{fig:lserr_conc1} shows the values
of {\it ldtheta} for each of the orbits - note the high peaks for the
island chain that are large enough to completely dwarf the values in
between the peaks. We used these plots to select the threshold for the
{\it ldtheta\_large} feature.

Next, we used a simple peak finder, and its converse as a valley
finder, to identify the peaks and valleys in the coarse and fine
resolutions of the two vectors. At each resolution, given the total
number of peaks and valleys in {\it lserr} (or {\it conc}), we
identified the fraction of the peaks and valleys of {\it lserr}
aligned with the peaks and valleys, respectively, of {\it conc}, and
the fraction that were misaligned, that is, the peaks/valleys of {\it
  lserr} were aligned with the valleys/peaks of {\it conc}. These
features allowed us to represent the behavior we see in
Figure~\ref{fig:lserr_conc1}.  In addition, to account for cases where
the peak and valley locations could be off by a small amount, we also
calculated an additional feature, the sign alignment; this is the
fraction of points where the sign of $\Delta${\it lserr} and
$\Delta${\it conc} is the same, that is, where {\it lserr} and {\it
  conc} are both increasing or both decreasing.  This fraction would
be large in orbits where the peaks and valleys in the two vectors are
aligned.

Table~\ref{tab:features} lists the 17 features that represent each
orbit; the first three (file name, total number of points in orbit,
and the number of points used in the instance) are not used in
classification but help to identify the provenance of an orbit.  The
remaining features were selected after initial experiments with
ensembles of decision trees to remove features that were not
discriminating.  For example, features related to {\it conc}, such as
its minimum, maximum, and mean across the windows, did not convey any
information about the class of an orbit, but the vector of {\it conc}
values was useful when combined with the {\it lserr} vector to
identify the relationship of peaks and valleys between the two
variables.

\renewcommand{\arraystretch}{1.2}
\begin{table}[htb]
  \begin{center}
    \begin{tabular}{|p{2.3cm} | p{12cm} |}
      \hline
      Feature name & Description \\
      \hline
      f0 & Orbit file name \\
      \hline
      f1 & Total number of points in orbit \\
      \hline 
      f2 & Number of points used in this instance \\
      \hline
      f3 & Maximum {\it lserr}, across windows, from polynomial fit in each window\\
      \hline
      f4 & Minimum {\it lserr}, across windows, from polynomial fit in each window\\
      \hline
      f5 & Mean {\it lserr}, across windows, from polynomial fit in each window \\
      \hline
      f6 & Standard deviation of {\it lserr}, across windows, from polynomial fit in each window \\
      \hline
      f7 & Mean spread, across windows \\
      \hline
      f8 & Maximum spread, across windows \\
      \hline
      f9 & Minimum spread, across windows \\
      \hline
      f10 & Fraction of peaks and valleys aligned at fine scale \\
      \hline
      f11 & Fraction of peaks  and valleys misaligned at fine scale \\
      \hline
      f12 & Fraction of valid points with sign alignment of $\delta${\it lserr} and $\delta${\it conc} at fine scale \\
      \hline
      f13 & Fraction of peaks and valleys aligned at coarse scale \\
      \hline
      f14 & Fraction of peaks and valleys misaligned at coarse scale \\
      \hline
      f15 & Fraction of valid points with sign alignment of $\delta${\it lserr} and $\delta${\it conc} at coarse scale \\
      \hline
      f16 & Any points where ldtheta is greater than 10 degrees (Boolean variable)\\
      \hline
    \end{tabular}
  \end{center}
  \vspace{-0.2cm}
  \caption{A description of the features extracted for the orbits. The first three 
  features are for identifying the provenance of each orbit and are not used in the 
  classification.}
  \label{tab:features}
\end{table}

\subsection{Feature Selection}
\label{sec:featsel}

Once we have identified and extracted the features for each orbit, it
can be helpful to determine if some features are more important than
others in classifying the orbits into the four classes. We considered
three feature selection methods, described in detail
in~\cite{cantupaz04:kdd}, and summarized briefly as follows:

\begin{itemize}

\item {\bf Distance filter:} Intuitively, a more discriminating
  feature will have greater distance between the histograms of the
  different classes. For each feature, we create a histogram of the
  values of the feature for each of the four classes and obtain the
  the Kullback-Leibler (KL) distance between these histograms. The
  features are ranked by sorting them in descending order of the
  distances.

\item {\bf Chi-square filter:} This method ranks features by sorting
  them in descending order of Chi-square statistics computed from
  their contingency tables~\cite{huang03:dimred}. As our features are
  all numeric (except for f16), they are first discretized using
  histograms, before the Chi-square statistic for each feature is
  calculated.

\item {\bf Stump filter:} A simple approach to ranking features by
  importance is to use the split criterion in decision trees that
  determine which feature to split on at any node of the tree.  We
  consider just the root node (hence the name of this method), where
  the entire training data set is used. In our work, we use the Gini
  index~\cite{breiman84:book} and rank the features according to the
  purity of their optimal split.
 
\end{itemize}

\subsection{Classification}
\label{sec:classification}

We have many options for a classifier to use for predicting the class
of an orbit. We selected decision trees mainly because they allowed us
to understand how decisions were made in assigning the class to an
orbit, helping us to improve iteratively, the correctness of class
labels and the features extracted for each orbit.  The specific
algorithm we used was ASPEN, a method to generate an ensemble of trees
based on approximate splits~\cite{kamath02:histogram}. We introduce
randomization at each node of the tree in two ways - first, by
randomly sampling the instances at a node and selecting a fraction (we
use 0.7) for further consideration, and second, by replacing the
sorting of feature values by a histogram.  At each node, we create a
histogram of feature values for the sampled instances, evaluate the
splitting criterion at the mid-point of each bin of the histogram,
identify the best bin, and then select the split point randomly in
this bin.  The use of the histograms and the smaller number of samples
speeds up the creation of each tree in the ensemble. We use two
different splitting criteria - Gini~\cite{breiman84:book} and
InfoGain~\cite{quinlan86:induction} for comparison.

%
\section{Experimental Results and Discussion}
\label{sec:results}
%

We next present the results of feature selection and classification
for the features listed in Table~\ref{tab:features}. We started with
264 orbits, and after removing the orbits with too few points, we
split them into individual orbits comprised of the first $1000, 1500,
\ldots$ points. This resulted in a training set of 1884 instances,
each represent by 14 features, f3 through f16, listed in
Table~\ref{tab:features}.

\begin{figure}[!htb]
\centering
\setlength\tabcolsep{1pt}
\begin{tabular}{c}
\includegraphics[width=0.5\textwidth]{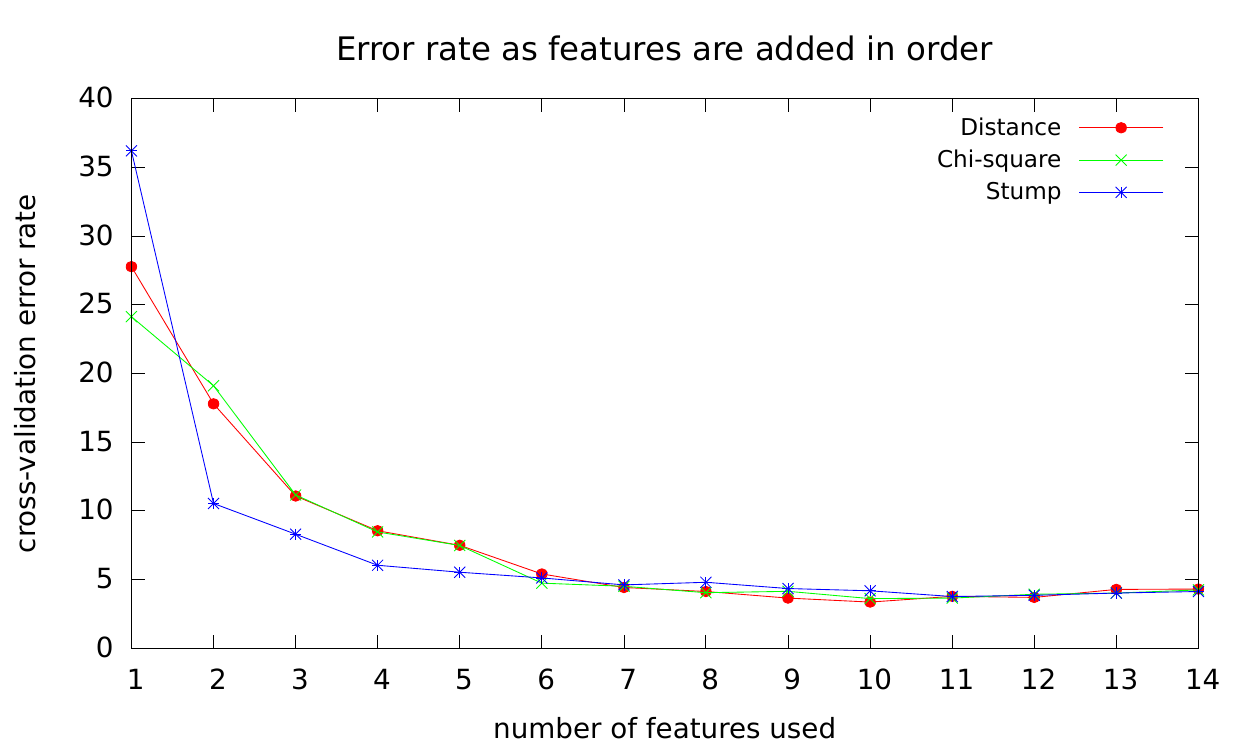} \\
\end{tabular}
\vspace{-0.4cm}
\caption{Error rate from five runs of five-fold cross validation for
  the top $k$ features identified as important by each of the three
  feature selection methods. The error stabilizes at 6-7 features. The
  Gini split criterion is used with eleven trees in the ensemble.  }
\label{fig:featsel}
\end{figure}

First, Table~\ref{tab:featsel} lists the features in descending order
of importance as identified by the three feature selection methods
when applied to the full data set.  Figure~\ref{fig:featsel} shows how
the error rate for five runs of five-fold cross validation varies when
we use only the first $k$ features in order of importance. The error
rate is for eleven trees using the Gini split criterion and the
ensemble approach described in Section~\ref{sec:classification}.

\renewcommand{\arraystretch}{1.1}
\begin{table}[!htb]
  \begin{center}
    \begin{tabular}{|l|l|l|l|}
      \hline
      Rank order & Distance filter &  Chi-square filter &  Stump filter \\
      \hline
      1   &  f3   &    f8   &   f16  \\
      \hline
      2   &  f8   &    f7   &   f3   \\
      \hline
      3   &  f5   &    f3   &   f8   \\
      \hline
      4   &  f6   &    f5   &   f5   \\
      \hline
      5   &  f7   &    f6   &   f6   \\
      \hline
      6   &  f16  &    f16  &   f7   \\
      \hline
      7   &  f15  &    f15  &   f4   \\
      \hline
      8   &  f4   &    f4   &   f14  \\
      \hline
      9   &  f12  &    f14  &   f15  \\
      \hline
      10  &  f10  &    f12  &   f9   \\
      \hline
      11  &  f14  &    f10  &   f12  \\
      \hline
      12  &  f11  &    f13  &   f10  \\
      \hline
      13  &  f13  &    f11  &   f13  \\
      \hline
      14  &  f9   &    f9   &   f11  \\
      \hline
    \end{tabular}
  \end{center}
  \vspace{-0.2cm}
  \caption{Ordering, by importance, of the  14 features representing the orbits, 
    using three different feature selection methods. Note that the first six 
    features identified by all three methods are the same, and based on 
    Figure~\ref{fig:featsel} reduce the error rate to $\approx$5\%.}
  \label{tab:featsel}
\end{table}

These results indicate that all three feature selection methods select
the same top six features as important, though the order changes with
the method as each uses a different metric to evaluate the importance.
Figure~\ref{fig:featsel} indicates that when these top six features
are considered, the error rate drops to 5\%, and stabilizes
thereafter. These six features are {\it ldtheta\_large}, the mean and
maximum spread across windows, and the maximum, mean, and standard
deviation of {\it lserr}, taken across windows, where {\it lserr} for
a window is the maximum error from a polynomial fit in that window.
Interestingly, the peak-valley features are ranked lower in
importance; this is likely because features indicating the angular
gaps, the radial spread, and the number of curves (one or two) in each
window, are more discriminating when the full data set is considered
as is the case in the feature selection methods.

\renewcommand{\arraystretch}{1.2}
\begin{table}[!htb]
  \begin{center}
    \begin{tabular}{|c| l| l | l|} 
      \hline
      & 5 trees & 11 trees & 21 trees \\
      \hline
      Gini & 4.55 (0.05) & 4.02 (0.13) & 3.98 (0.07) \\
      \hline
      InfoGain & 3.89 (0.08) & 3.60 (0.15) & 3.46 (0.13) \\
      \hline
    \end{tabular}
  \end{center}
  \vspace{-0.2cm}
  \caption{Error rate for five runs of five-fold cross-validation using Gini and InfoGain split criteria. The values in parenthesis are the standard error across the five folds.}
  \label{tab:xvalid}
\end{table}

\begin{table}[!tb]
  \centering
  \renewcommand{\arraystretch}{1.2}
  %
  %
  \begin{subtable}[h]{0.45\textwidth}
  \centering
  \begin{tabular}{ll|l|l|l|l|}   
    \multicolumn{2}{c}{}&   \multicolumn{4}{c}{Predicted Classes}\\
    \multicolumn{2}{c}{} & 
    \multicolumn{4}{c}{{\rotatebox[origin=c]{0}{Q \ \quad}
      } {\rotatebox[origin=c]{0}{S \ \quad}
      } {\rotatebox[origin=c]{0}{I \ \quad}
      } {\rotatebox[origin=c]{0}{R}
      }}\\
    \cline{3-6}
    \multirow{4}{1em}{{\rotatebox[origin=c]{90}{Actual Classes}}} & Q & 757 & 15 & 3 & 3 \\ 
    \cline{3-6}
    &   S & 28 & 212 & 2 & 8 \\ 
    \cline{3-6}
    &   I & 6 & 2 & 432& 0 \\ 
    \cline{3-6}
    &   R & 2 & 6 & 6 & 402 \\ 
    \cline{3-6}
  \end{tabular}
  \vspace{0.3cm}
  \subcaption{Five trees, Gini criterion, 4.3\% error.}
  \end{subtable}
  \vspace*{0.5cm} 
  \begin{subtable}[h]{0.45\textwidth}
  \centering
  \begin{tabular}{ll|l|l|l|l|}   
    \multicolumn{2}{c}{}&   \multicolumn{4}{c}{Predicted Classes}\\
    \multicolumn{2}{c}{} & 
    \multicolumn{4}{c}{{\rotatebox[origin=c]{0}{Q \ \quad}
      } {\rotatebox[origin=c]{0}{S \ \quad}
      } {\rotatebox[origin=c]{0}{I \ \quad}
      } {\rotatebox[origin=c]{0}{R}
      }}\\
    \cline{3-6}
    \multirow{4}{1em}{{\rotatebox[origin=c]{90}{Actual Classes}}} & Q & 762 & 8 & 3 & 5 \\ 
    \cline{3-6}
    &   S & 18 & 221 & 2 & 9 \\ 
    \cline{3-6}
    &   I & 4 & 3 & 433 & 0 \\ 
    \cline{3-6}
    &   R & 0 & 6 & 6 & 404 \\ 
    \cline{3-6}
  \end{tabular}
  \vspace{0.3cm}
  \subcaption{Five trees, InfoGain criterion, 3.4\% error.}
  \end{subtable}
  %
  %
  \begin{subtable}[h]{0.45\textwidth}
  \centering
  \begin{tabular}{ll|l|l|l|l|}   
    \multicolumn{2}{c}{}&   \multicolumn{4}{c}{Predicted Classes}\\
    \multicolumn{2}{c}{} & 
    \multicolumn{4}{c}{{\rotatebox[origin=c]{0}{Q \ \quad}
      } {\rotatebox[origin=c]{0}{S \ \quad}
      } {\rotatebox[origin=c]{0}{I \ \quad}
      } {\rotatebox[origin=c]{0}{R}
      }}\\
    \cline{3-6}
    \multirow{4}{1em}{{\rotatebox[origin=c]{90}{Actual Classes}}} & Q & 762 & 12 & 2 & 2 \\ 
    \cline{3-6}
    &   S & 21 & 220 & 2 & 7 \\ 
    \cline{3-6}
    &   I & 6 & 2 & 432& 0 \\ 
    \cline{3-6}
    &   R & 0 & 6 & 6 & 404 \\ 
    \cline{3-6}
  \end{tabular}
  \vspace{0.3cm}
  \caption{Eleven trees, Gini criterion, 3.5\% error.}
  \end{subtable} 
  \vspace*{0.5cm} 
  \begin{subtable}[h]{0.45\textwidth}
  \centering
  \begin{tabular}{ll|l|l|l|l|}   
    \multicolumn{2}{c}{}&   \multicolumn{4}{c}{Predicted Classes}\\
    \multicolumn{2}{c}{} & 
    \multicolumn{4}{c}{{\rotatebox[origin=c]{0}{Q \ \quad}
      } {\rotatebox[origin=c]{0}{S \ \quad}
      } {\rotatebox[origin=c]{0}{I \ \quad}
      } {\rotatebox[origin=c]{0}{R}
      }}\\
    \cline{3-6}
    \multirow{4}{1em}{{\rotatebox[origin=c]{90}{Actual Classes}}} & Q & 763 & 9 & 3 & 3 \\ 
    \cline{3-6}
    &   S & 19 & 220 & 2 & 9 \\ 
    \cline{3-6}
    &   I & 4 & 2 & 434 & 0 \\ 
    \cline{3-6}
    &   R & 0 & 8 & 6 & 402 \\ 
    \cline{3-6}
  \end{tabular}
  \vspace{0.3cm}
  \caption{Eleven trees, InfoGain criterion, 3.45\% error.}
  \end{subtable}
%
%
  %
  %
  \begin{subtable}[h]{0.45\textwidth}
  \centering
  \begin{tabular}{ll|l|l|l|l|}   
    \multicolumn{2}{c}{}&   \multicolumn{4}{c}{Predicted Classes}\\
    \multicolumn{2}{c}{} & 
    \multicolumn{4}{c}{{\rotatebox[origin=c]{0}{Q \ \quad}
      } {\rotatebox[origin=c]{0}{S \ \quad}
      } {\rotatebox[origin=c]{0}{I \ \quad}
      } {\rotatebox[origin=c]{0}{R}
      }}\\
    \cline{3-6}
    \multirow{4}{1em}{{\rotatebox[origin=c]{90}{Actual Classes}}} & Q & 765 & 9 & 2 & 2 \\ 
    \cline{3-6}
    &   S & 21 & 221 & 2 & 6 \\ 
    \cline{3-6}
    &   I & 7 & 2 & 431 & 0 \\ 
    \cline{3-6}
    &   R & 0 & 7 & 6 & 403 \\ 
    \cline{3-6}
  \end{tabular}
  \vspace{0.3cm}
  \caption{Twenty one trees, Gini criterion, 3.4\% error.}
  \end{subtable}
  \begin{subtable}[h]{0.45\textwidth}
  \centering
  \begin{tabular}{ll|l|l|l|l|}   
    \multicolumn{2}{c}{}&   \multicolumn{4}{c}{Predicted Classes}\\
    \multicolumn{2}{c}{} & 
    \multicolumn{4}{c}{{\rotatebox[origin=c]{0}{Q \ \quad}
      } {\rotatebox[origin=c]{0}{S \ \quad}
      } {\rotatebox[origin=c]{0}{I \ \quad}
      } {\rotatebox[origin=c]{0}{R}
      }}\\
    \cline{3-6}
    \multirow{4}{1em}{{\rotatebox[origin=c]{90}{Actual Classes}}} & Q & 762 & 11  & 2 & 3 \\ 
    \cline{3-6}
    &   S & 16 & 226 & 3 & 5 \\ 
    \cline{3-6}
    &   I & 4 & 2 & 434 & 0 \\ 
    \cline{3-6}
    &   R & 0 & 6 & 6 & 404 \\ 
    \cline{3-6}
  \end{tabular}
  \vspace{0.3cm}
  \caption{Twenty one trees, InfoGain criterion, 3.08\% error.}
  \end{subtable}
  \caption{Confusion matrix obtained using the leave-one-out approach, with the Gini 
   and InfoGain split criteria, and five, eleven, and twenty one trees in the ensemble. 
   The confusion matrix is a summary of the actual and predicted classes of each 
   instance in the data set.}
\label{tab:conf_matrix}
\end{table}

Next, in Table~\ref{tab:xvalid}, we show the results of five runs of
five-fold cross validation as the number of trees in the ensemble is
varied.  This table indicates that the InfoGain split criterion gives
slightly better accuracy than the Gini split criterion.  To obtain
greater insight, especially as there are unequal numbers of orbits of
each of the four types in the training data set, we consider the
leave-one-out metric that gives us the types of mis-classification
when a model, built with all but one instance, is used to predict the
class of that instance.  Table~\ref{tab:conf_matrix} lists the
confusion matrix for five, eleven, and twenty one trees in the
ensemble for the Gini and InfoGain split criteria. These results
indicate that both the number of trees and the split criterion have
relatively little effect on the overall accuracy of prediction.  It
also indicates that the largest numbers of mis-classifications are when
a separatrix orbit is mis-classified as a quasiperiodic orbit, and
vice-versa; this would likely be the case when the orbit is very thin
or has mild stochasticity. The use of InfoGain split criterion results
in fewer such mis-classifications.

Finally, Figure~\ref{fig:tree} shows a single tree created using all the 1884
instances in the training set, using the InfoGain split criterion. We
make several observations based on this tree:

\begin{itemize}

\item First, we observe that the initial splits on features f16 and f6
  create three distinct parts of the tree: the top part is a mix of
  quasiperiodic and separatrix orbits, followed by a part with a mix
  of separatrix and stochastic orbits, with the island chain orbits at
  the bottom of the tree. This confirms our observations that very
  thin separatrix orbits look like quasiperiodic at coarse scale and
  many of the separatrix orbits have some level of stochasticity.

\item The tree selects f16 and f6 as the most important features.
  Feature f16, indicating large angular gaps in the points, is an
  obvious choice as it is critical to identifying island chains. The
  choice of f6, which is the standard deviation of {\it lserr} across
  windows around valid points, allows quasiperiodic and very thin
  separatrix orbits, both of which have small values of f6, to be
  differentiated from other non-island orbits.

\item We also observe that some leaf nodes have a large number of
  instances of the majority class at the node. Two leaf nodes, with
  531 and 151 instances, account for 87.6\% of the quasiperiodic
  orbits. Similarly, two leaf nodes with 261 and 146 instances account
  for 92.5\% of the island chain orbits, and one leaf node with 347
  instances accounts for 83.4\% of the stochastic orbits. However, for
  the separatrix orbits, the instances are more scattered through out
  the tree, with the largest number at a leaf node being 99, or just
  40\% of the total separatrix orbits. This confirms that separatrix
  orbits have the greatest overlap with the other orbits, as we have
  observed visually.

\item The relatively small size of the tree, along with the leaf nodes with a
  large percentage of each of the four classes of orbits, indicate
  that the features used in the training set are able to discriminate
  among the classes successfully.

\item By focusing on the leaf nodes with a large number of orbits of
  one class, and the decisions made in the path to these nodes, we can
  understand how many of the labels are assigned. For example, a large
  number (531+17) of Q orbits are identified by very small values ($<$
  0.004) of f6, the standard deviation of {\it lserr} across windows
  around valid points. A majority of the stochastic orbits are
  identified by larger values of both f6 ($>$ 0.0254) and f8 ($>$
  0.8125), which is the maximum spread across windows. The key
  identifying feature for island chains is obviously f16. It is the
  separatrix orbits that take a deeper path through the tree as they
  share many similarities with both quasiperiodic and stochastic
  orbits.

\item The highly ranked features in Table~\ref{tab:featsel}, which
  were identified using the full data set, occur at levels closer to
  the root of the tree, suggesting they remain important when the subsets
  of the data at the nodes are considered.

\item The feature f14, identifying the fraction of peaks and valleys
  misaligned at coarse scale, does not appear in the tree, though its
  counterpart, f11, at fine scale, appears at a leaf node to
  distinguish a small number of quasiperiodic from island chain
  orbits. Both f11 and f14 are lower ranked features by order of
  importance in Table~\ref{tab:featsel} when the entire training data
  set is considered.

\end{itemize}

\begin{figure}[!htb]
\begin{center}
\begin{footnotesize}
\begin{verbatim}

                  f16 = 0:
                  :  f6 < 0.0254451:
                  :  :  f6 < 0.00402432:
                  :  :  :  f13 < 0.875: class Q (531/0)         <---------- Q
                  :  :  :  f13 >= 0.875:
                  :  :  :     f11 < 0.0618687: class Q (17/0)  <---------- Q 
                  :  :  :     f11 >= 0.0618687: class I (7/1)
                  :  :  f6 >= 0.00402432:
                  :  :     f4 < 0.000891132:
                  :  :     :  f5 < 0.0561031:
                  :  :     :  :  f12 < 0.654276:
                  :  :     :  :  :  f4 < 0.00048728:
                  :  :     :  :  :  :  f12 < 0.515885: class S (41/0)  <---------- S 
                  :  :     :  :  :  :  f12 >= 0.515885: class S (4/3)
                  :  :     :  :  :  f4 >= 0.00048728:
                  :  :     :  :  :     f12 < 0.416771:
                  :  :     :  :  :     :  f7 < 0.334634: class S (14/1)  <---------- S 
                  :  :     :  :  :     :  f7 >= 0.334634: class Q (3/0)
                  :  :     :  :  :     f12 >= 0.416771: class Q (8/0)
                  :  :     :  :  f12 >= 0.654276: class Q (21/0)    <---------- Q 
                  :  :     :  f5 >= 0.0561031: class R (8/0)
                  :  :     f4 >= 0.000891132:
                  :  :        f15 < 0.363317:
                  :  :        :  f3 < 0.0837068: class Q (4/1)
                  :  :        :  f3 >= 0.0837068: class S (12/0)   <---------- S
                  :  :        f15 >= 0.363317:
                  :  :           f6 < 0.0142841: class Q (151/1)   <---------- Q
                  :  :           f6 >= 0.0142841:
                  :  :              f5 < 0.0638765:
                  :  :              :  f10 < 0.1125: class S (4/0)
                  :  :              :  f10 >= 0.1125:
                  :  :              :     f3 < 0.101987: class Q (10/0)
                  :  :              :     f3 >= 0.101987: class S (3/1)
                  :  :              f5 >= 0.0638765: class Q (23/0)    <---------- Q
                  :  f6 >= 0.0254451:
                  :     f8 < 0.8125:
                  :     :  f4 < 0.00613825:
                  :     :  :  f7 < 0.195314: class I (5/0)
                  :     :  :  f7 >= 0.195314:
                  :     :  :     f15 < 0.481: class S (99/0)   <---------- S
                  :     :  :     f15 >= 0.481:
                  :     :  :        f8 < 0.6875:
                  :     :  :        :  f9 < 0.125: class S (34/0)  <---------- S
                  :     :  :        :  f9 >= 0.125: class S (9/3)
                  :     :  :        f8 >= 0.6875: class R (7/0)
                  :     :  f4 >= 0.00613825:
                  :     :     f13 < 0.166667: class S (14/3)
                  :     :     f13 >= 0.166667:
                  :     :        f8 < 0.625: class S (8/0)
                  :     :        f8 >= 0.625:
                  :     :           f3 < 0.501971: class S (2/3)
                  :     :           f3 >= 0.501971: class R (43/1)  <---------- R
                  :     f8 >= 0.8125: class R (347/1)     <---------- R          
                  f16 = 1:
                     f10 < 0.121403: class I (261/0)  <---------- I
                     f10 >= 0.121403:
                        f5 < 0.239159: class I (146/4)   <---------- I
                        f5 >= 0.239159:
                           f9 < 0.125: class R (5/0)
                           f9 >= 0.125: class I (19/1)    <---------- I

\end{verbatim}
\end{footnotesize}
\end{center}
\caption{Single tree created using the entire data set, InfoGain split
  criterion. The data set has 1884 instances - 778 quasiperiodic, 250
  separatrix, 440 island chains, and 416 stochastic (indicated by R)
  orbits. The node leaves with a large number of orbits of one class
  have been highlighted with the class initial. Together, they
  comprise 743 (95\%) of the quasiperiodic orbits, 200 (80\%) of
  separatrix orbits, 426 (97\%) of the island chain orbits, and 386
  (94\%) of the stochastic orbits. }
\label{fig:tree}
\end{figure}

\afterpage{\clearpage}

%
\section{Conclusions}
\label{sec:conc}
%

In this paper, we considered the task of assigning a class label to
orbits in a Poincar\'e plot, where each orbit is represented by the
coordinates of a set of two-dimensional points and the class label
indicates the shape formed by the points.  Though the data set is
small, and the task is ideally suited for a machine learning solution,
creating a high-quality training data set is challenging.  We
presented an approach that iteratively refines the class assignment
and the features extracted to represent each orbit.  Our decision tree
based approach results in less than 5\% error rate, demonstrating that
an automated machine learning approach can replace a tedious,
error-prone, subjective visual classification of orbits, providing
plasma physicists a useful tool for analysis of simulation data.

%
\section{Acknowledgment}
\label{sec:ack}
%

I would like to thank Scott Klasky, from ORNL, for introducing us to
this problem that appeared deceptively simple, but posed several
unique challenges to a solution. This work would not have been
possible without the data and domain expertise provided by Joshua
Breslau from PPPL.  Erick Cant\'u-Paz proposed the use of polar
coordinates to magnify the radial variation in an orbit. Siddharth
Manay created the Poincar\'e plot schematic in
Figure~\ref{fig:pplot_schematic}(a), as well as the initial codes for
fitting the second degree polynomial.

LLNL-TR-834167 This work performed under the auspices of the U.S.
Department of Energy by Lawrence Livermore National Laboratory under
Contract DE-AC52-07NA27344. This work was part of the SDM Scidac-II
Center, funded by the Office of Science, US Department of Energy.

\clearpage
\bibliographystyle{acm}
\bibliography{ms_arxiv}

\end{document}